\documentclass{raa}
\usepackage{graphicx,times}             
\usepackage{natbib}
\usepackage{amssymb,amsmath}
\usepackage{mathrsfs}
\bibpunct{(}{)}{;}{a}{}{,}
\usepackage[a4paper=true,dvipdfm=true,pagebackref=true]{hyperref}
\hypersetup{colorlinks = true, linkcolor = red, anchorcolor = red, citecolor = blue, filecolor = red, pagecolor = pink, urlcolor = red}

\begin{document}

   \title{The Origin of The Soft X-Ray Excess in the Seyfert 1.5 Galaxy ESO 362-G18}

   \volnopage{Vol.0 (20xx) No.0, 000--000}      
   \setcounter{page}{1}          

   \author{Xiao-Gu Zhong
      \inst{1,2,3}
   \and Jian-Cheng Wang
      \inst{1,2}
     }

   \institute{Yunnan Observatories, Chinese Academy of Sciences,  Kunming 650216, China; {\it guqian29@ynao.ac.cn}\\
        \and
        Key Laboratory for the Structure and Evolution of Celestial Objects, Chinese Academy of Sciences,  Kunming 650216, China\\
        \and
        University of Chinese Academy of Sciences, Beijing 100049, China\\
\vs\no
  }

\abstract{
We review the Seyfert 1.5 Galaxy ESO 362-G18 for exploring the origin of the soft X-ray excess. The Warm Corona and Relativistic Reflection models are two main scenarios to interpret the soft X-ray excess in AGNs at present. We use the simultaneous X-ray observation data of XMM-Newton and NuSTAR on Sep. 24th, 2016 to perform spectral analysis in two steps. First, we analyze the time-average spectra by using Warm Corona and Relativistic Reflection models. Moreover, we also explore the Hybrid model, Double Reflection model and Double Warm Corona model. We find that both of Warm Corona and Relativistic Reflection models can interpret the time-average spectra well but cannot be distinguished easily based on the time-averaged spectra fit statistics. Second, we add the RMS and covariance spectra to perform the spectral analysis with time-average spectra. The result shows that the warm corona could reproduce all of these spectra well. The the hot, optical thin corona and neutral distant reflection will increase their contribution with the temporal frequency, meaning that the corona responsible for X-ray continuum comes from the inner compact X-ray region and the neutral distant reflection is made of some moderate scale neutral clumps.
}

   \authorrunning{X.-G. Zhong and J.-C. Wang}            
   \titlerunning{The Origin of The Soft Excess in ESO 362-G18}  

   \maketitle

\section{Introduction}           
\label{sect:intro}
Matter accreting into super massive black holes (SMBHs) at the center of active galaxies is one of the most efficient mechanisms to emit electromagnetic radiation from gravitational potential energy, which covers the broad band spectrum from radio to gamma-ray. Particularly, X-ray can be used to probe the physical process of most internal regions of the accretion disk near the black hole. Generally, the X-ray continuum is considered to be from the inverse Compton scattering of the soft photons emitted by accretion disk in the hot corona. But, the details of the geometry of the corona and the radiation mechanism of disk-corona are not well understood. Especially, below 2 keV in the X-ray band, the X-ray data are above the low energy extrapolation of the best fitting continuum within 3-10 kev power law by index about 2, this is so-called soft X-ray excess. The soft X-ray excess is a major component of many active galactic nuclei (AGNs). \citet{2004A&A...422...85P} found that about $90\%$ of the quasars in their sample exhibit significant soft X-ray excess. However, its origin has been debated over the years, and therefore, the study of soft X-ray excess is significant to discover the detail of the radiation mechanism within the innermost region of the accretion disk. Historically, the soft X-ray excess was first believed to be the hard tail of UV black-body emission from the accretion disk \citep{1999ApJS..125..317L}. However, this explanation was ruled out because the temperature is difficult to maintain within the range of 0.1-0.2 keV for many AGNs with different masses and accretion rates.

Currently, there are some models to explain the soft X-ray excess, for example, warm corona \citep{1987ApJ...321..305C, 2007ASPC..373..121D, 2009MNRAS.398L..16J, 2018A&A...611A..59P}, relativistic reflection \citep{2002MNRAS.331L..35F, 2006MNRAS.365.1067C, 2020MNRAS.497.4213G, 2021MNRAS.501..916J}, warm absorbtion \citep{2016MNRAS.457..875P} and magnetic reconnection \citep{2013ApJ...773...23Z}, in which the warm corona and relativistic reflection models are favored by many previous works.

In the warm corona case, the accretion disk photons are Comptonized by a warm (the electron temperature is below 1 keV), optical thick corona (the optical depth is between about 10 and 40.), which is different from the hot (the typical value of electron temperature is 300 keV \citep{2001MNRAS.328..501P}), optical thin corona (the optical depth is far less than one) that are responsible for the primary X-ray continuum. The soft X-ray excess is the high energy tail of the resulting Comptonized spectrum. Generally, the primary X-ray continuum is a cutoff power-law spectra with a typical index, $\Gamma \sim1.8$, in unobscured AGNs. The physical origin of the warm corona is unknown so far. The magnetic field could play a crucial role to transport the substantial amounts of energy from the disk to the warm corona vertically, then heat up the electrons in the warm corona \citep{2000MNRAS.313..193M, 2006ApJ...640..901H, 2015ApJ...809..118B}. Recently, the radiative transfer computation in hydrostatic and radiative equilibrium showed that the magnetic dynamo could heat the upper layers of the accretion flow up to a few keV for optical depths as great as ten \citep{2020A&A...633A..35G}. \citet{2019ApJ...871...88G} put forward that the photoelectric absorption is dominant in optically thick regions by considering the coronal and photoionization equilibrium. Therefore, a forest of absorption lines are predicted in the soft X-ray spectra if the soft X-ray excess comes from the warm corona. If true, the warm corona would invalidate as an origin of the soft X-ray excess. However, \citet{2020A&A...634A..85P} and \citet{2020MNRAS.491.3553B} propose that these absorption lines could be eliminated by providing sufficient internal mechanical heating into the warm corona because the electron scattering is dominant instead of absorption opacity.

In the relativistic reflection case, a multitude of fluorescent atomic lines produced in the disk which is illuminated by the hot corona are relativistically blurred due to the proximity to the black hole. For example, the broad Fe K$\alpha$ line is the prominent result of the 'blurring' effect. These relativistic broadening or 'blurring' lines below 2 keV will generate soft X-ray excess spontaneously and they are generally associated with the innermost stable orbit radius of disk which is determined by the black hole spin. Therefore, the relativistic reflection could be used to study the black hole spin. Another associated prominent reflection feature is the 'Compton hump' which comes from the Compton down-scattering of coronal high-energy photons reprocessed in the accretion disk or distant matter \citep{1991MNRAS.248..760N} and achieves the peak in the reflection spectrum at about 30keV depended on the column density as well as the geometry and inclination angle. The relativistic reflection model generally requires extreme values for the spin and hot corona compactness as well as high ionization degree. The large densities of the disk surface will enhance the soft X-ray excess \citep{2019MNRAS.489.3436J, 2019MNRAS.484.1972J}, which also significantly reduce the inferred iron abundances compared to typical (e.g., with density generally equal to $10^{15}$ cm$^{-3}$) reflection models (e.g. \citet{2018ApJ...855....3T}).

In this work, we review the Seyfert 1.5 galaxy ESO 362-G18 (a.k.a. MCG 05-13-17) with redshift $z=0.012$ \citep{2006A&A...456..953B}. \citet{2014MNRAS.443.2862A} (hereafter AG14) reported the black hole with a mass of $(4.5\pm1.5)\times 10^7M_\odot$ inferred by the virial relationship and the disk inclination of $i=53^{\circ}\pm5^{\circ}$ obtained by using a partially ionized reflection model convolved with the $KERRCONV$ relativistic kernel. However, \citet{2021ApJ...913...13X} (hereafter Xu21) report that the inclination is a lower value by using another relativistic reflection model, a sub-version of $relxill$ \citep{2014ApJ...782...76G, 2014MNRAS.444L.100D} assuming a lamppost geometry. The contradiction of the disk inclination could come from parameters degeneracy because of using different models. They propose that the black hole has a high spin$\sim0.998$ inferred by the different relativistic reflection model. But Xu21 put forward that the inner disk has a high density ($log[n_e/cm^{-3}]>18.3$) which will enhance the soft X-ray excess. AG14 propose that the X-ray radiation region locates within 50 gravitational radii ($r_g=GM/c^2$) by analyzing the absorption variability with 2005-2010 multi-epoch X-ray observations. Xu21 analyzed the properties of X-ray within 0.3-79 keV bands using the simultaneous X-ray observation data of XMM-Newton and NuSTAR on Sep. 24th, 2016. They find that the warm corona and the relativistic reflection model cannot be distinguished easily based on time-averaged spectra fit statistics (the same situations are also seen in, for example these recent works, \citet{2019ApJ...871...88G} for Mrk509 and \citet{2020ApJ...896..160L} for Fairall 9). However, they consider that the relativistic reflection scenario could be accepted due to its reasonable parameters, for example, the X-ray continuum index was inferred as 1.74 which is agreed with the typical value $\sim 1.8$ for unobscured AGNs and without a hypothetical component such as the warm corona. They also find that the spectrum will be decoupled outside the coverage of 0.5-20 keV, and then suggest that improving theoretical models and future observations covering wider energy bands with high resolution are essential for exploring the nature of the soft X-ray excess.

The paper is organized as follows. In Section 2, we present the observational data. In Section 3, we analyse the time properties. In Section 4, we perform the spectral analysis. Finally, we discuss our results and conclude in Section 5.

\section{Observation Data}
\label{sec:Observation Data}
We use the simultaneous X-ray observation data of NuSTAR \citep{2013ApJ...770..103H} and XMM-Newton \citep{2001A&A...365L...1J} on Sep. 24th, 2016 to perform the spectral analysis for ESO 362-G18. The observation ID are 60201046002 for NuSTAR with 100ks net exposure and 0790810101 for XMM-Newton with 120ks net exposure. The observation information we used in the paper are shown in Table 1.

\begin{table*}
\begin{center}
\begin{tabular}{ccccc}
  \hline
  \hline
  Telescope   &Instrument         &Obs. ID           &Date         &Net Exp. (ks)  \\
  \hline
  NuSTAR      &FPMA/B             &60201046002       &2016-09-24   &102            \\
  XMM-Newtion &EPIC               &0790810101        &2016-09-24   &121            \\
  \hline
\end{tabular}
\caption{Observations Log for ESO 362-G18.}
\end{center}
\end{table*}

\subsection{NuSTAR Data Reduction}
The reduction of the NuSTAR data is conducted following the standard procedures using the NuSTAR Data Analysis Software (NUSTARDAS v.2.0.0). We use $nupipeline$ and the 20210427 version of NuSTAR CALDB to produce clean event files. We use the 'nucalcsaa' module to calculate South Atlantic Anomaly (SAA) and set the input parameter 'saacalc=2', 'saamode=STRICT' to choose the strict mode. The circular source region is a radius of 60 arcsec. In order to avoid the background dominance, we select eight circular regions with repeating attempts to extract the background spectra . Fig.1 shows the NuSTAR FPMA event image for extracting the source and background spectrum, and the total area of background is larger than the source. We then utilize the task package $nuproducts$ to extract the source and background light curves (shown in top two panels of Fig.2) and spectra. The spectra are grouped to at least 30 counts in each bin in order to have high signal-to-noise ratio.

\begin{figure}
\centerline{
    \includegraphics[scale=0.45,angle=0]{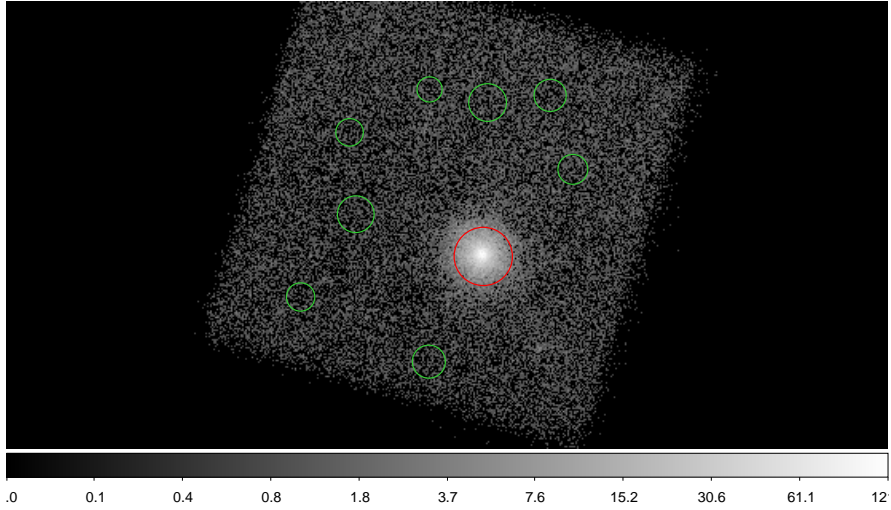}}
  \caption{NuSTAR FPMA event image of observation ID 60201046002 (ESO 362-G18) displayed in ds9. The source spectrum is extracted from the red circle region and the background spectrum is extracted from these eight green circles shown in ds9.}
\end{figure}

\begin{figure}
\centerline{
    \includegraphics[scale=0.45,angle=0]{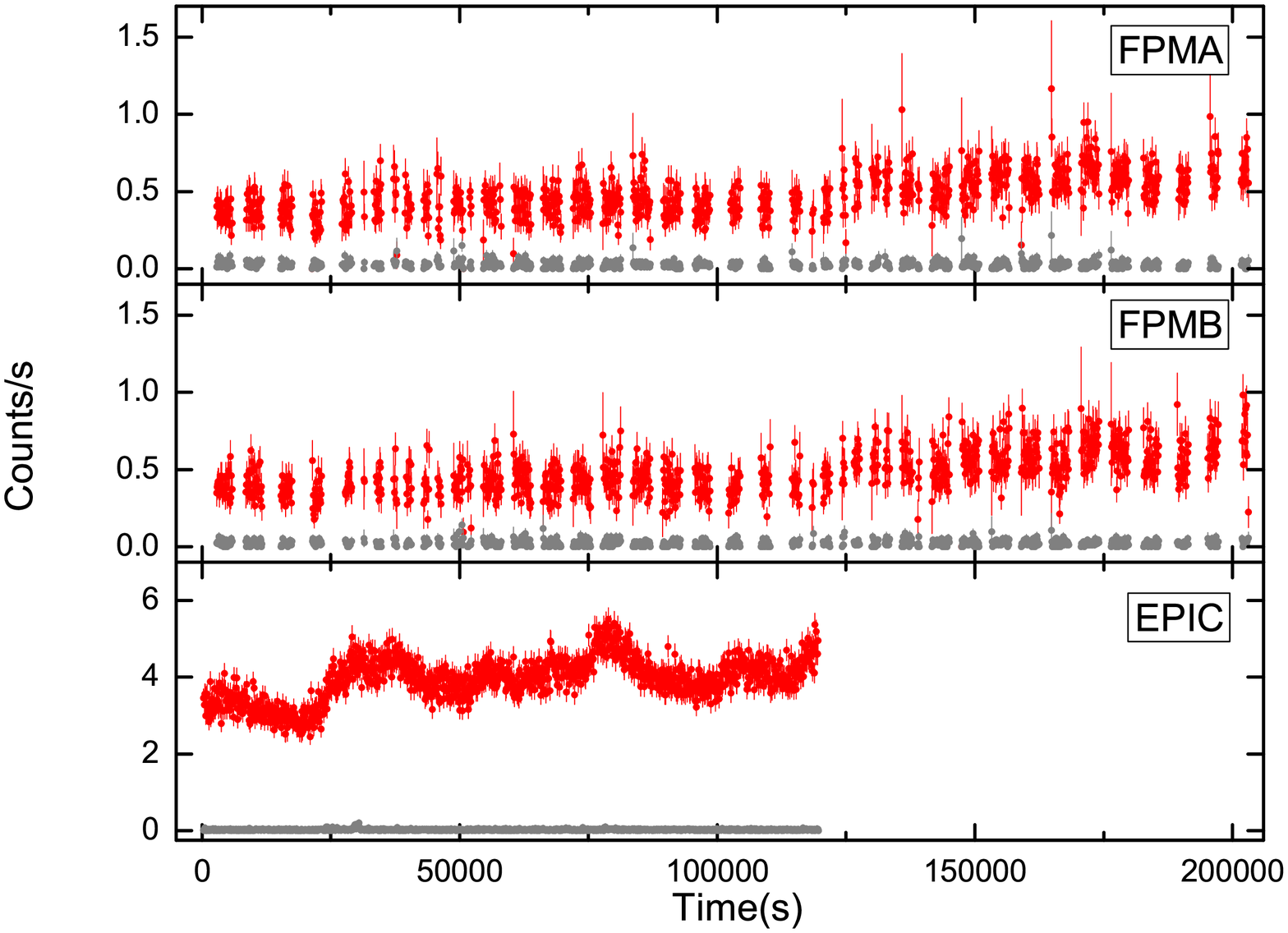}}
  \caption{The light curve of NuSTAR FPMA, FPMB and XMM-Newton EPIC from top to bottom panel. The red and grey data represent source and background light curves,  respectively.}
\end{figure}

\subsection{XMM-Newton Data Reduction}
For XMM-Newton, we only use EPIC-pn data in 0.3-10 keV to do the spectral analysis. The raw data are processed from Observation Data Files following standard procedures based on the Science Analysis System (SAS v16.1.0) and the latest calibration files. The spectra and light curve are extracted using tool $evselect$ with default pattern selected. We extract the source spectra and light curve from a circular region with the radius of 35 arcsec centered on the source. The background spectra are taken from a circular region with the same size near the source but excluding source photons. $rmfgen$ and $arfgen$ are used to produce response matrices. Source spectra are rebinned by $grppha$ with a minimum of 30 counts per bin. $epiclccorr$ is used to correct the light curve (shown in bottom panel of Fig.2). We test the pile-up for these data by using $epatplot$ and find that the pile-up is not important to ESO 362-G18.
\section{Time Analysis}
We analyse the variability properties by only using the XMM-Newton observation data because NuSTAR is a near-earth satellite with the altitude $\sim$650 km $\times$ 610 km, its data is discontinuous due to the occlusion by Earth every 5 ks (seen in top two panel of Fig.2) to distort the power spectral density (PSD). We calculate PSD by choosing a bin time of $400s$ to ensure that there are no zero-count bins contained in the light curves \citep{2021MNRAS.500.2475J}. The PSD can be calculated from the periodogram \citep{1997ApJ...474L..43V, 1999ApJ...510..874N},

\begin{equation}
|X_n|^2 = X_n^{*} X_n,
\end{equation}
where $X_n$ is the discrete Fourier transform (DFT),
\begin{equation}
X_n = \sum_{k = 0}^{N - 1} x_k \exp\left( 2\pi i n k / N \right),
\end{equation}
at each Fourier frequency, $f_n=n/(N\Delta t)$, where $n=1,2,3...N/2$, N is the number of tine bins of width $\Delta t$. The maximum frequency is the Nyquist frequency, $f_{max}=1/(2\Delta t)$. $x_k$ is the kth value of the light curve. The asterisk denotes complex conjugation. Then, the normalised PSD could be calculated as follows:
\begin{equation}
P_n=\frac{2\Delta t}{\langle x \rangle^{2} N} |X_n|^{2},
\end{equation}
where $\langle x \rangle$ is the average count rate in the light curve. We then calculate the time lag between the soft band 0.3-1 keV and hard band 1-4 keV by estimating the Fourier cross spectrum between the two energy band light curves $x(t)$ and $y(t)$ with DFTs, $X_n$ and $Y_n$ \citep{2010MNRAS.401.2419Z, 2013MNRAS.434.1129K}, as follows:
\begin{equation}
C_{XY,n}=X_n^{*} Y_n,
\end{equation}
where $C_{XY,n}$ is the Fourier cross spectrum. The time lag between the two energy bands is estimated by
\begin{equation}
\tau(\nu_{j}) = \phi(\nu_{j})/(2\pi \nu_{j}),
\end{equation}
where $\phi(\nu_{j})=arg[C_{XY,n}]$ is the phase angle of the Fourier cross spectrum and $\nu_{j}$ is the centre value of the frequency bin. The time lag between the soft band 0.3-1 keV and hard band 1-4 keV is shown in Fig.3.

\begin{figure}
\centerline{
  \includegraphics[scale=0.45,angle=0]{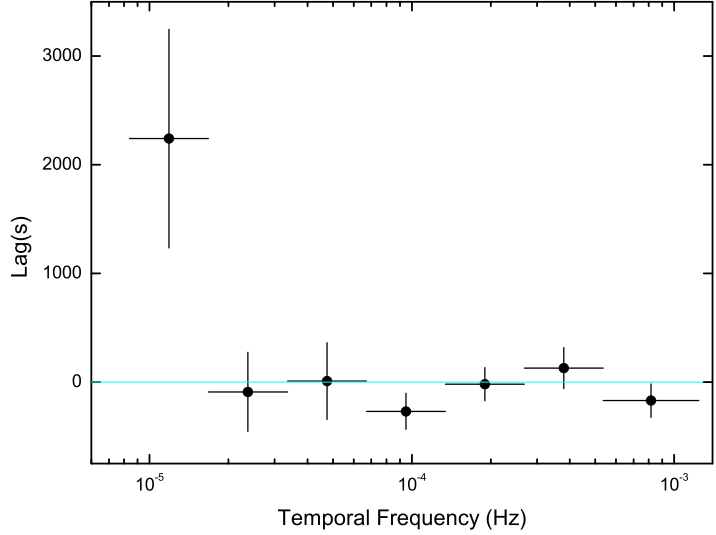}}
  \caption{The frequency-dependent lags between the hard band at 1-4 $keV$ and the soft band at 0.3-1 $keV$.}
\end{figure}

We find that the hard band lags behind the soft band, called as hard-lag, at lower frequencies. However, the soft band lags behind the hard band, called as soft-lag, at higher frequency except $[1.3-5.4]\times10^{-4}$ Hz, but this lag could be unreal due to its low signal-noise ratio. The soft-lag at $[0.5-1.3]\times10^{-4}$ Hz is distinct, which is consistent with the work of AG14 (seen Fig.9 in AG14). It is noted that the lag is significant when two energy bands have prominent correlation. Therefore, we test the coherence of two energy bands , $\gamma^2$, defined as \citep{1999ApJ...510..874N, 2010MNRAS.401.2419Z}:
\begin{equation}
\gamma^{2}(\nu_{j}) = \frac{|\bar{C}_{XY}(\nu_{j})|^{2}}{\bar{P}_{X} (\nu_{j}) \bar{P}_{Y}(\nu_{j})},
\end{equation}
where the bar above parameters means the average value with the frequency bin. The coherence between the two energy bands is shown in Fig.4, we select lager frequency bin at the lowest frequency bin to ensure statistical efficiency and we find that there are good correlation at low frequency bands, especially $\gamma^2 \sim 1$ at the lowest frequency bin, which implies a small error bars. In contrast, the higher frequency have bad correlation, especially $\gamma^2 \sim 0$ at the highest frequency bin, which also implies a small error bars.

\begin{figure}
\centerline{
  \includegraphics[scale=0.45,angle=0]{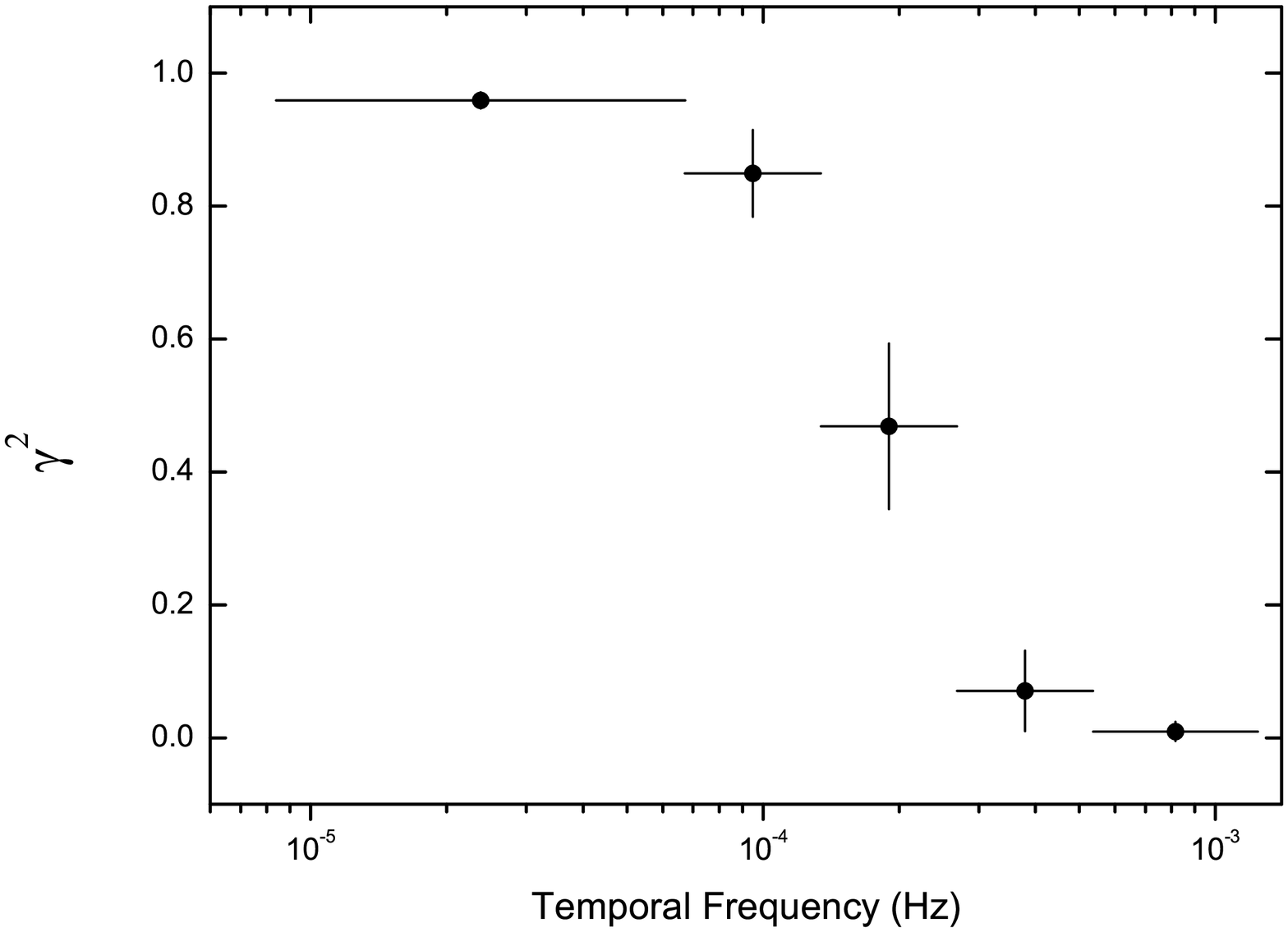}}
  \caption{The frequency-dependent coherence between the hard band at 1-4 $keV$ and the soft band at 0.3-1 $keV$.}
\end{figure}

In order to analyze the energy-dependent variability in the general, we focus on three frequencies, the low frequency ($[1-2]\times10^{-5}$Hz), the middle frequency ($[0.5-1.3]\times10^{-4}$Hz) and the high frequency ($[0.5-1.3]\times10^{-3}$Hz). The energy bin we adopt is 100 eV if we don't state otherwise. The Fig.5 shows the energy-dependent coherence of low, middle and high frequencies for each energy bin, the reference energy band we selected is the whole energy band of XMM-Newton data, $0.3$-$10$ keV, but excluding the interested energy bin themselves. It displays that the correlation decreases from low to high frequency and from low to high energy band in low and middle frequencies. The high frequency has a poor correlation.

\begin{figure}
\centerline{
  \includegraphics[scale=0.23,angle=0]{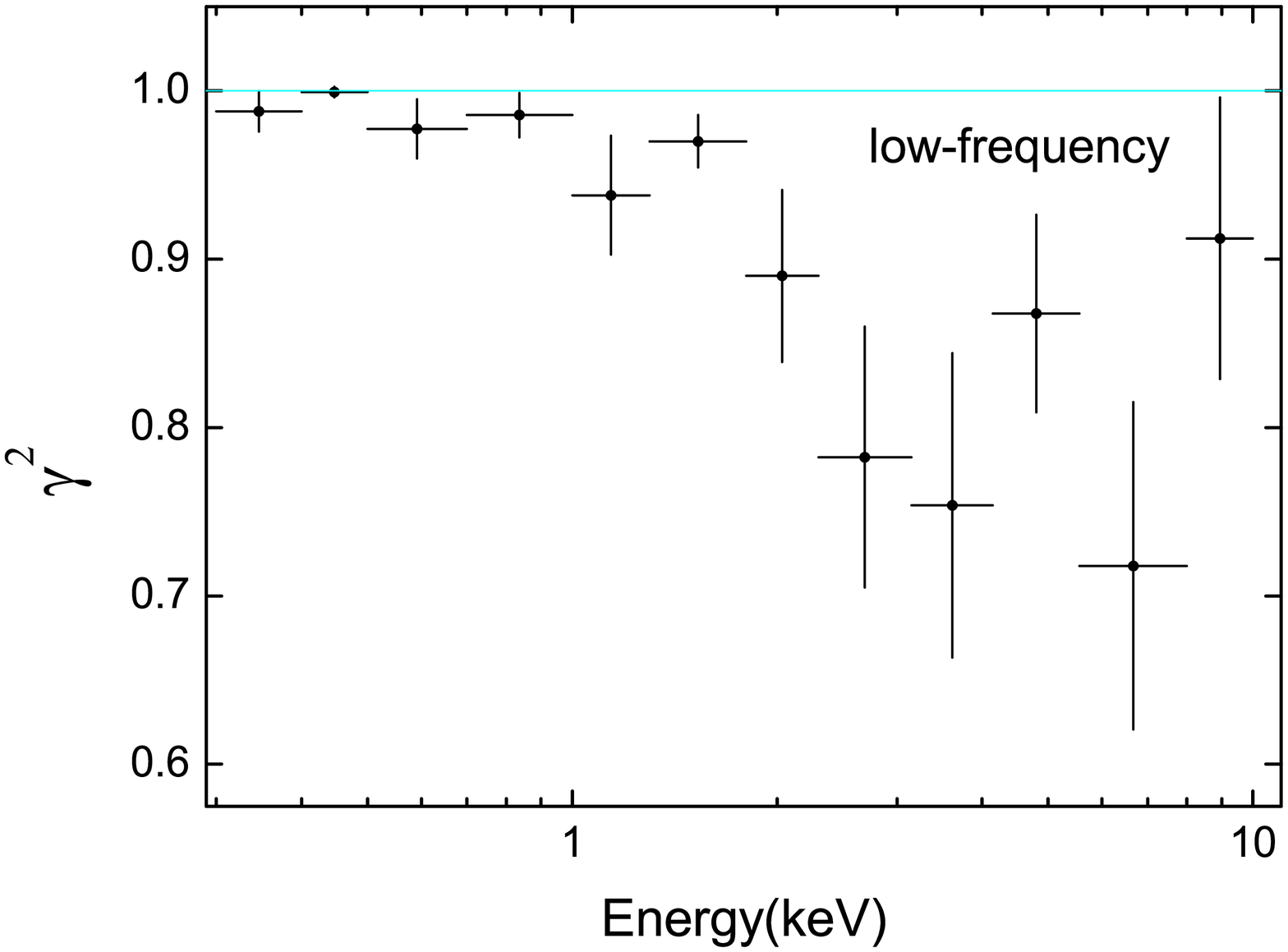}
  \includegraphics[scale=0.23,angle=0]{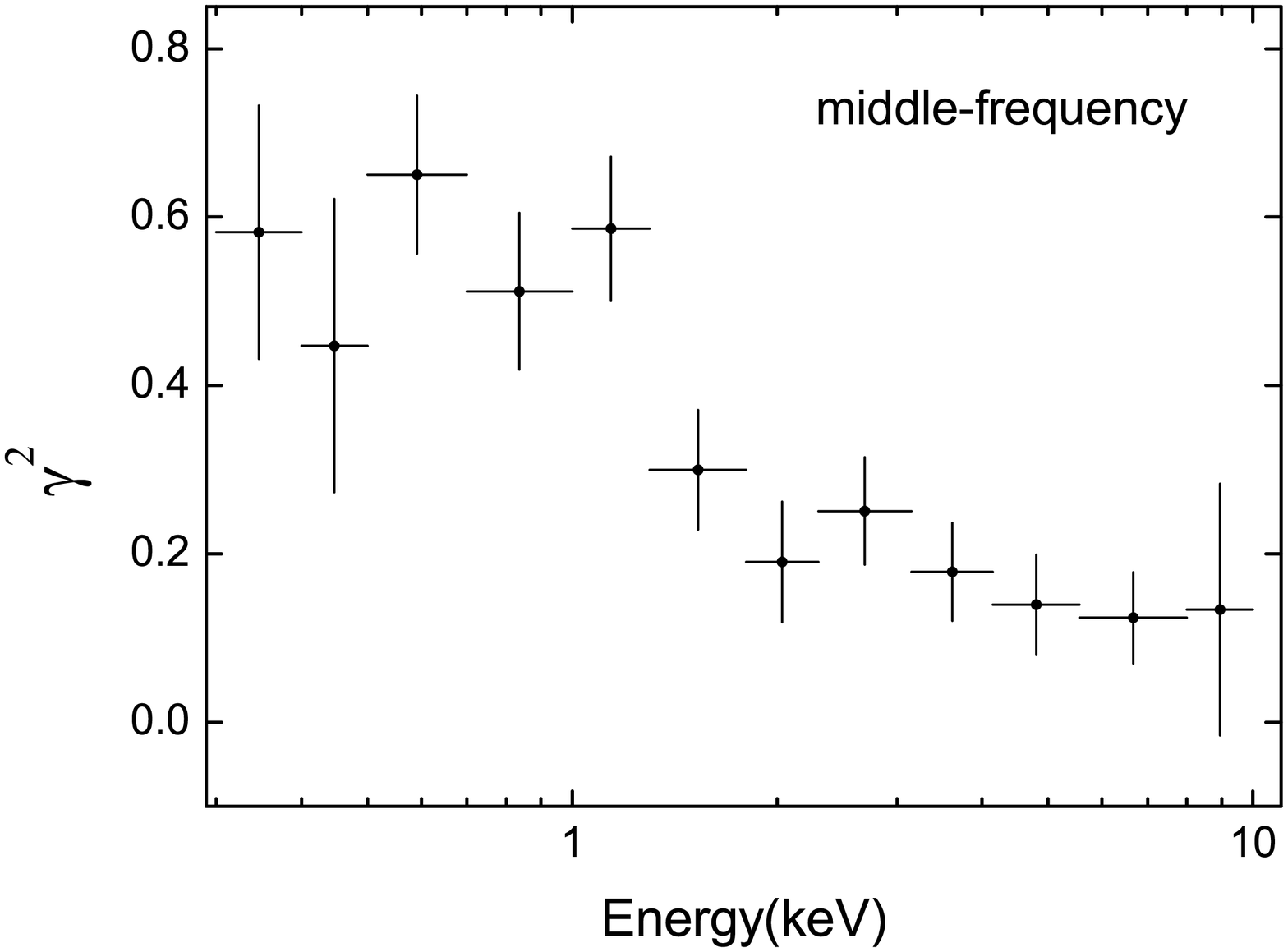}
  \includegraphics[scale=0.23,angle=0]{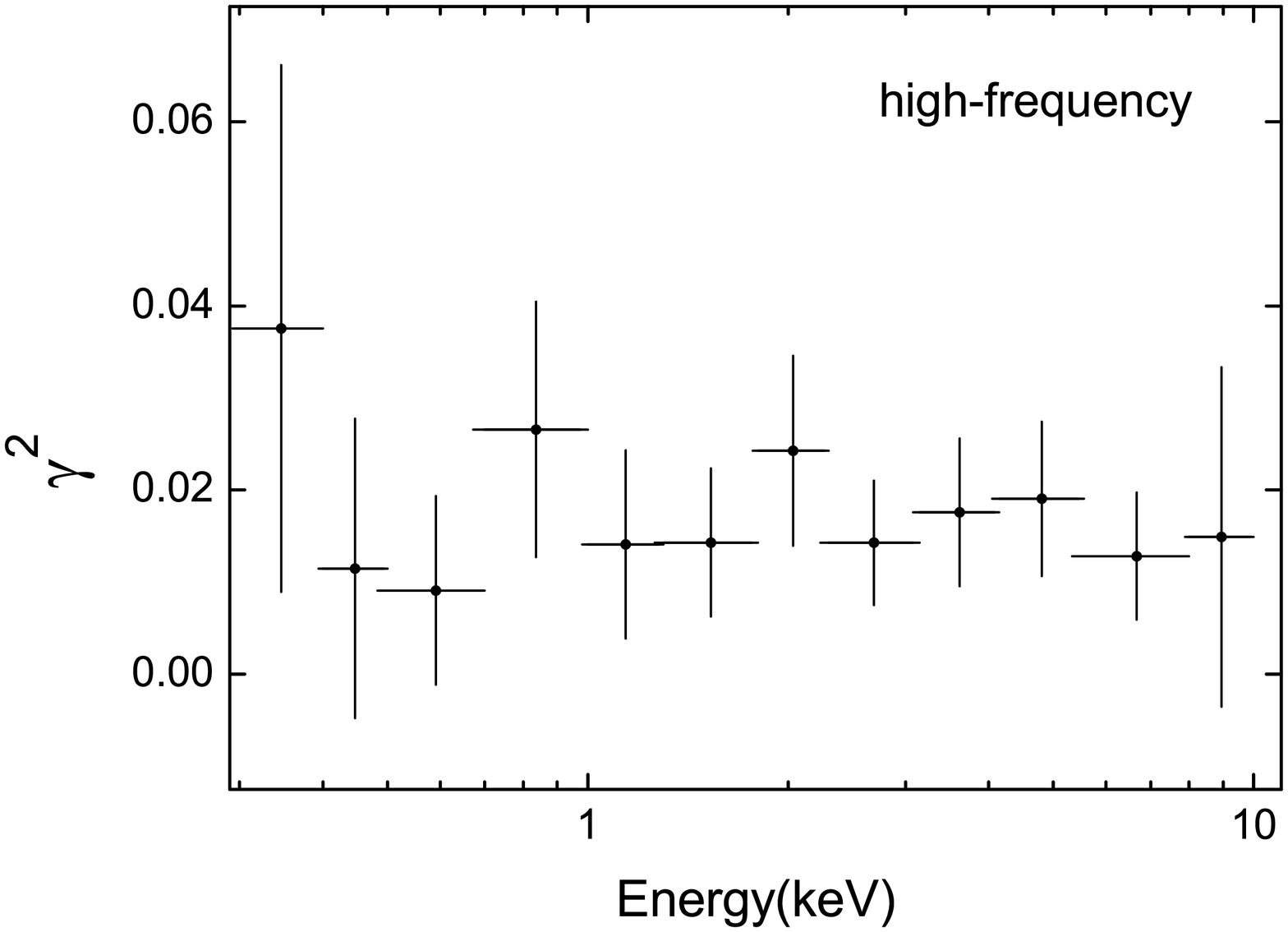}}
  \caption{The energy-dependent coherence of low, middle and high frequencies for each energy bin from left to right panel. The reference energy band is the whole energy band of XMM-Newton data, 0.3-10 keV, but excluding the interested energy bin themselves to reduce self-correlation.}
\end{figure}

Fig.6 shows more detail about the time lags of low, middle and high frequencies for each energy bin, the reference energy band is also the whole energy band of XMM-Newton data, $0.3$-$10$ keV, but excluding the interested energy bin themselves to reduce self-correlation. It displays a clear hard-lag at the low frequency. It is indicated that the variability is propagated from soft to hard energy band in a larger scale. There are no clear lags below 3 keV in middle frequency, but it has some leading at 3-5 keV and then has some lag at a high energy, about 8-10 keV, which may show the reverberation characteristics because 3-5 keV energy band represents the X-ray continuum radiated from hot corona to illuminate the disk, and the lag at 8-10 keV could be a tail of 'Compton hump' that is a reflection feature. The high frequency doesn't show any significant lags.

\begin{figure}
\centerline{
  \includegraphics[scale=0.23,angle=0]{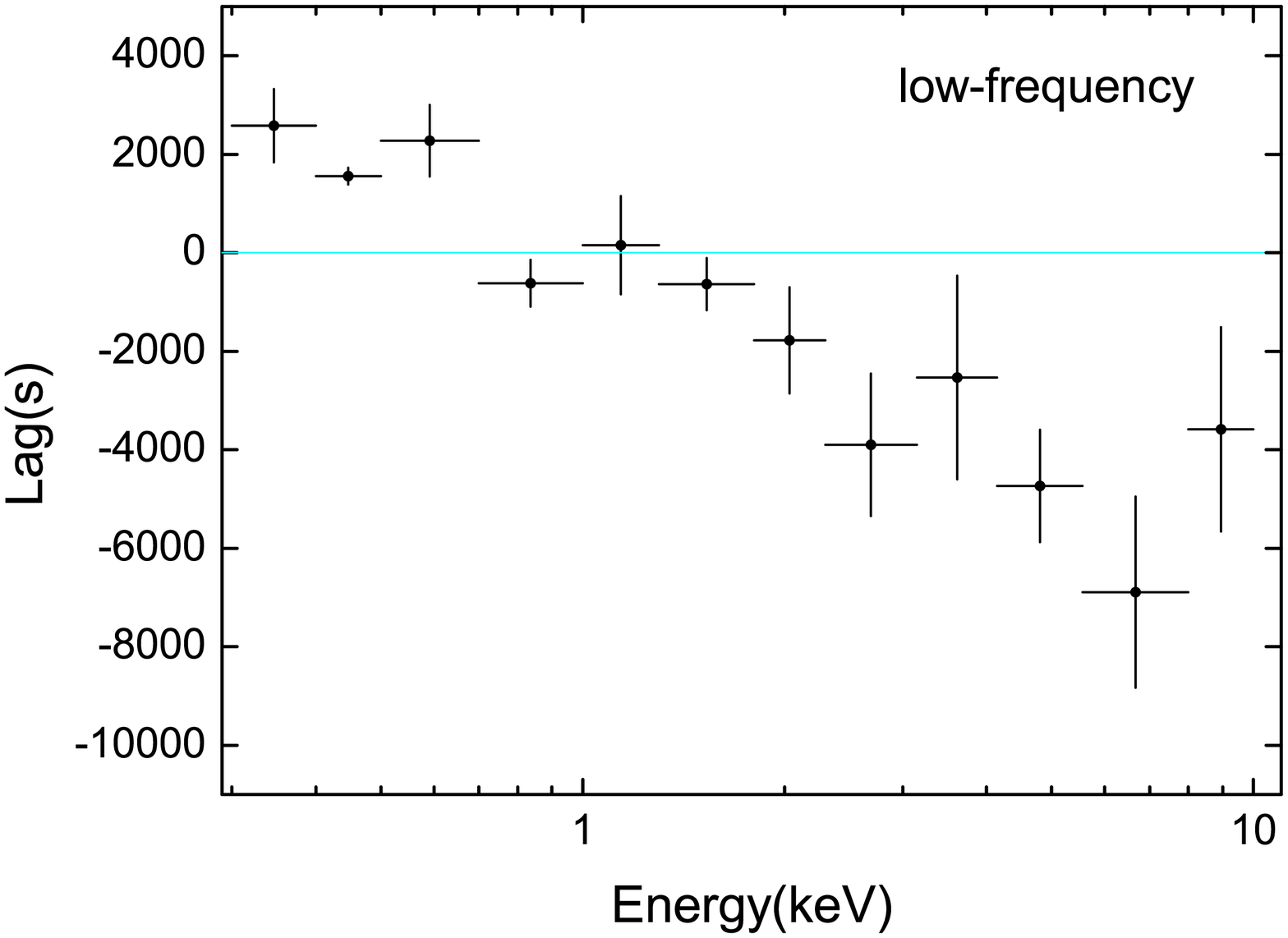}
  \includegraphics[scale=0.23,angle=0]{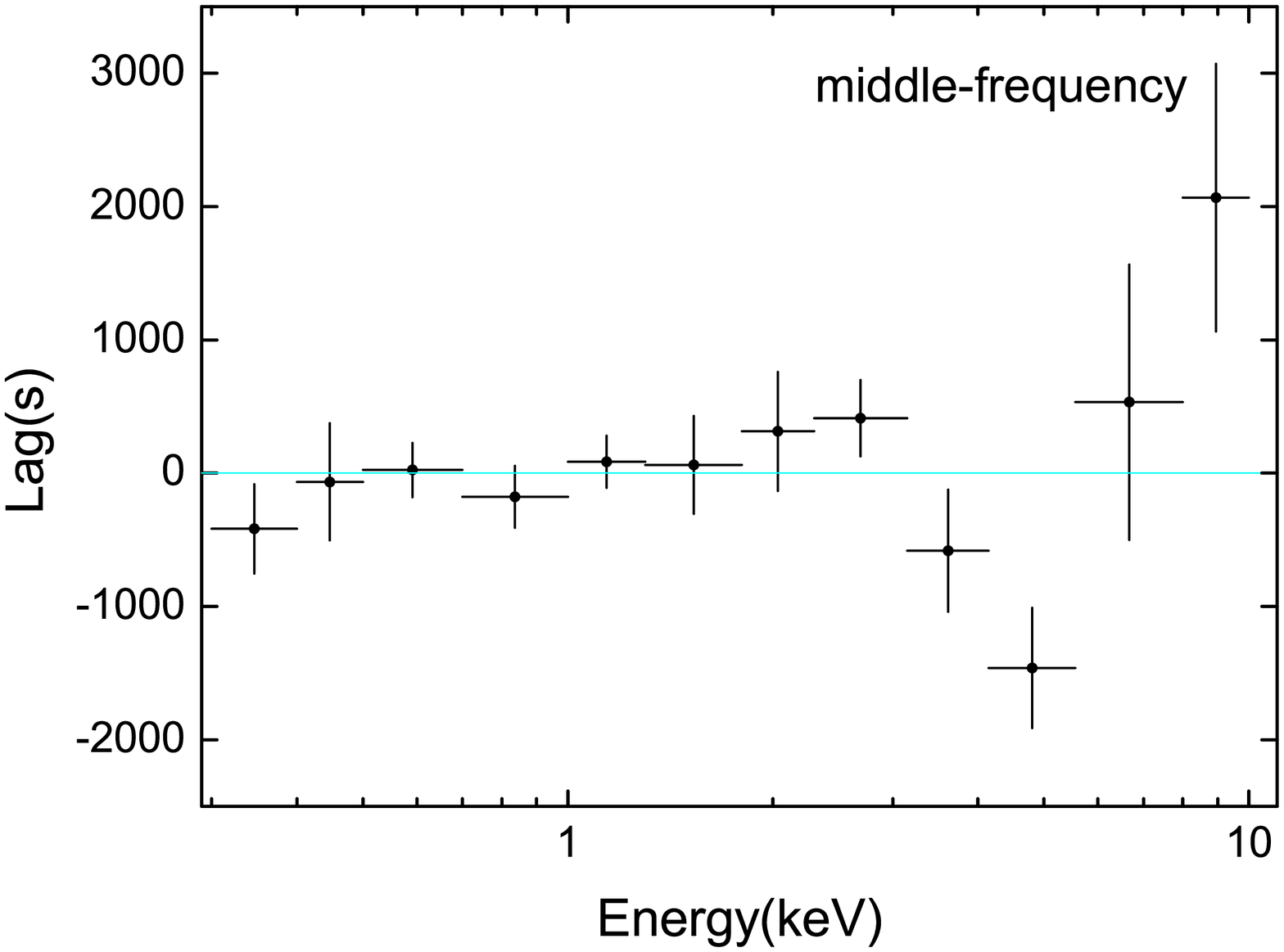}
  \includegraphics[scale=0.23,angle=0]{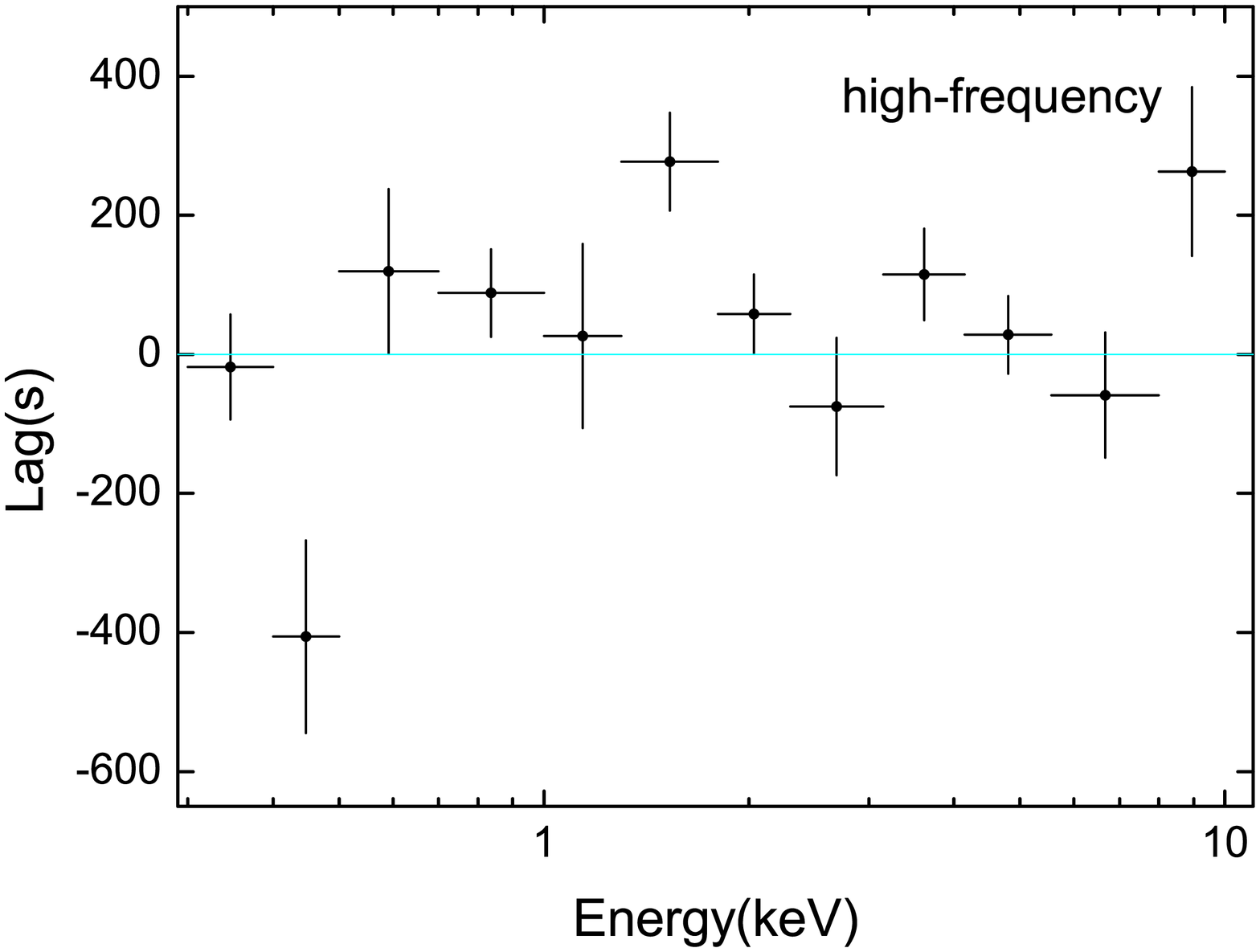}}
  \caption{The energy-dependent lags of low, middle and high frequencies for each energy bin from left to right panel. The reference energy band is the whole energy band of XMM-Newton data, 0.3-10 keV, but excluding the interested energy bin themselves to reduce self-correlation.}
\end{figure}

The root mean square (rms) describes the degree of the variability and is classified into fractional rms and absolute rms (hereafter we use the capital, RMS, represents the absolute rms), where fractional rms is defined as  \citep{2003MNRAS.345.1271V}:
\begin{equation}
fractional \ rms(\nu_{j})=\sqrt{(P(\nu_{j}) - P_{noise})\Delta \nu_{j}},
\end{equation}
where $P_{\rm noise}=\langle \Delta x^{2} \rangle / \left( \langle x \rangle^{2} f_{\rm Nyq} \right) $, $\langle \Delta x^{2} \rangle$ is the average of the squared error-bars of the light curve, $f_{\rm Nyq}$ is the Nyquist frequency. $\Delta \nu_{j}$ is the frequency bin. RMS could be obtained from fractional rms multiplied by the average count rate, $\langle x \rangle$. Fig.7 shows the energy-dependent fractional rms of low, middle and high frequencies for each energy bin, which displays different situation of variability between each frequency. Fractional rms slightly increases below 1keV and then decreases slowly for the low frequency. For the middle frequency, fractional rms keeps unchanged below 1 keV and then slightly increase for middle frequency. For the high frequency, fractional rms keeps unchanged below 0.7 keV, then increases rapidly to 2 keV and keep unchanged again from 3-7 keV finally, which implies that there are different components between the soft and high energy bands.

\begin{figure}
\centerline{
  \includegraphics[scale=0.23,angle=0]{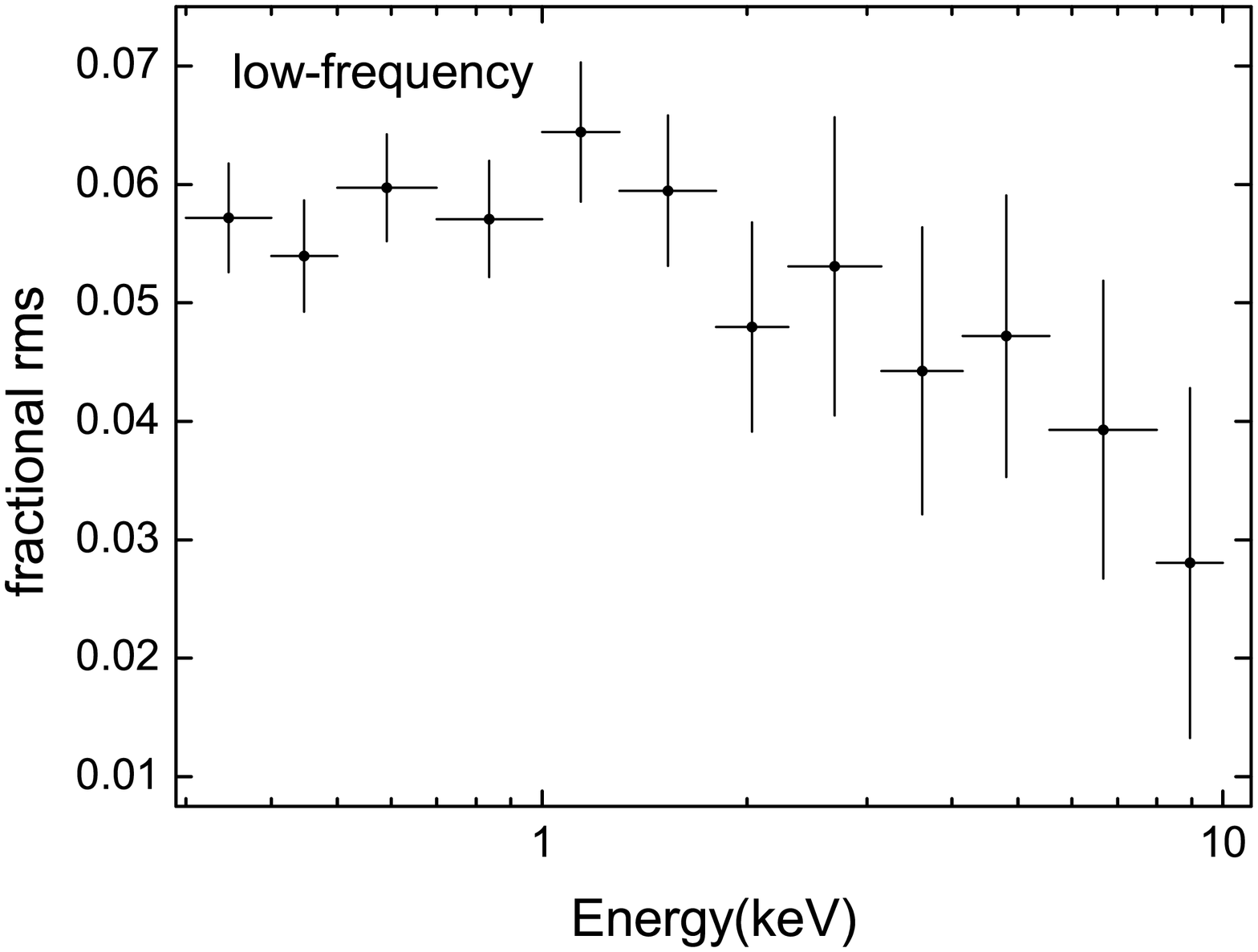}
  \includegraphics[scale=0.23,angle=0]{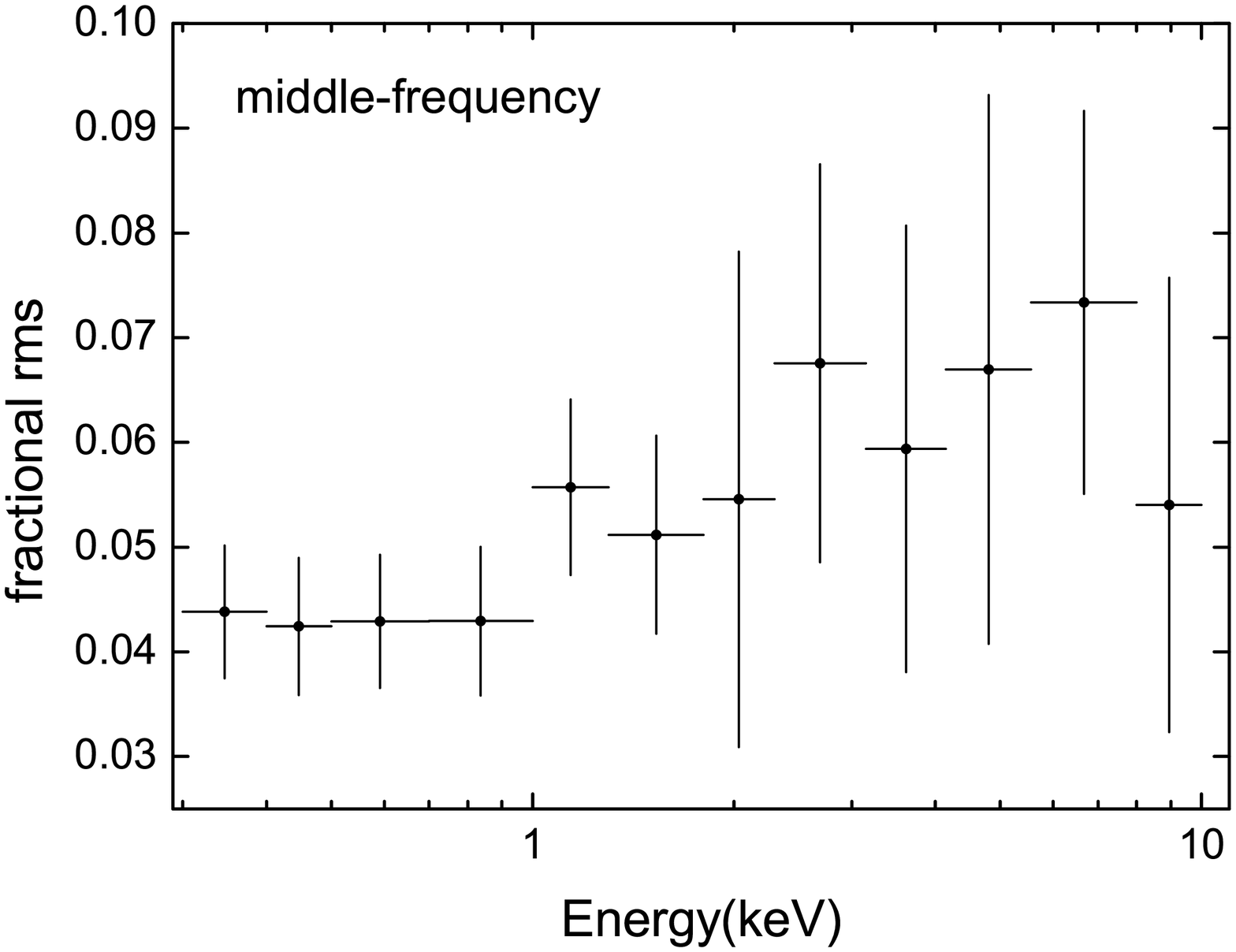}
  \includegraphics[scale=0.23,angle=0]{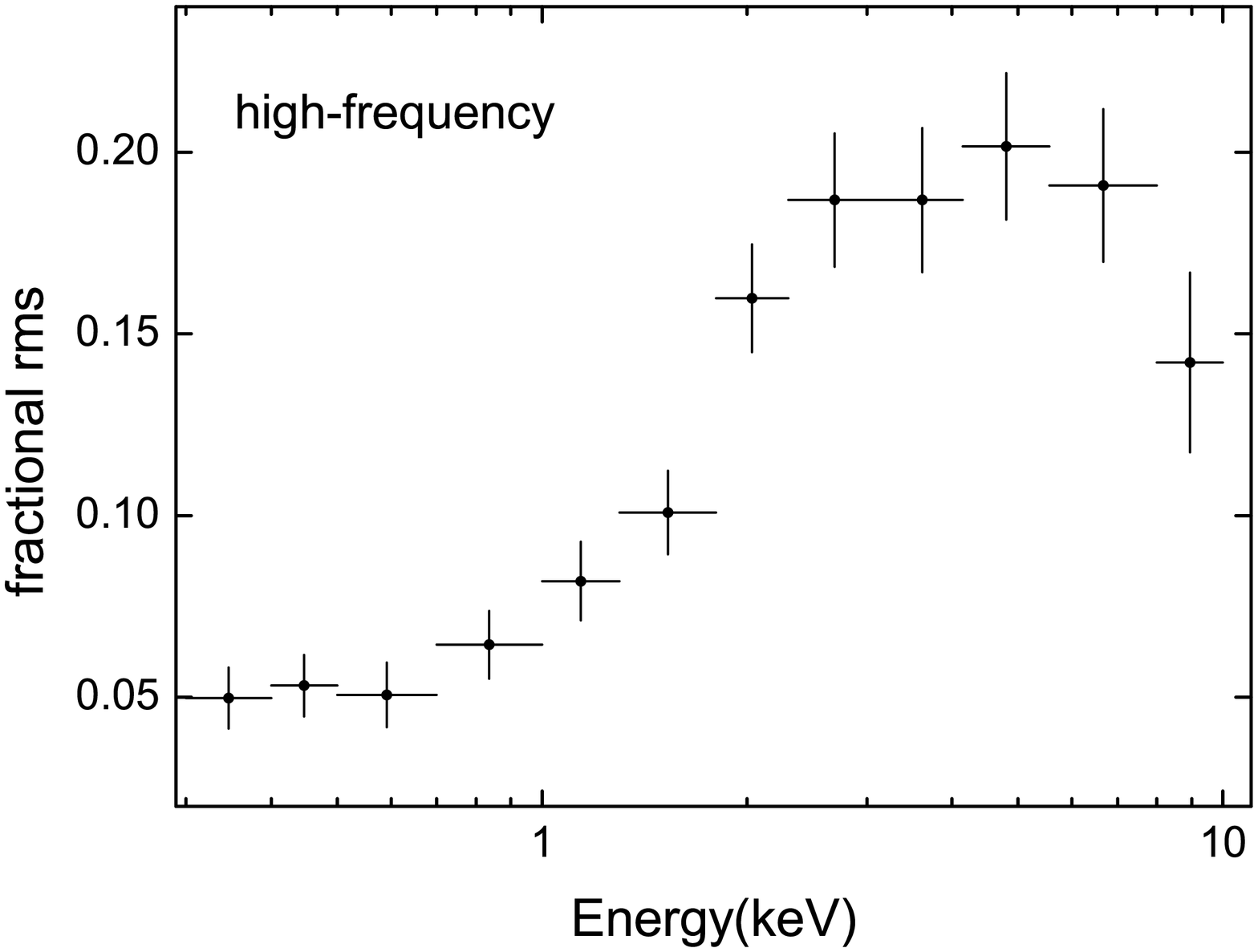}}
  \caption{The energy-dependent fractional rms of low, middle and high frequencies for each energy bin from left to right panel.}
\end{figure}

The covariance is the cross-spectral counterpart of the RMS, which measures the rms amplitude of variability as a detailed function of energy \citep{2009MNRAS.397..666W}:
\begin{equation}
\label{eqn:covspec}
Cv(\nu_{j})=\langle x \rangle \sqrt{\frac{\Delta \nu_{j} \left(|\bar{C}_{XY}(\nu_{j})|^{2}\right)}{\bar{P}_{Y}(\nu_{j}) - P_{Y,\rm noise}}},
\end{equation}
where $\bar{P}_{Y}$ and $P_{Y,\rm noise}$ are the reference band and noise of PSD, respectively, which are adopted at the whole energy band from 0.3-10 keV and removed the light curve of the interested energy bin for each channel to avoid the self-correlation by themselves just like mentioned above. Fig.8 and Fig.9 show the energy-dependent RMS and covariance of low, middle and high frequencies for each energy bin. The covariance displays the same shape as the RMS. However, the signal-to-noise of the covariance is better than that of RMS, since the reference band light curve is effectively used as a `matched filter' to pick out the correlated variations in each energy bin. It is noted that we adopt wider energy bins above 2.3 keV, which are larger than 100 ev stating above, so the RMS and covariance spectra can be extended to 10 keV and avoid the zero-count bin contained in the light curves in both of Fig.8 and Fig.9. However, based on the definition of the RMS and covariance mentioned above, both of them are depended on the average count rate of the light curve, $\langle x \rangle$, which relies on the scale of energy bin. Therefore, the RMS and covariance have the deviation above 2.3 keV because the inhomogeneous energy bins are adopted to avoid the zero-count contained in the light curves. If the time bin we adopt is larger than 400 s, it helps to extend the energy band range and ensure the homogeneous energy bins for the RMS and covariance spectra, but it lost the information of high frequency. In order to cover the whole energy band and ensure the homogeneous energy bins for the RMS and covariance spectra, the scale of energy bin could be adopted as the energy bin of the highest energy band at about 8-10 keV, which will ensure each energy bin does not include zero-count because the count rate of the highest energy band is more less than lower energy band. But it will lost the detailed information of the soft energy band (0.3-1 keV) because one energy bin could cover the soft energy band. We will analyze the RMS and covariance in detail in following section of spectra analysis.
\begin{figure}
\centerline{
  \includegraphics[scale=0.23,angle=0]{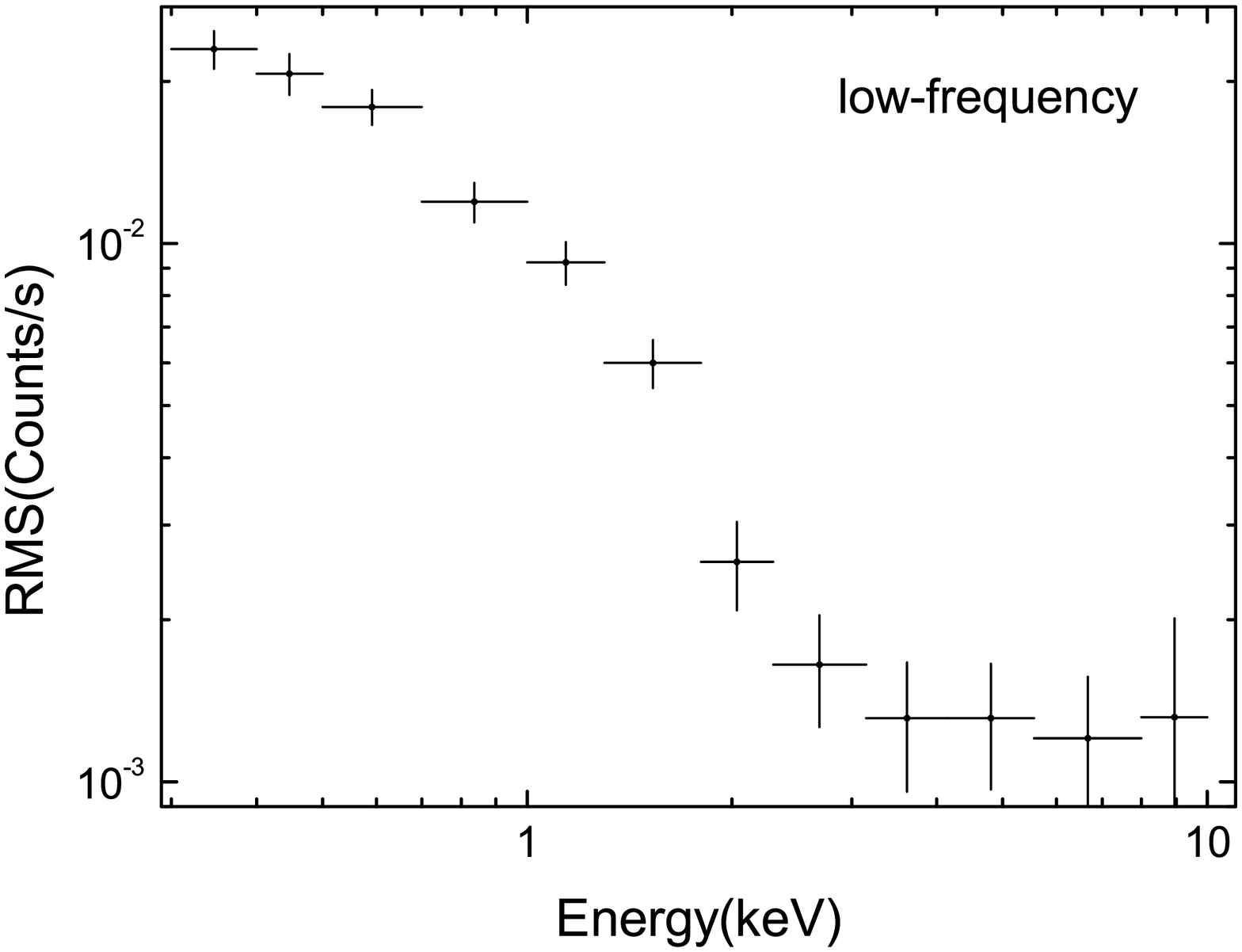}
  \includegraphics[scale=0.23,angle=0]{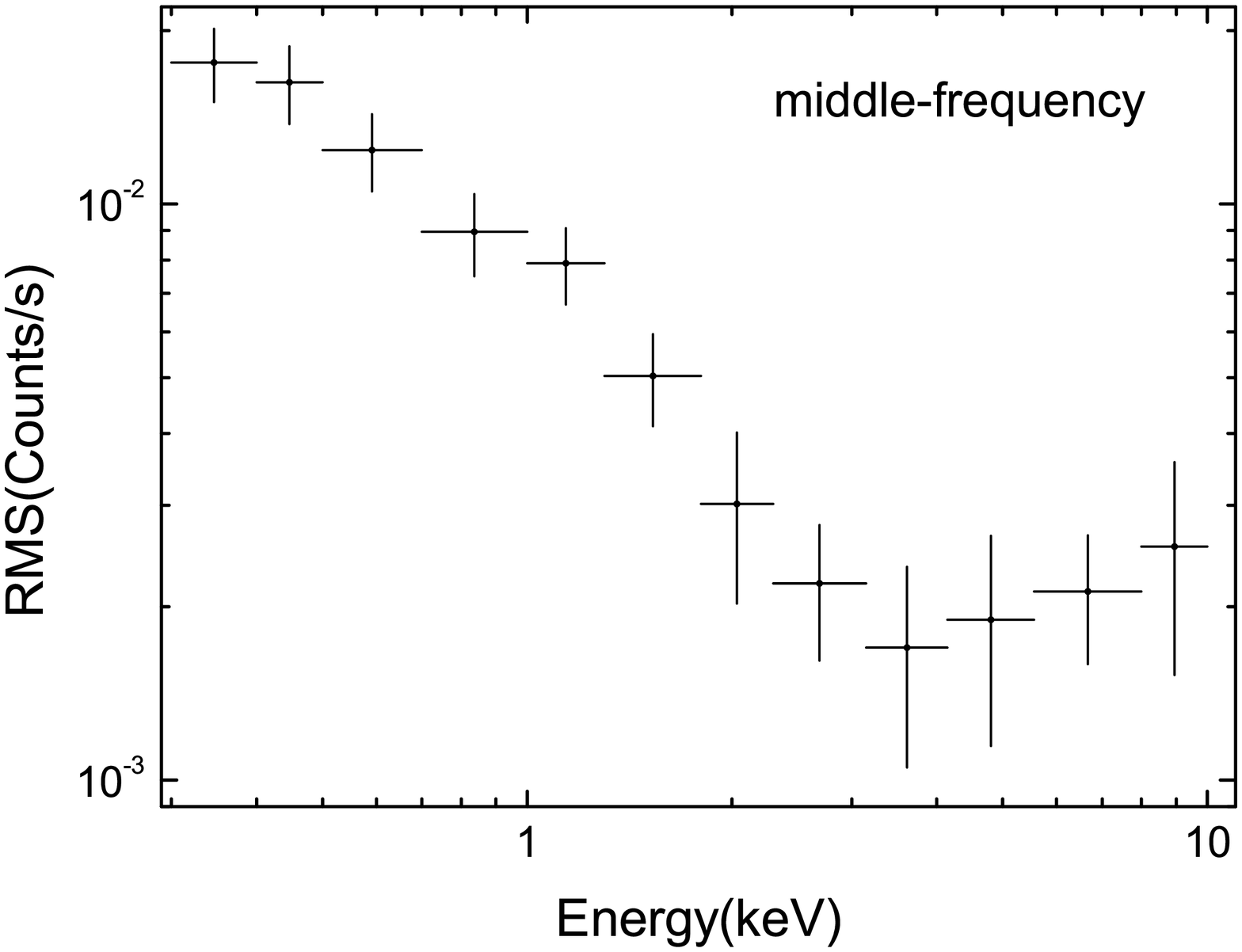}
  \includegraphics[scale=0.23,angle=0]{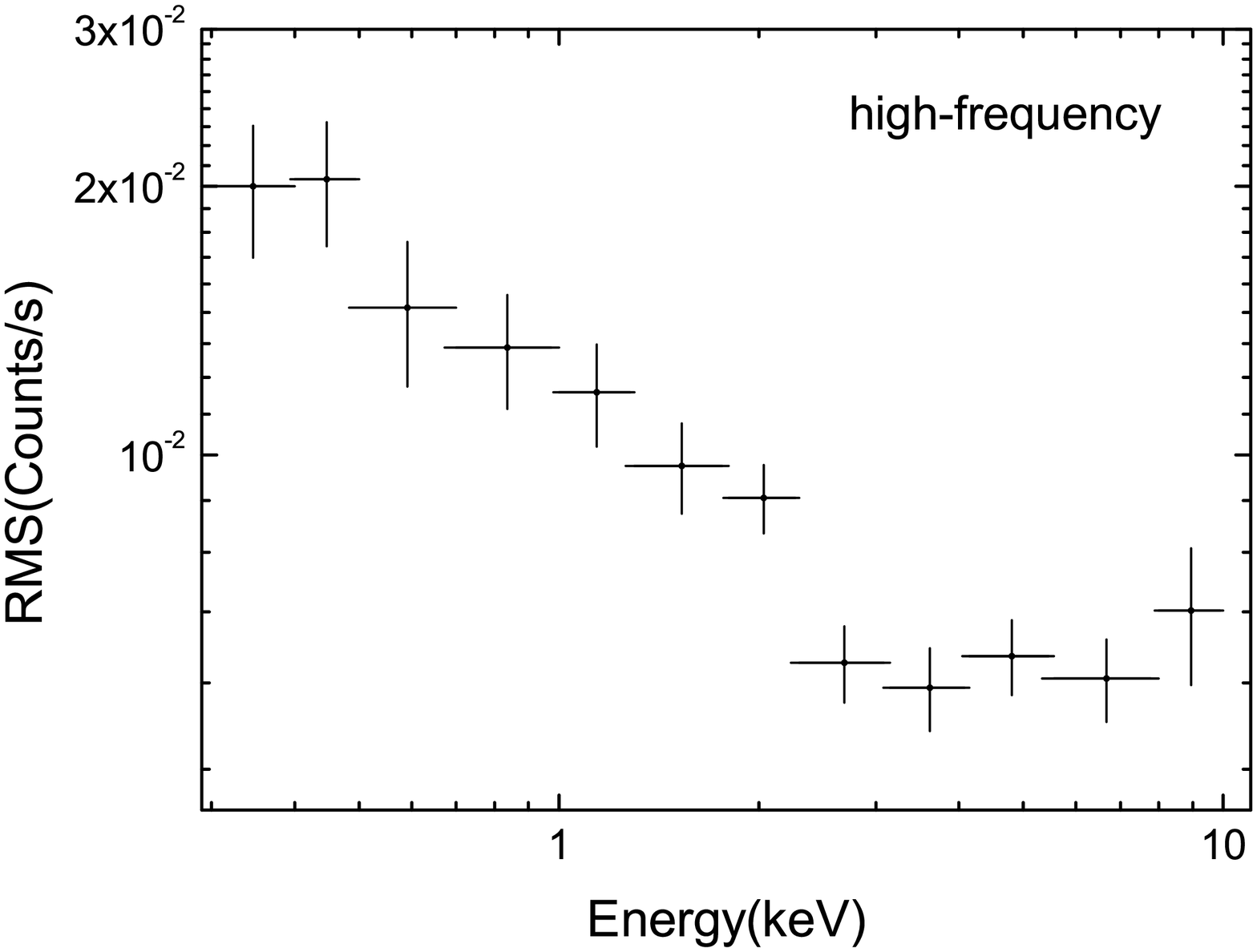}}
  \caption{The energy-dependent RMS of low, middle and high frequencies for each energy bin from left to right panel.}
\end{figure}

\begin{figure}
\centerline{
  \includegraphics[scale=0.23,angle=0]{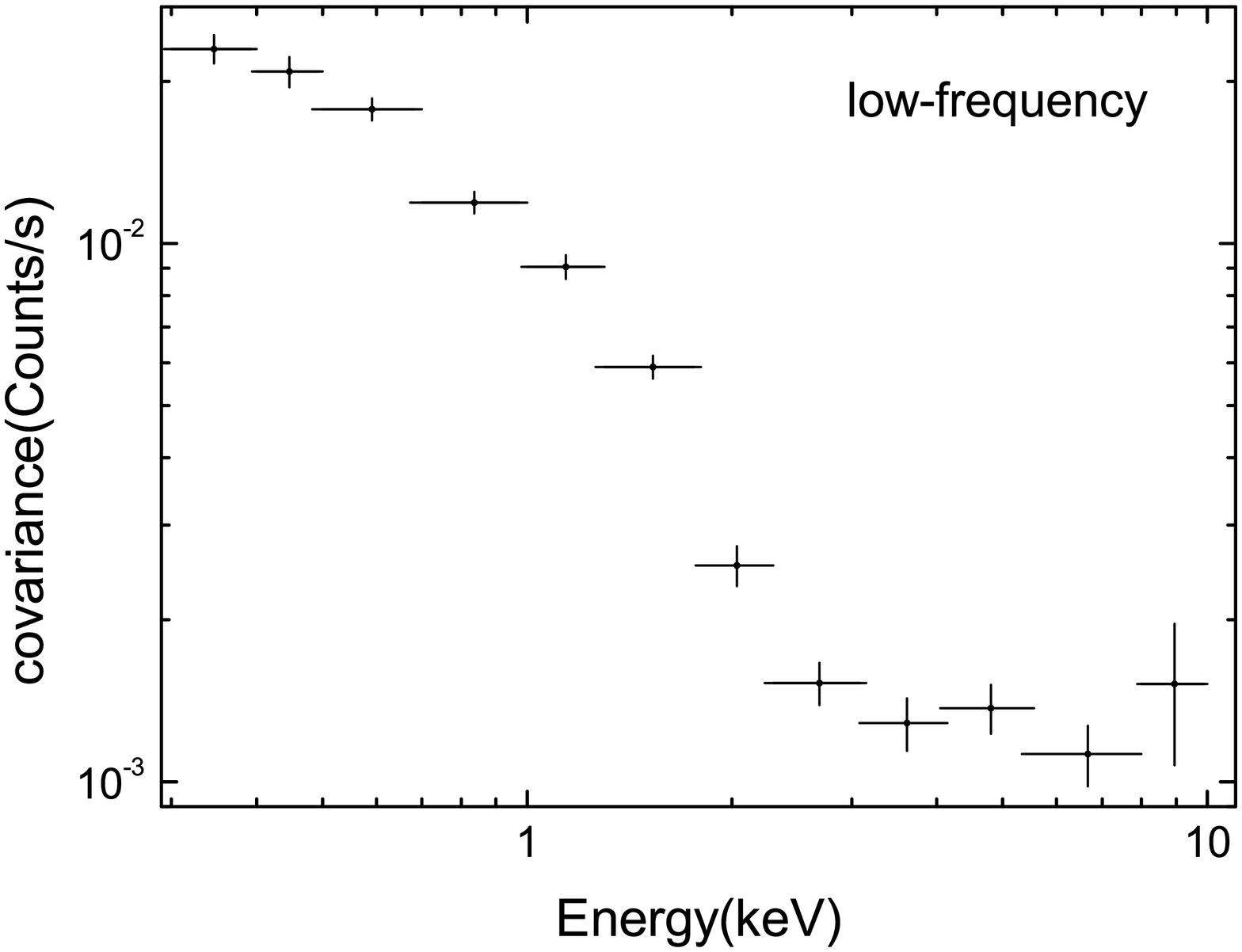}
  \includegraphics[scale=0.23,angle=0]{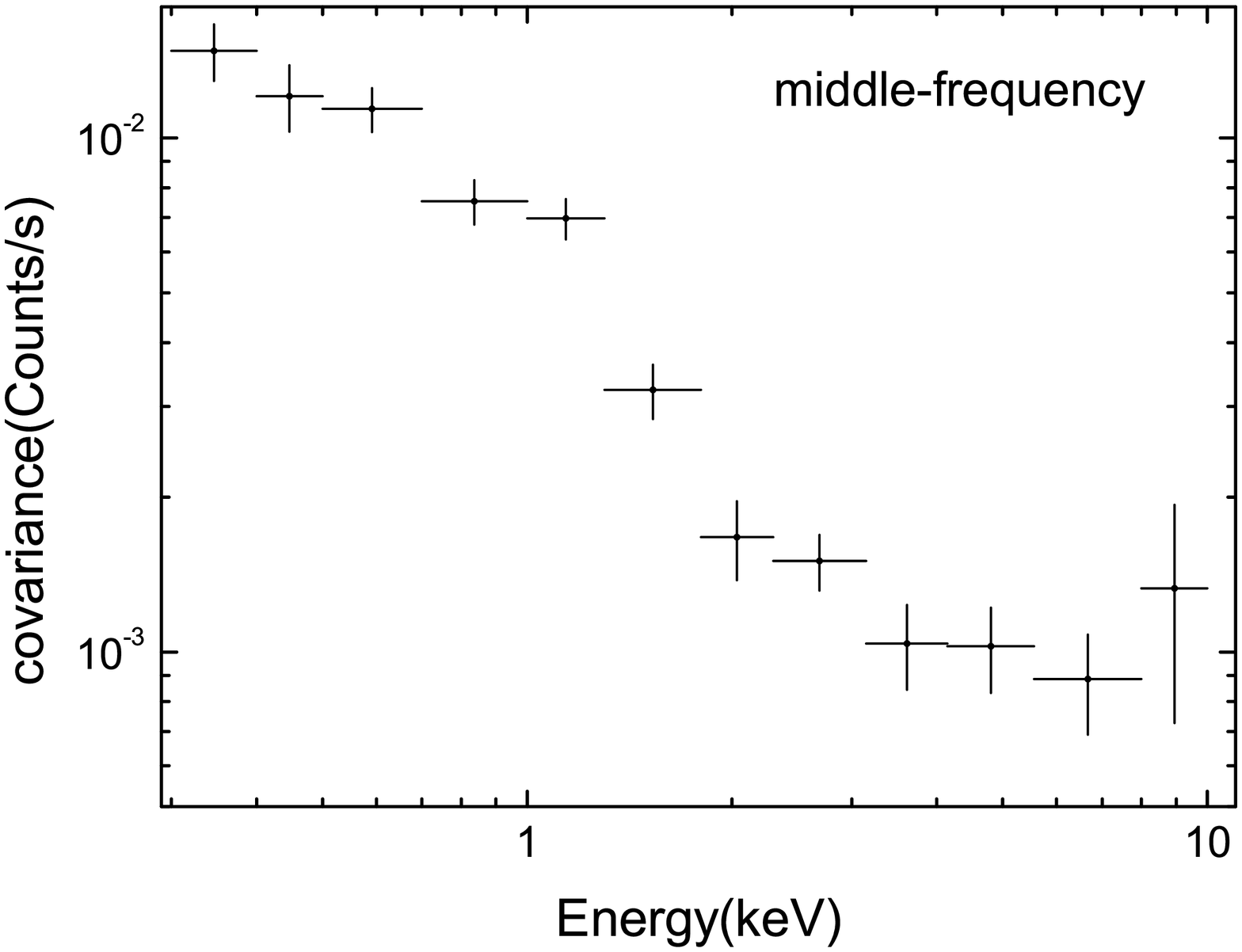}
  \includegraphics[scale=0.23,angle=0]{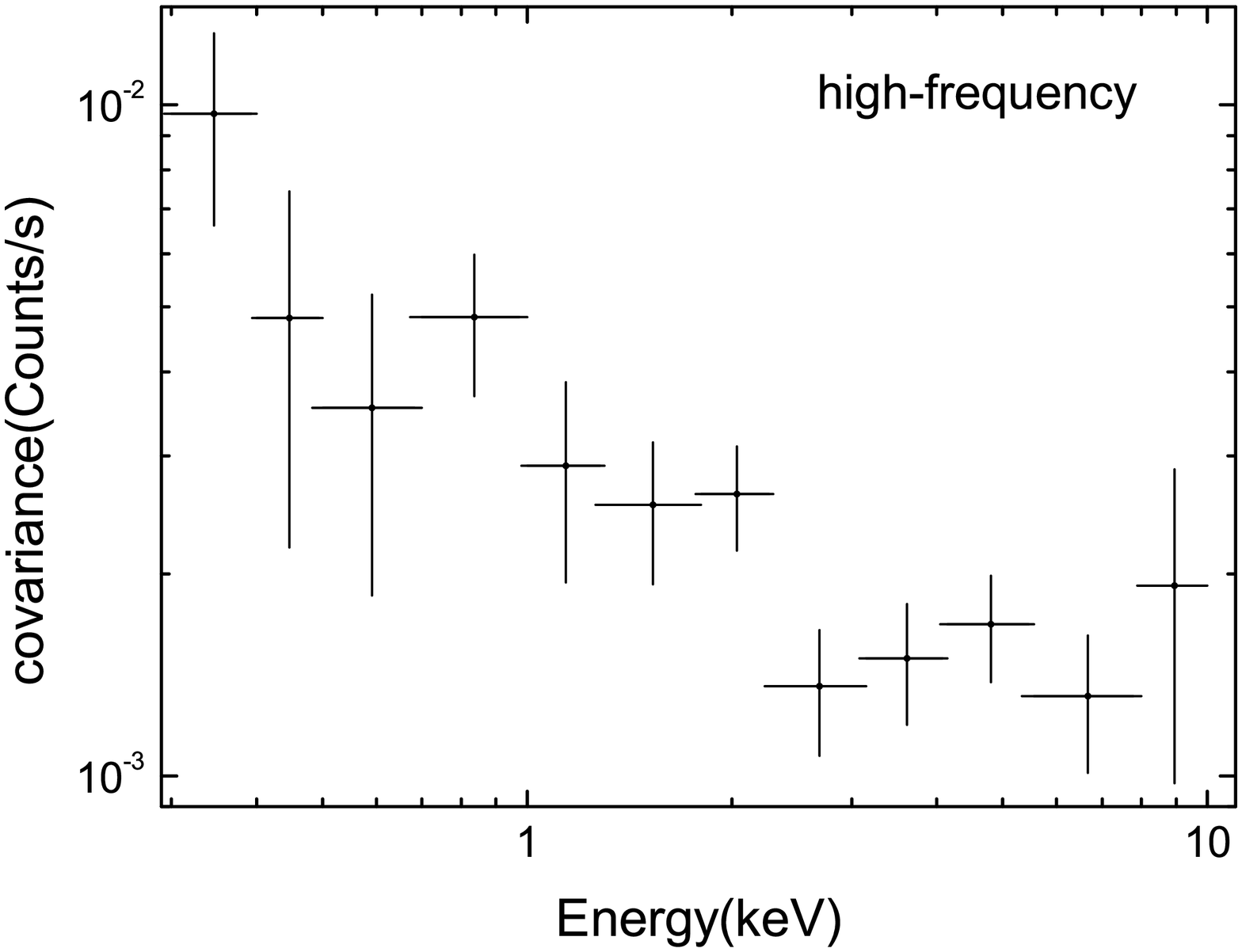}}
  \caption{The energy-dependent covariance of low, middle and high frequencies for each energy bin from left to right panel.}
\end{figure}

\section{Spectral Analysis}
  We use the XSPEC software (v12.11.1) \citep{arn96} to analyse the energy spectrum of ESO 362-G18. The summary of the models used in the paper is shown in Table 2. The first two models are the most favored competing models, Warm Corona and Relativistic Reflection. We also explore the Hybrid model which includes the warm corona and relativistic reflection components, Double Reflection model which includes two relativistic reflection components and the Double Warm corona model which includes two warm corona components. The Cross-normalization, $constant$, is fixed as one for the XMM-Newtion data and set free for NuSTAR FPMA/B. We use two absorption model, $tbabs$ \citep{2000ApJ...542..914W}, to describe the Galactic absorption with setting the Galactic absorption hydrogen column density at $1.75\times10^{20}$cm$^{-2}$ \citep{2005A&A...440..775K} during the spectral fitting with abundance set to $wilms$ \citep{2000ApJ...542..914W} with $vern$ cross-section \citep{1996ApJ...465..487V}, and the version with redshift, $ztbabs$, to describe the neutral absorption component for the host of galaxy of ESO 362-G18 with fixing redshift at $z=0.012$ and a free absorption hydrogen column density. $zxipcf$ \citep{2008MNRAS.385L.108R} is used to fit the warm absorber proposed by AG14 and Xu21. We use the thermal comptonisation model, $nthcomp$ \citep{1999MNRAS.309..561Z}, to fit the X-ray continuum radiated from hot corona. The seed photon temperature of disk, $kT_{disk}$, is fixed at 3 eV. The physical torus model, $borus12$ \citep{2019RNAAS...3..173B}, be used to fit the distant reflection just like Xu21 did and we fix the the cosine of the torus inclination angle $cos\theta_{inc}=0.8$ or $=37^{\circ}$. The photon index, $\Gamma$, and electron temperature, $kT_{e}$, are linked with the hot corona radiation, $nthcomp$. The Spectral Analysis has two steps as follows: The first step is fitting the time-average spectra simultaneously by using above models. The second step is not only the same fitting recipes like as the first step, but also add the RMS and covariance spectrum into the fit simultaneously with the time-average spectra. The RMS and covariance data are just adopted from 0.3-2.3 keV due to avoid the non-uniform energy bin influenced by the zero-count contained in the light curves, which will bring the deviation to the result of spectral analysis. We will plot the fitting results of the time-average spectra and the RMS and covariance spectrum on the same figure for the second step with different unit, in which the photon flux unit, photons cm$^{-2}$ s$^{-1}$ keV$^{-1}$, is used for the time-average spectra and the count rate unit, counts s$^{-1}$, is used for the RMS and covariance spectrum. The two units actually differ by one normalized factor because the response file could be assumed a unit diagonal matrix for the RMS and covariance spectrum, and the uniform energy bin is adopted.
\begin{table*}
\begin{center}
\begin{tabular}{cc}
  \hline
  \hline
  Model               &Components     \\
  \hline
  Warm Corona         &$constant*ztbabs*tbabs*zxipcf(nthcomp1+nthcomp2+borus12)$\\
  Reflection          &$constant*ztbabs*tbabs*zxipcf(nthcomp+relxilllpd+borus12)$\\
  Hybrid              &$constant*ztbabs*tbabs*zxicpf(nthcomp1+nthcomp2+borus12+relxillcp)$\\
  Double Reflection   &$constant*ztbabs*tbabs*zxipcf(nthcomp+relxilllpd+relxillcp+borus12)$\\
  Double Warm Corona  &$constant*ztbabs*tbabs*zxipcf(nthcomp1+nthcomp2+borus12+nthcomp3)$\\
  \hline
\end{tabular}
\caption{The summary of the model used in the paper.}
\end{center}
\end{table*}
\subsection{Warm Corona Model}
We add a second thermal comptonisation component, $nthcomp$, to fit the soft X-ray excess. The seed photon temperature of the warm corona component also is also fixed at 3 eV. The fitting results are shown in Table 3. and Fig.10.
\begin{figure}
\centerline{
  \includegraphics[scale=0.3,angle=0]{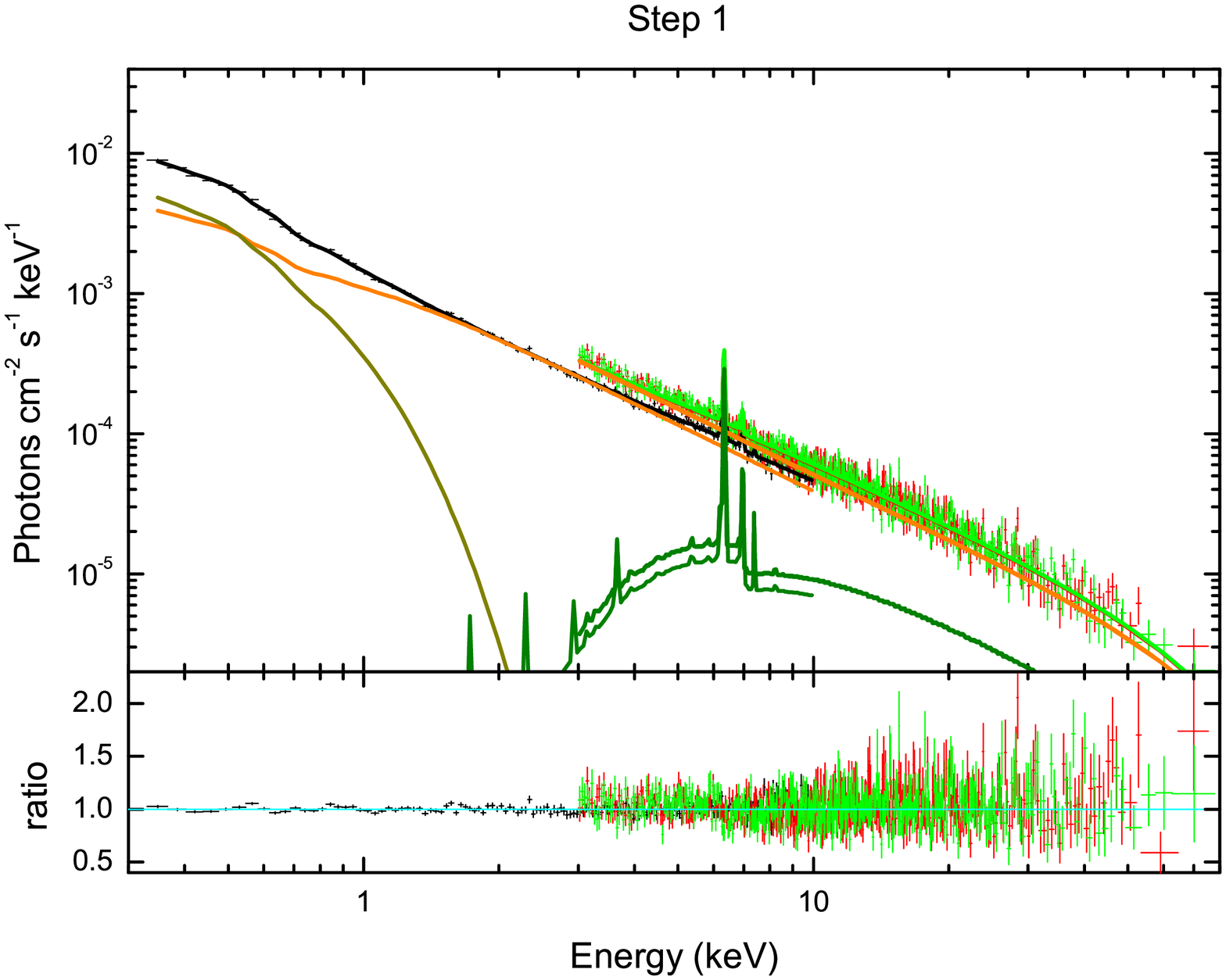}
  \includegraphics[scale=0.3,angle=0]{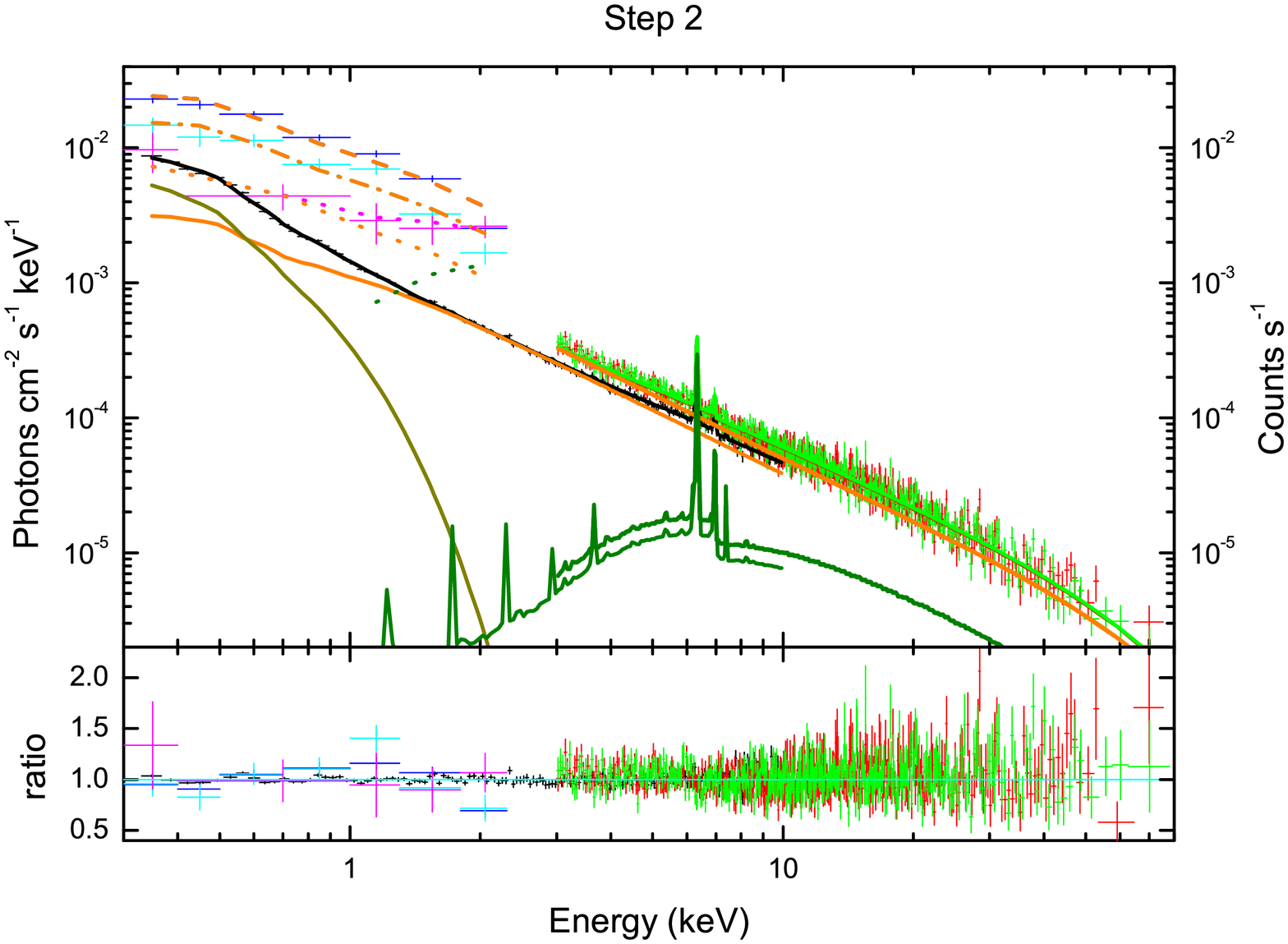}}
  \caption{The fitting results of the Warm Corona model for step 1 and 2 shown in left and right panel, respectively. Black, red and green data points are the time-average spectrum of XMM-Newton, NuSTAR FPMA, and FPMB data, respectively. Blue, cyan and magenta data points are the covariance spectrum of low, middle and high frequency, respectively. The corresponding colour lines are the best fitting results. Orange, dark yellow and olive lines represent the radiation of hot corona, warm corona and the distant material reflection. The dash, dash-dot and dot lines represent the fitting results of covariance spectrum of low, middle and high frequency with each model component which is as same color as the corresponding solid lines. The photon flux unit, photons cm$^{-2}$ s$^{-1}$ keV$^{-1}$, is used for the time-average spectra and the count rate unit, counts s$^{-1}$, is used for the covariance spectrum in the right panel. The bottom panel is the data-to-model ratio.}
\end{figure}

\begin{figure}
\centerline{
  \includegraphics[scale=0.4,angle=0]{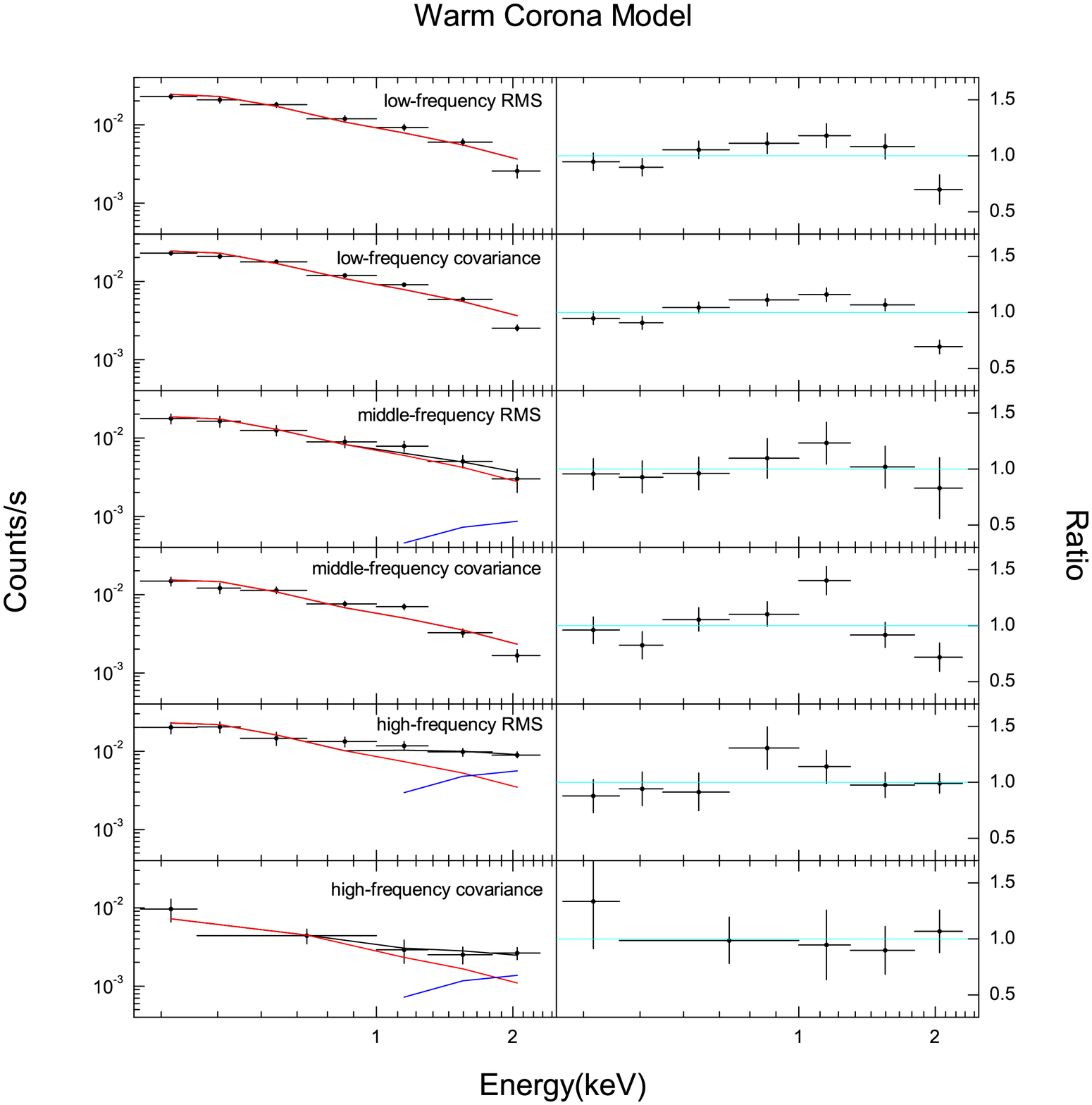}}
  \caption{The fitting results of the the RMS and covariance spectrum shown in the left panel by using the Warm Corona model. The black, red and blue line represent the fitting results, the comptonisation radiation of hot corona and the reflection of distant material. The data-to-model ratio is shown in the right panel.}
\end{figure}

\begin{table*}
\begin{center}
\begin{tabular}{ccccc}
  \hline
  \hline
  Description            &Component     &Parameter                   &Step 1             &Step 2  \\
  \hline
  Cross-normalization    &$constant$    &NuSTAR FPMA                 &$1.29\pm0.01$      &$1.29\pm0.01$ \\
                         &              &NuSTAR FPMB                 &$1.31\pm0.01$      &$1.31\pm0.01$ \\
  \hline
  Host galaxy Absorption &$ztbabs$      &$N_{H}(10^{20}cm^{-2})$     &$<1.09$            &$1.47\pm0.42$  \\
  \hline
  Warm Absorption        &$zipcf$       &$N_{H}(10^{22}cm^{-2})$     &$0.41\pm0.13$      &$0.38\pm0.14$ \\
                         &$zipcf$       &$log\xi$                    &$-0.24\pm0.11$     &$-0.32\pm0.08$\\
                         &$zipcf$       &$C_{F}$                     &$0.46\pm0.08$      &$0.41\pm0.06$\\
  \hline
  Hot Corona             &$nthcomp1$    &$\Gamma$                    &$1.63\pm0.02$      &$1.63\pm0.01$ \\
                         &$nthcomp1$    &$kT_{e}(keV)$               &$20.66\pm4.27$     &$21.43\pm4.10$  \\
  \hline
  Warm Corona            &$nthcomp2$    &$\Gamma$                    &$1.81\pm0.27$      &$2.15\pm0.12$   \\
                         &$nthcomp2$    &$kT_{e}(keV)$               &$0.16\pm0.01$      &$0.17\pm0.01$   \\
  \hline
  Distant Reflection     &$boru12$     &$log[N_{H,tor}/cm^{-2}]$     &$23.34\pm0.17$     &$23.36\pm0.12$ \\
                         &$boru12$     &$C_{tor}$                    &$0.88\pm0.17$      &$0.82\pm0.10$ \\
                         &$boru12$     &$A_{Fe}$                     &$1.06\pm0.22$      &$0.97\pm0.29$  \\
  \hline
  $\chi^2/d.o.f$         &              &                            &$1061/959$        &$1171/977$     \\
  \hline
\end{tabular}
\caption{Fitting results of Warm Corona model for step 1 and 2.}
\end{center}
\end{table*}
\subsubsection{the first step}
The Chi-square values/d.o.f of fitting result is 1061/959 (reduced $\chi^2=1.11$). The hydrogen column density of the host galaxy absorption is less than $1.09\times10^{20}$cm$^{-2}$. The ionization degree, $log\xi$ of warm absorber, $zxipcf$, is very low, which means that the radiation is absorbed by neutral material. The column density of warm absorber, $N_{H}$ is order of $10^{21}$cm$^{-2}$, which is an order of magnitude larger than that of Xu21, but consistent with the result of AG14 who also used the $zxipcf$ to model the warm absorber, it could be using different warm absorption model. The X-ray continuum index, $\Gamma=1.63\pm0.02$, and the electron temperature of hot corona, $kT_{e}=20.66\pm4.27$ keV, are consistent with the result of Xu21, however, it is not reasonable because the optical depth is optical thick. The photon index of warm corona is $1.81\pm0.27$ instead of the fixed value of 2.5 adopted by Xu21. The electron temperature of warm corona is $0.16\pm0.01$ keV which is consistent with the result of Xu21. But the corresponding optical depth of the warm corona is about 50, which is larger than the prediction of \citet{2020A&A...634A..92U} ($10\sim40$).
\subsubsection{the second step}
The Chi-square values/d.o.f of fitting result is 1171/977 (reduced $\chi^2=1.19$), which is worse than that of the first step. The hydrogen column density of the host galaxy absorption is obtained about an order of $10^{20}cm^{-2}$ as same as the Galactic absorption. In the right panel of Fig.10, we add the covariance spectrum of low, middle and high frequency into plot for comparison with the time-average spectrum. Fig.11 shows the detailed fitting result of variability spectrum (RMS and covariance spectrum) and the deviation mainly comes from the variability spectrum fitting, which displays that the X-ray continuum radiated from hot corona is the main contributor, but the reflection of distant material, $borus12$, increases the contribution at high frequency. The Warm model can't interpret the residual error shown in the right panel of Fig.11.
\subsection{Reflection Model}
Just like Xu21, we implement the relativistic reflection with variable disk density, $relxilllpd$ to be as the soft X-ray excess component with the lamp post geometry \citep{2014MNRAS.444L.100D, 2014ApJ...782...76G, 2016MNRAS.462..751G}. We set the disk inclination to be free and the inner/outer disk radius to be the default values. The incident spectral index of the relativistic reflection is linked with the X-ray continuum. We fix the electron temperature of hot corona, $kT_{e}=300$keV. The fitting results are shown in Table 4. and Fig.12.

\begin{figure}
\centerline{
  \includegraphics[scale=0.3,angle=0]{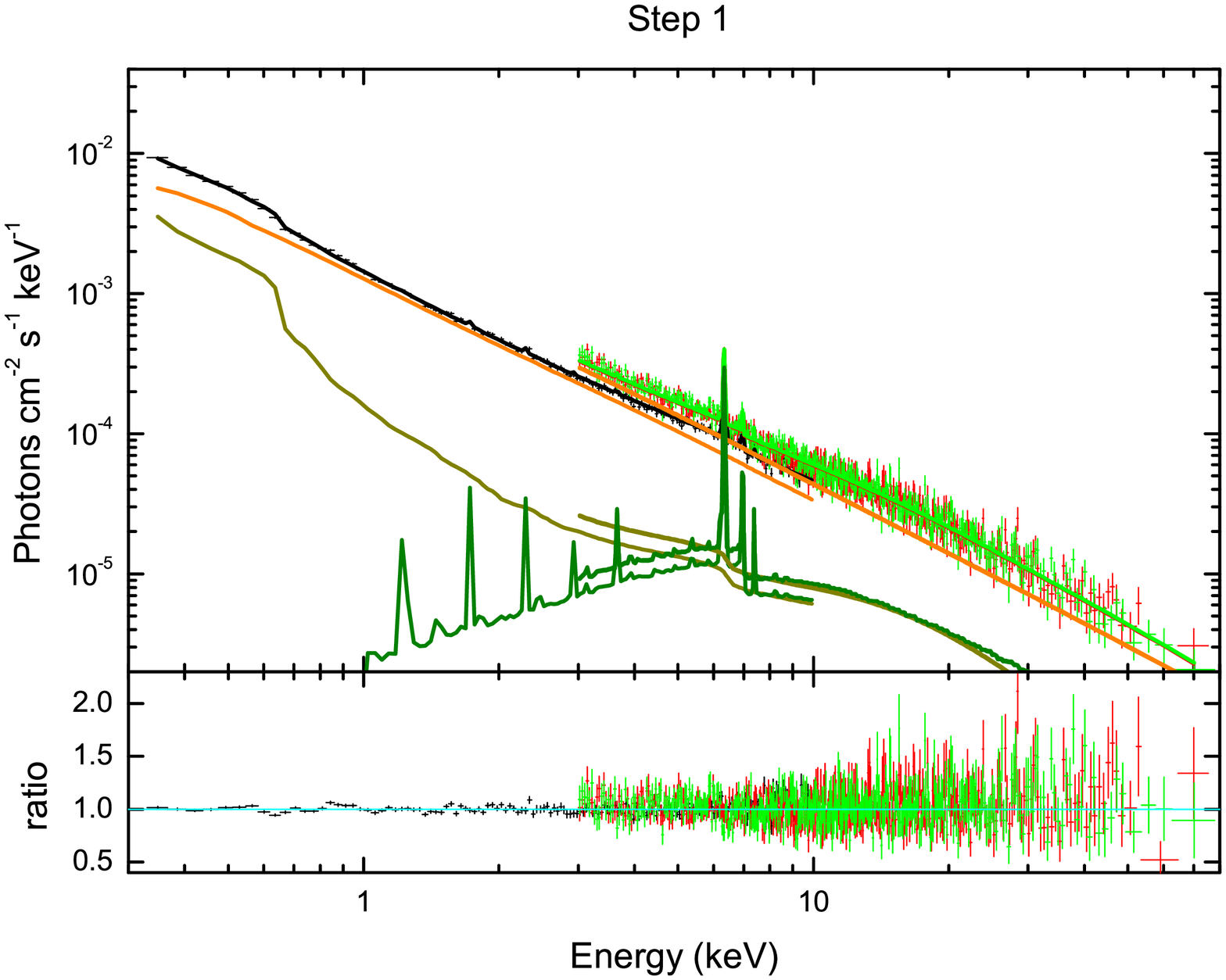}
  \includegraphics[scale=0.3,angle=0]{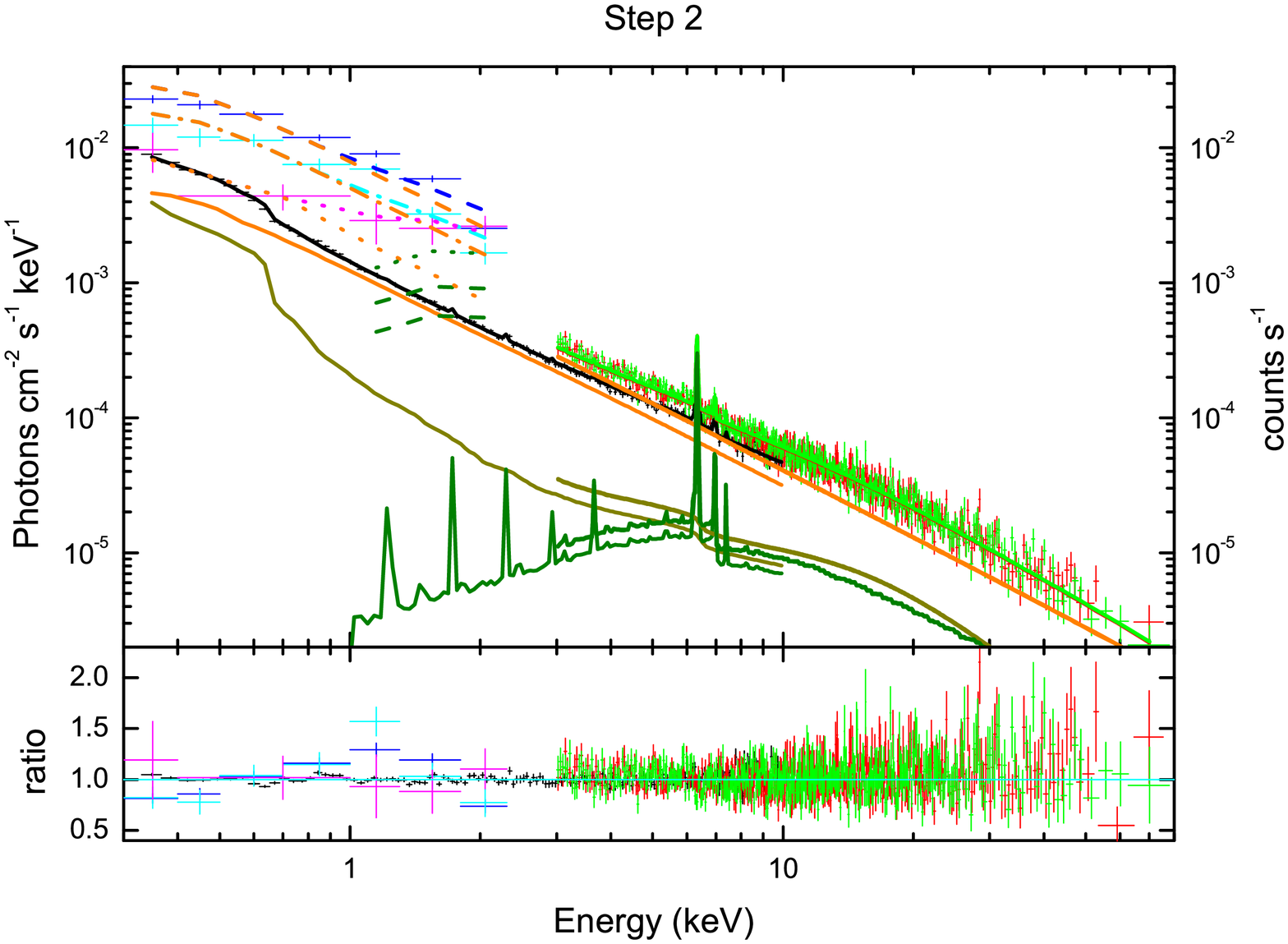}}
  \caption{The fitting results of the Reflection model for step 1 and 2 shown in left and right panel, respectively. Black, red and green data points are the time-average spectra of XMM-Newton, NuSTAR FPMA, and FPMB data, respectively. Blue, cyan and magenta data points are the covariance spectrum of low, middle and high frequency, respectively. The corresponding colour lines are the best fitting results. Orange, dark yellow and olive lines represent the radiation of hot corona, relativistic reflection and the distant material reflection. The dash, dash-dot and dot lines represent the fitting results of covariance spectrum of low, middle and high frequency with each model component which is as same color as the corresponding solid lines. The photon flux unit, photons cm$^{-2}$ s$^{-1}$ keV$^{-1}$, is used for the time-average spectra and the count rate unit, counts s$^{-1}$, is used for the covariance spectrum in the right panel. The bottom panel is the data-to-model ratio.}
\end{figure}

\begin{figure}
\centerline{
  \includegraphics[scale=0.4,angle=0]{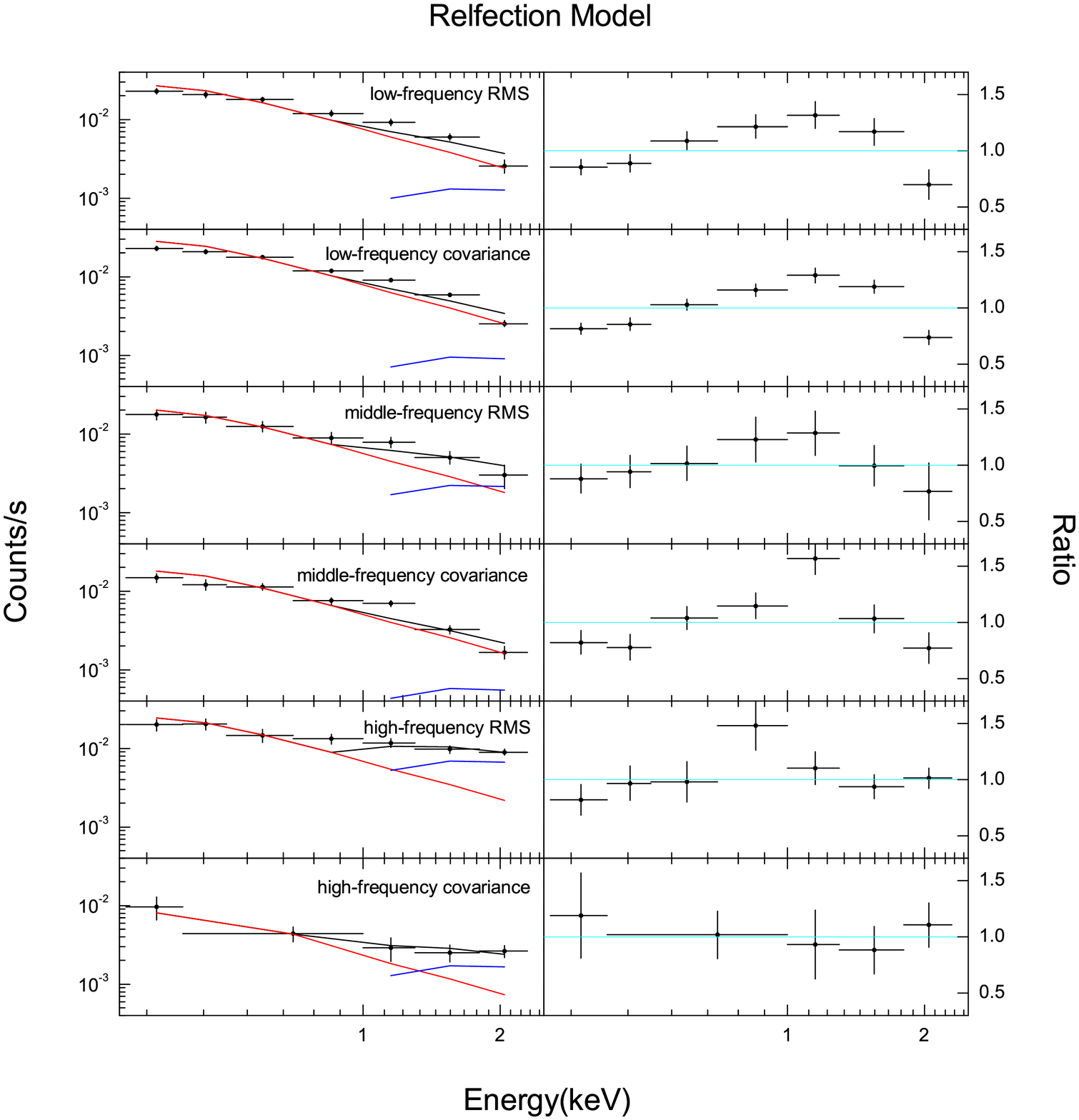}}
  \caption{The fitting result of the RMS and covariance spectrum shown in the left panel by using the Reflection model. The black, red and blue line represent the fitting result, the comptonisation radiation of hot corona and the reflection of distant material. The data-to-model ratio is shown in the right panel.}
\end{figure}

\begin{table*}
\begin{center}
\begin{tabular}{ccccc}
  \hline
  \hline
  Description            &Component     &Parameter                   &Step 1             &Step 2       \\
  \hline
  Cross-normalization    &$constant$    &NuSTAR FPMA                 &$1.29\pm0.01$      &$1.29\pm0.01$ \\
                         &              &NuSTAR FPMB                 &$1.31\pm0.01$      &$1.31\pm0.01$ \\
  \hline
  Host galaxy Absorption &$ztbabs$      &$N_{H}(10^{20}cm^{-2})$     &$<0.73$            &$1.07\pm0.34$  \\
  \hline
  Warm Absorption        &$zipcf$       &$N_{H}(10^{22}cm^{-2})$     &$3.98\pm1.73$      &$3.65\pm2.25$ \\
                         &$zipcf$       &$log\xi$                    &$<1.86$            &$<2.37$\\
                         &$zipcf$       &$C_{F}$                     &$0.16\pm0.11$      &$0.14\pm0.10$\\
  \hline
  Hot Corona             &$nthcomp$    &$\Gamma$                     &$1.68\pm0.05$      &$1.68\pm0.04$ \\
                         &$nthcomp$    &$kT_{e}(keV)$                &$300$              &$300$  \\
  \hline
  Ionized Reflection     &$relxilllpd$  &$h(r_g)$                    &$2.53\pm1.49$      &$2.58\pm0.88$ \\
                         &$relxilllpd$  &$a$                         &$0.998\pm0.69$     &$0.998\pm0.51$ \\
                         &$relxilllpd$  &$\Theta_{disk}(degree)$     &$<67.59$           &$<34.09$  \\
                         &$relxilllpd$  &$log\xi$                    &$2.67\pm0.61$      &$2.67\pm0.52$  \\
                         &$relxilllpd$  &$A_{Fe}^{disk}$             &$1.02^{*}$         &$0.88\pm0.46$  \\
                         &$relxilllpd$  &$log[n_{e}/cm^{-3}]$        &$17.85\pm1.06$     &$17.83\pm0.77$ \\
  \hline
  Distant Reflection     &$boru12$     &$log[N_{H,tor}/cm^{-2}]$     &$23.81\pm0.46$     &$23.72\pm0.35$ \\
                         &$boru12$     &$C_{tor}$                    &$<0.42$            &$<0.36$ \\
                         &$boru12$     &$A_{Fe}$                     &$1.06\pm0.48$      &$1.00\pm0.52$  \\
  \hline
  $\chi^2/d.o.f$         &              &                            &$1073/956$        &$1268/974$     \\
  \hline
\end{tabular}
\caption{Fitting results of Reflection model for step 1 and 2. The asterisk represents that the parameter can't be constrained in our fitting and the value is pegged.}
\end{center}
\end{table*}
\subsubsection{the first step}
The Chi-square values/d.o.f of fitting result is 1073/956 (reduced $\chi^2=1.12$), which cannot be distinguished easily compared with the Warm corona model in the first step. The hydrogen column density of the host galaxy absorption has a upper limit value, $N_{H}(10^{20}cm^{-2})<0.73$. The X-ray continuum index, $\Gamma=1.68\pm0.05$, is less than the typical index, $\Gamma\sim1.8$ but reasonable because it is according with the physical property of optical thin for hot corona with the the electron temperature of 300$keV$. The hight of corona is, $h=2.53\pm1.49r_{g}$, which means that the hot corona is compact just as the prediction of microlensing (e.g. \citet{2010ApJ...709..278D, 2013ApJ...769...53M}). The fitting result shows the black hole has a high spin, $a=0.998$. The disk inclination is less than $67.59^{\circ}$. The ionization degree of disk is obtained a higher value than obtained by Xu21. The iron abundances cannot be constrained in the fitting and is pegged at $A_{Fe}^{disk}=1.02$ and the disk density is obtained at $log[n_{e}/cm^{-3}]=17.85\pm1.06$. However, the iron abundances and the disk density could be degenerated \citep{2018ApJ...855....3T}.
\subsubsection{the second step}
The Chi-square values/d.o.f of fitting result is 1268/974 (reduced $\chi^2=1.30$), which is worse than that of the first step and the Warm corona model in the second step. The hydrogen column density of the host galaxy absorption is obtained a moderate value $N_{H}(10^{20}cm^{-2})=1.07\pm0.34$. The X-ray continuum index, $\Gamma=1.68\pm0.04$, which is less than the typical index. The hight of corona is, $h=2.58\pm0.88r_{g}$ consistent with the first step. The black hole spin also has a high value. The disk inclination is obtained a upper value, $<34.09^{\circ}$. In the right panel of Fig.12, we add the covariance spectrum of low, middle and high frequency into plot for comparison with the time-average spectrum. Fig.13 shows the detailed fitting result of the variability spectrum. The X-ray continuum radiated from hot corona and the reflection of distant material are the main contributor, but still can't interpret the residual error.
\subsection{Hybrid Model}
Hybrid model considers that the soft X-ray excess comes from both of warm corona and relativistic reflection. We adopt the standard relativistic reflection model, $relxillcp$, which assume the incident radiation is an comptonization continuum. Just like the set of Warm Corona model, we set the seed photon temperature to be 3 ev for the second $nthcomp$ component. We set the index of emissivity to be $q_{1}=q_{2}=3$, the inner radius of the disk is $50r_{g}$ and the outer of disk is the default value, assuming that the relativistic reflection component is out of the warm corona, so the relativistic effects could be weak. The electron temperature is also fixed at 300 keV like as the Reflection model, the disk inclination is set to be free. The photon index of the relativistic reflection is linked with the hot corona. The fitting results are shown in Table 5. and Fig.14.

\begin{table*}
\begin{center}
\begin{tabular}{ccccc}
  \hline
  \hline
  Description            &Component     &Parameter                   &Step 1             &Step 2       \\
  \hline
  Cross-normalization    &$constant$    &NuSTAR FPMA                 &$1.29\pm0.01$      &$1.29\pm0.01$ \\
                         &              &NuSTAR FPMB                 &$1.31\pm0.01$      &$1.31\pm0.01$ \\
 \hline
  Host galaxy Absorption &$ztbabs$      &$N_{H}(10^{20}cm^{-2})$     &$<0.48$            &$1.60\pm0.44$  \\
  \hline
  Warm Absorption        &$zipcf$       &$N_{H}(10^{22}cm^{-2})$     &$0.50\pm0.18$      &$0.37\pm0.10$ \\
                         &$zipcf$       &$log\xi$                    &$-0.37\pm0.08$     &$-0.49\pm0.06$\\
                         &$zipcf$       &$C_{F}$                     &$0.51\pm0.11$      &$0.47\pm0.05$\\
  \hline
  Hot Corona             &$nthcomp1$    &$\Gamma$                    &$1.75\pm0.02$      &$1.74\pm0.01$ \\
                         &$nthcomp1$    &$kT_{e}(keV)$               &$300$              &$300$  \\
  \hline
  Warm Corona            &$nthcomp2$    &$\Gamma$                    &$2.08\pm0.10$      &$2.49\pm0.16$   \\
                         &$nthcomp2$    &$kT_{e}(keV)$               &$0.19\pm0.03$      &$0.19\pm0.01$   \\
  \hline
  Ionized Reflection     &$relxillcp$   &$q_{1}=q_{2}$               &$3$                &$3$ \\
                         &$relxillcp$   &$R_{in}(r_g)$               &$50$               &$50$ \\
                         &$relxillcp$   &$a$                         &$0.97\pm0.13$      &$0.93\pm0.24$ \\
                         &$relxillcp$   &$\Theta_{disk}(degree)$     &$<69$              &$<70$ \\
                         &$relxillcp$   &$log\xi$                    &$1.37\pm0.10$      &$1.46\pm0.11$  \\
                         &$relxillcp$   &$A_{Fe}$                    &$0.53\pm0.43$      &$0.73\pm0.42$   \\
 \hline
  Distant Reflection     &$boru12$     &$log[N_{H,tor}/cm^{-2}]$     &$23.28\pm0.14$     &$23.39\pm0.11$ \\
                         &$boru12$     &$C_{tor}$                    &$0.81\pm0.09$      &$0.58\pm0.19$ \\
                         &$boru12$     &$A_{Fe}$                     &$0.49\pm0.11$      &$0.40\pm0.08$  \\
  \hline
  $\chi^2/d.o.f$         &              &                            &$1031/955$        &$1144/967$     \\
  \hline
\end{tabular}
\caption{Fitting results of Hybrid model for step 1 and 2.}
\end{center}
\end{table*}

\begin{figure}
\centerline{
  \includegraphics[scale=0.3,angle=0]{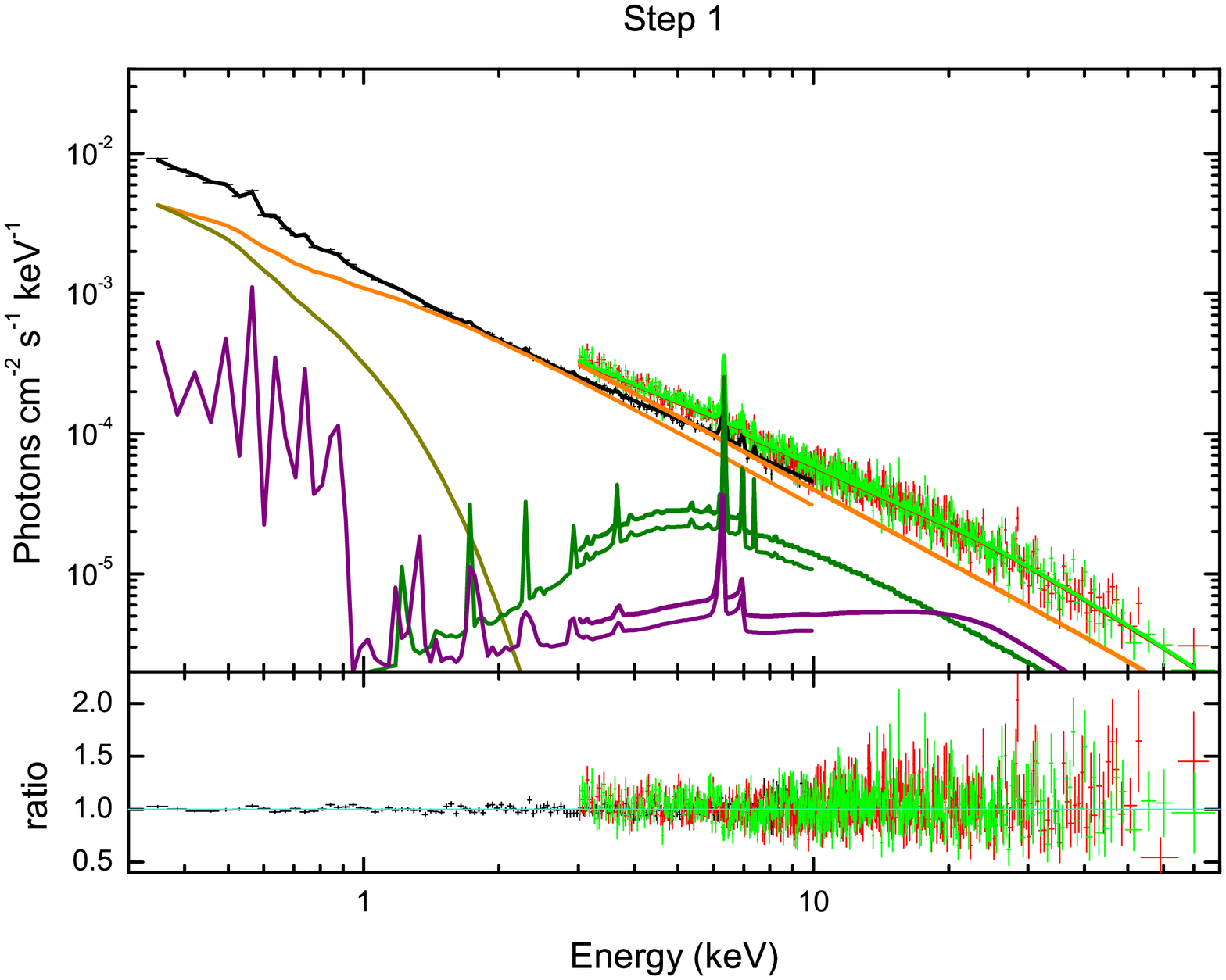}
  \includegraphics[scale=0.3,angle=0]{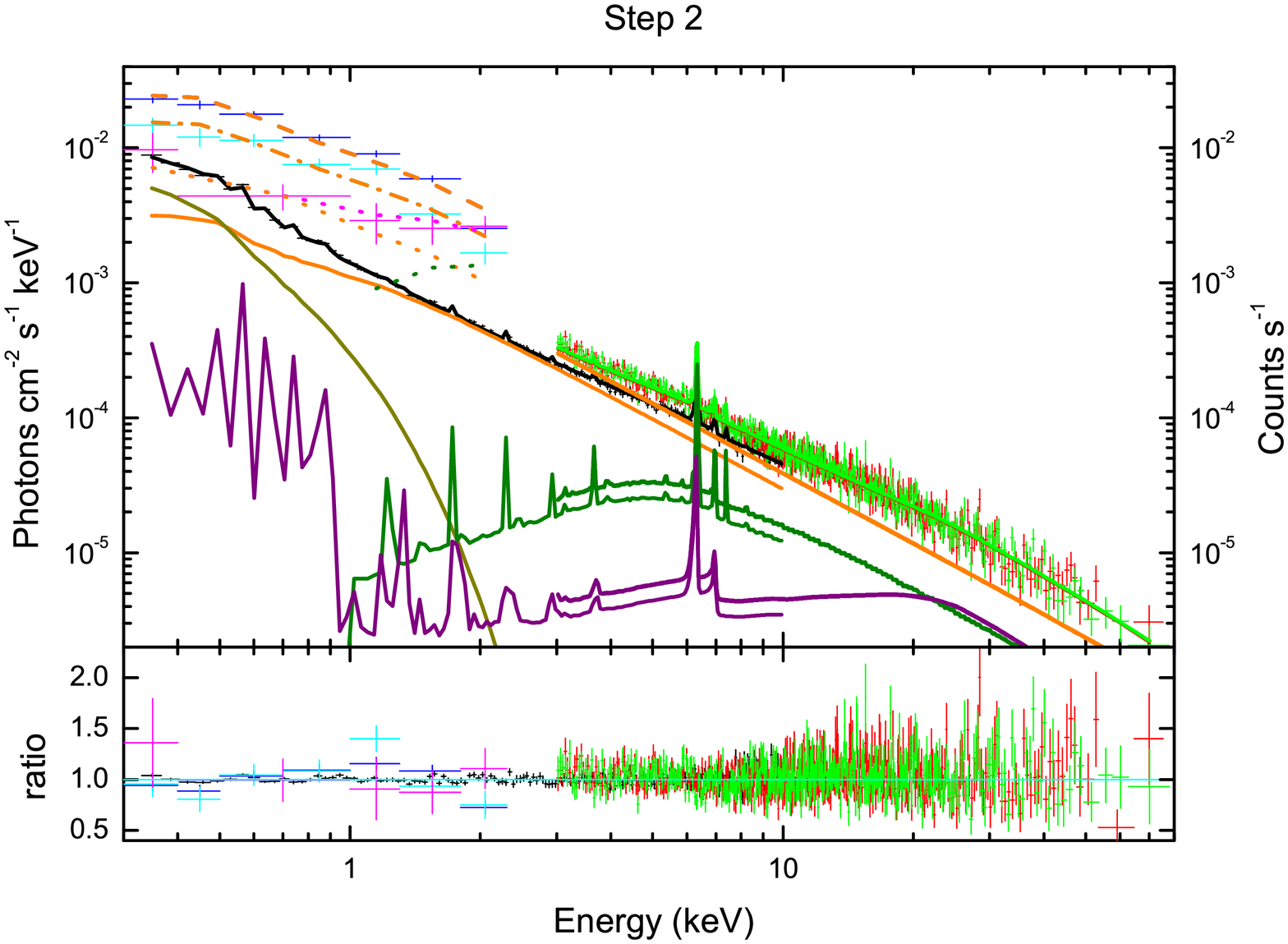}}
  \caption{The fitting results of the Hybrid model for step 1 and 2 shown in left and right panel, respectively. Black, red and green data points are the time-average spectra of XMM-Newton, NuSTAR FPMA, and FPMB data, respectively. Blue, cyan and magenta data points are the covariance spectrum of low, middle and high frequency, respectively. The corresponding colour lines are the best fitting results. Orange, dark yellow, olive and purple lines represent the radiation of hot corona, warm corona, the reflection of distant material and the relativistic reflection. The dash, dash-dot and dot lines represent the fitting results of covariance spectrum of low, middle and high frequency with each model component which is as same color as the corresponding solid lines. The photon flux unit, photons cm$^{-2}$ s$^{-1}$ keV$^{-1}$, is used for the time-average spectra and the count rate unit, counts s$^{-1}$, is used for the covariance spectrum in the right panel. The bottom panel is the data-to-model ratio.}
\end{figure}

\begin{figure}
\centerline{
  \includegraphics[scale=0.4,angle=0]{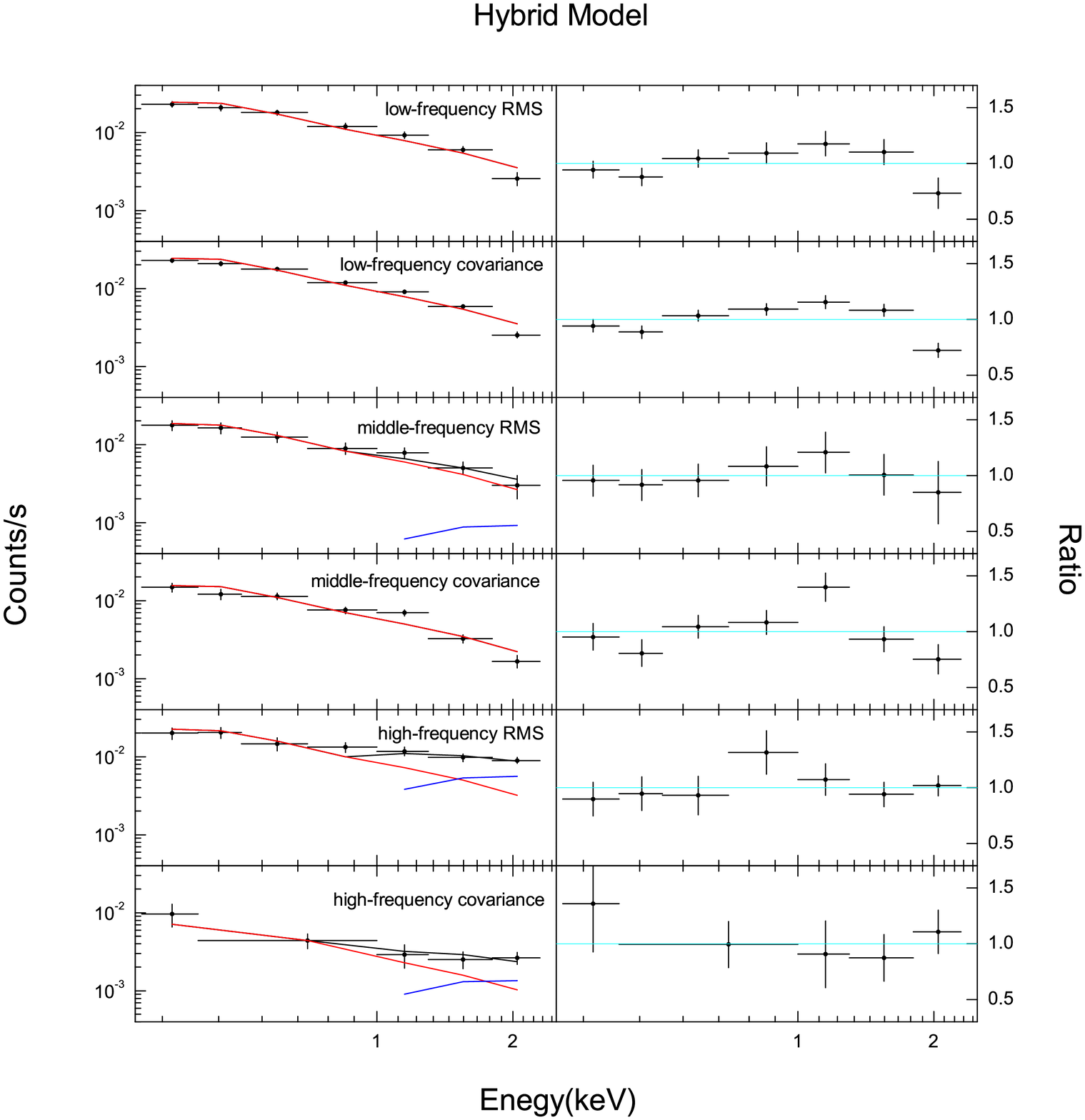}}
  \caption{The fitting results of the the RMS and covariance spectrum shown in the left panel by using the Hybrid model. The black, red and blue line represent the fitting results, the comptonisation radiation of hot corona and the reflection of distant material. The data-to-model ratio is shown in the right panel.}
\end{figure}

\subsubsection{the first step}
The Chi-square values/d.o.f of fitting result is 1031/955 (reduced $\chi^2=1.07$), which obtains the best fitting compared with the other four models in the first step. The hydrogen column density of the host galaxy absorption has upper limit value, $N_{H}(10^{20}cm^{-2})<0.48$. The X-ray continuum index, $\Gamma=1.75\pm0.02$, is consistent with the typical index. The photon index and electron temperature of warm corona are $2.08\pm0.10$ and $0.19\pm0.03$ keV, where the optical depth of the warm corona is about 40 agreed with the prediction of \citet{2020A&A...634A..92U}. The black hole spin also get a high value, $a=0.97$. The ionization degree of disk takes a moderate value ,$log\xi=1.37\pm0.10$. The iron abundances has a sub-solar value $A_{Fe}=0.49\pm0.11$ consistent with the result of Xu21.
\subsubsection{the second step}
The Chi-square values/d.o.f of fitting result is 1144/967 (reduced $\chi^2=1.18$), which is worse than that of the first step. The hydrogen column density of the host galaxy absorption is obtained about an order of $10^{20}cm^{-2}$ as same as the Galactic absorption. Other parameters obtained by the fitting result are consistent with that of the first step. In the right panel of Fig.14, we add the covariance spectrum of low, middle and high frequency into plot for comparison with the time-average spectrum. The deviation mainly comes from the variability spectrum fitting (Fig.15), which displays that the X-ray continuum radiated from hot corona is the main contributor, the reflection of distant material, $borus12$, increases the contribution at high frequency. But it still also can't interpret the residual error in the right panel of Fig.15.
\subsection{Double Reflection Model}
Double relativistic reflection components are considered to be the mechanism of the soft X-ray excess and other reflection features. We adopt the first reflection model with lamp post geometry and variable disk density, $relxilllpd$, which describes the reflection process of inner compact region and the second reflection model, $relxillcp$, with coronal geometry that assumes a density of $10^{15}$cm$^{-2}$, which is different from the first reflection component and outside of inner compact region of black hole. Therefore, we set the index of emissivity to be $q_{1}=q_{2}=3$, the inner radius of the disk is $50r_{g}$ and the outer of disk is the default value. We still fix the electron temperature at 300 keV. The photon index of two reflection components are linked to the hot corona. The iron abundances and disk inclination are linked with each other. The fitting results are shown in Table 6. and Fig.16.

\begin{table*}
\begin{center}
\begin{tabular}{ccccc}
  \hline
  \hline
  Description            &Component     &Parameter                   &Step 1             &Step 2       \\
  \hline
  Cross-normalization    &$constant$    &NuSTAR FPMA                 &$1.29\pm0.01$      &$1.29\pm0.01$ \\
                         &              &NuSTAR FPMB                 &$1.31\pm0.01$      &$1.31\pm0.01$ \\
  \hline
  Host galaxy Absorption &$ztbabs$      &$N_{H}(10^{20}cm^{-2})$     &$<1.23$            &$<1.94$  \\
  \hline
  Warm Absorption        &$zipcf$       &$N_{H}(10^{22}cm^{-2})$     &$<0.15$            &$<0.20$ \\
                         &$zipcf$       &$log\xi$                    &$<1.86$            &$<0.76$\\
                         &$zipcf$       &$C_{F}$                     &$<0.32$            &$0.28^{*}$\\
  \hline
  Hot Corona             &$nthcomp1$    &$\Gamma$                    &$1.71\pm0.02$      &$1.72\pm0.02$ \\
                         &$nthcomp1$    &$kT_{e}(keV)$               &$300$              &$300$  \\
  \hline
  Ionized Reflection     &$relxilllpd$  &$h(r_{g})$                  &$2.63\pm1.74$      &$2.62\pm1.49$ \\
                         &$relxilllpd$  &$a$                         &$0.99\pm0.10$      &$0.97\pm0.18$ \\
                         &$relxilllpd$  &$\theta_{disk}(degree)$     &$<67$              &$<60$  \\
                         &$relxilllpd$  &$log\xi$                    &$2.54\pm0.98$      &$2.27\pm0.27$  \\
                         &$relxilllpd$  &$A_{Fe}^{disk}$             &$0.67\pm0.65$      &$0.51\pm0.36$  \\
                         &$relxilllpd$  &$log[n_{e}/cm^{-3}]$        &$18.09\pm1.58$     &$18.68\pm0.22$ \\
                         &$relxillcp$   &$q_{1}=q_{2}$               &$3$                &$3$ \\
                         &$relxillcp$   &$R_{in}(r_g)$               &$50$               &$50$ \\
                         &$relxillcp$   &$log\xi$                    &$1.38\pm0.14$      &$1.47\pm0.12$ \\
 \hline
  Distant Reflection     &$boru12$     &$log[N_{H,tor}/cm^{-2}]$     &$23.03\pm0.12$     &$23.02\pm0.07$ \\
                         &$boru12$     &$C_{tor}$                    &$0.83\pm0.08$      &$0.82\pm0.04$ \\
                         &$boru12$     &$A_{Fe}$                     &$0.49\pm0.08$      &$0.55\pm0.11$  \\
  \hline
  $\chi^2/d.o.f$         &              &                            &$1058/954$        &$1189/966$     \\
  \hline
\end{tabular}
\caption{Fitting results of Double Reflection model for step 1 and 2. The asterisk represents that the parameter can't be constrained in our fitting and the value is pegged.}
\end{center}
\end{table*}

\begin{figure}
\centerline{
  \includegraphics[scale=0.3,angle=0]{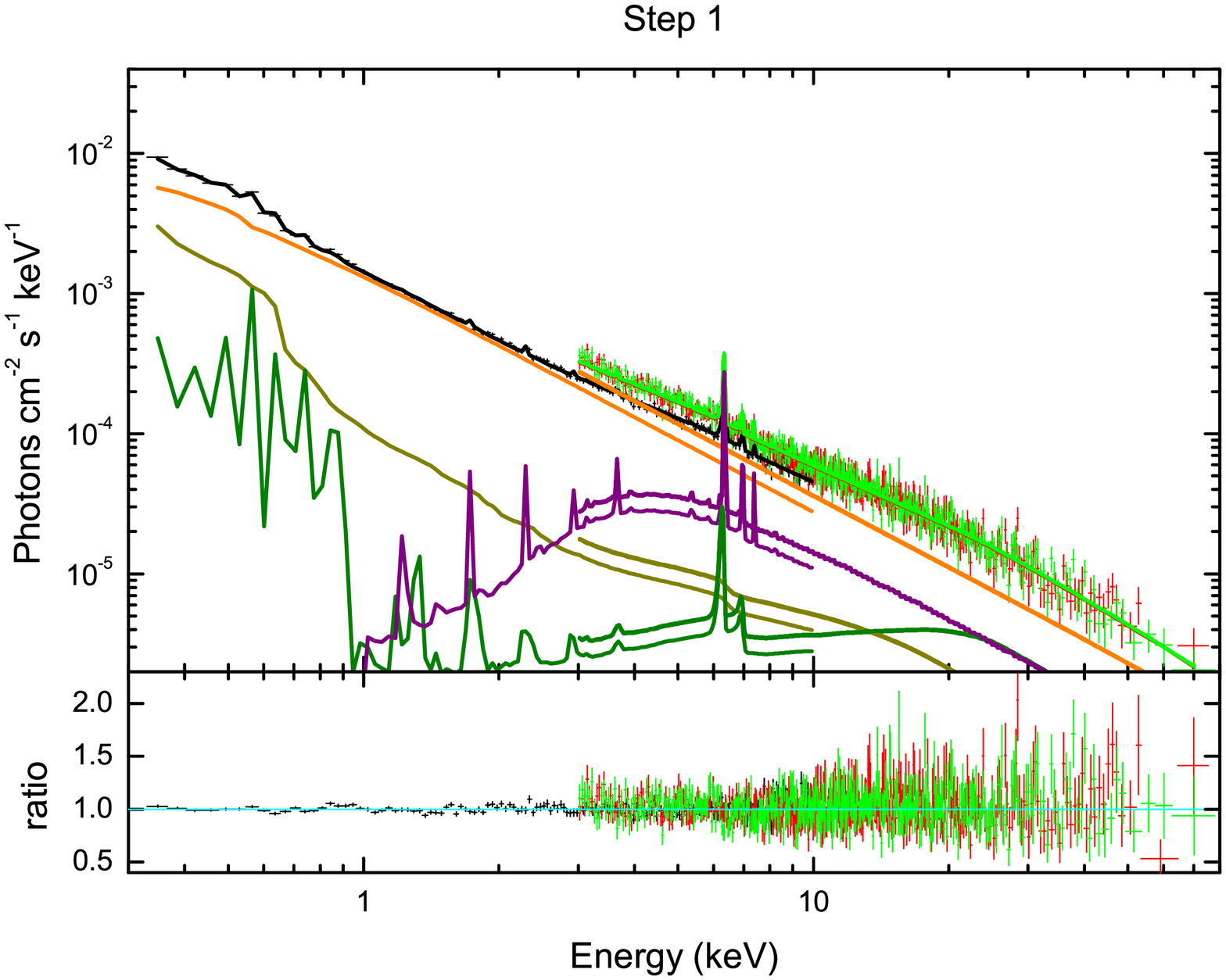}
  \includegraphics[scale=0.3,angle=0]{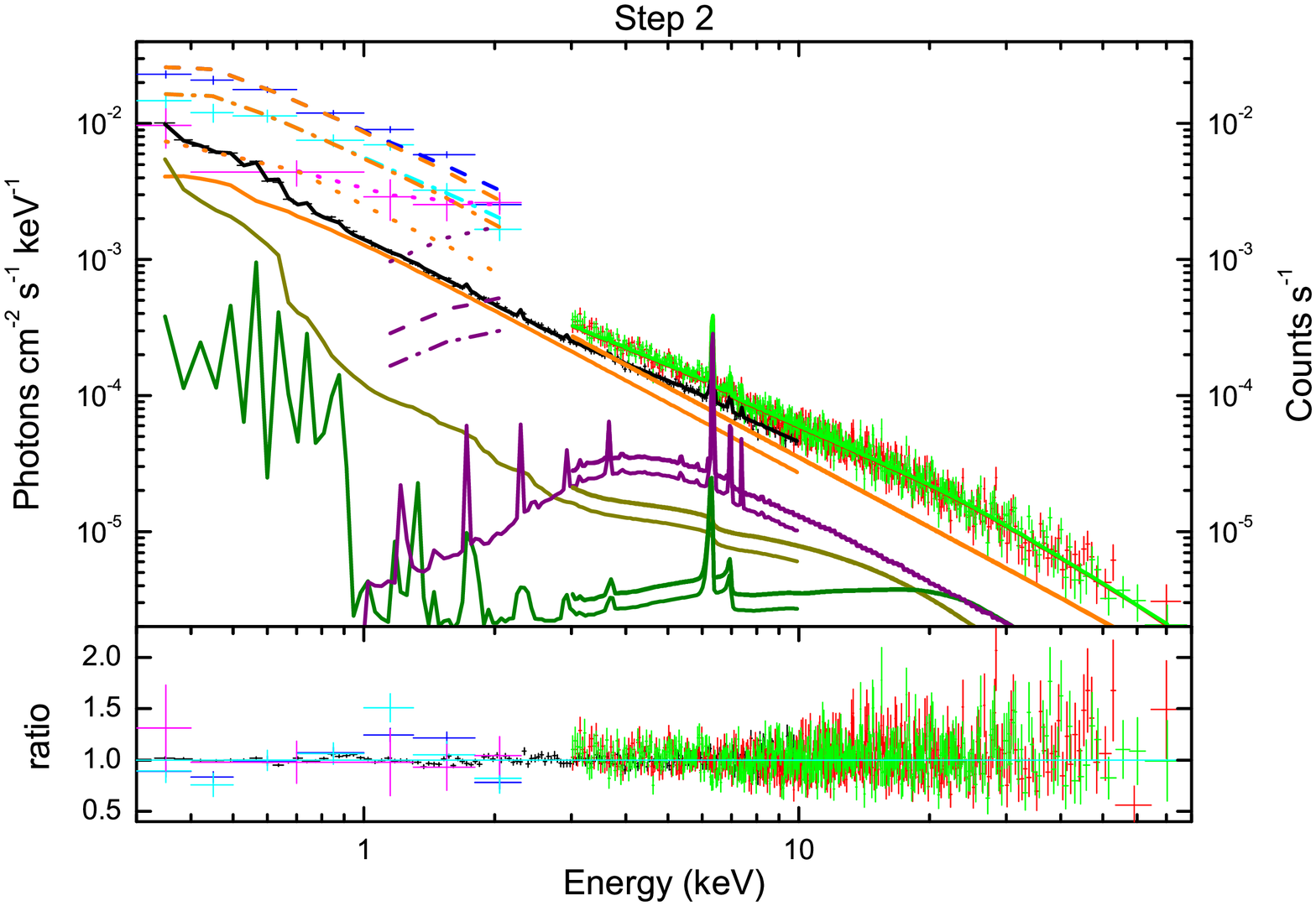}}
  \caption{The fitting results of the Double Reflection model for step 1 and 2 shown in left and right panel, respectively. Black, red and green data points are the time-average spectrum of XMM-Newton, NuSTAR FPMA, and FPMB data, respectively. Blue, cyan and magenta data points are the covariance spectrum of low, middle and high frequency, respectively. The corresponding colour lines are the best fitting results. Orange, dark yellow, olive and purple lines represent the radiation of hot corona, the first reflection, $relxilllpd$, the second reflection, $relxillcp$, and the reflection of distant material. The dash, dash-dot and dot lines represent the fitting results of covariance spectrum of low, middle and high frequency with each model component which is as same color as the corresponding solid lines. The photon flux unit, photons cm$^{-2}$ s$^{-1}$ keV$^{-1}$, is used for the time-average spectra and the count rate unit, counts s$^{-1}$, is used for the covariance spectrum in the right panel. The bottom panel is the data-to-model ratio.}
\end{figure}

\begin{figure}
\centerline{
  \includegraphics[scale=0.4,angle=0]{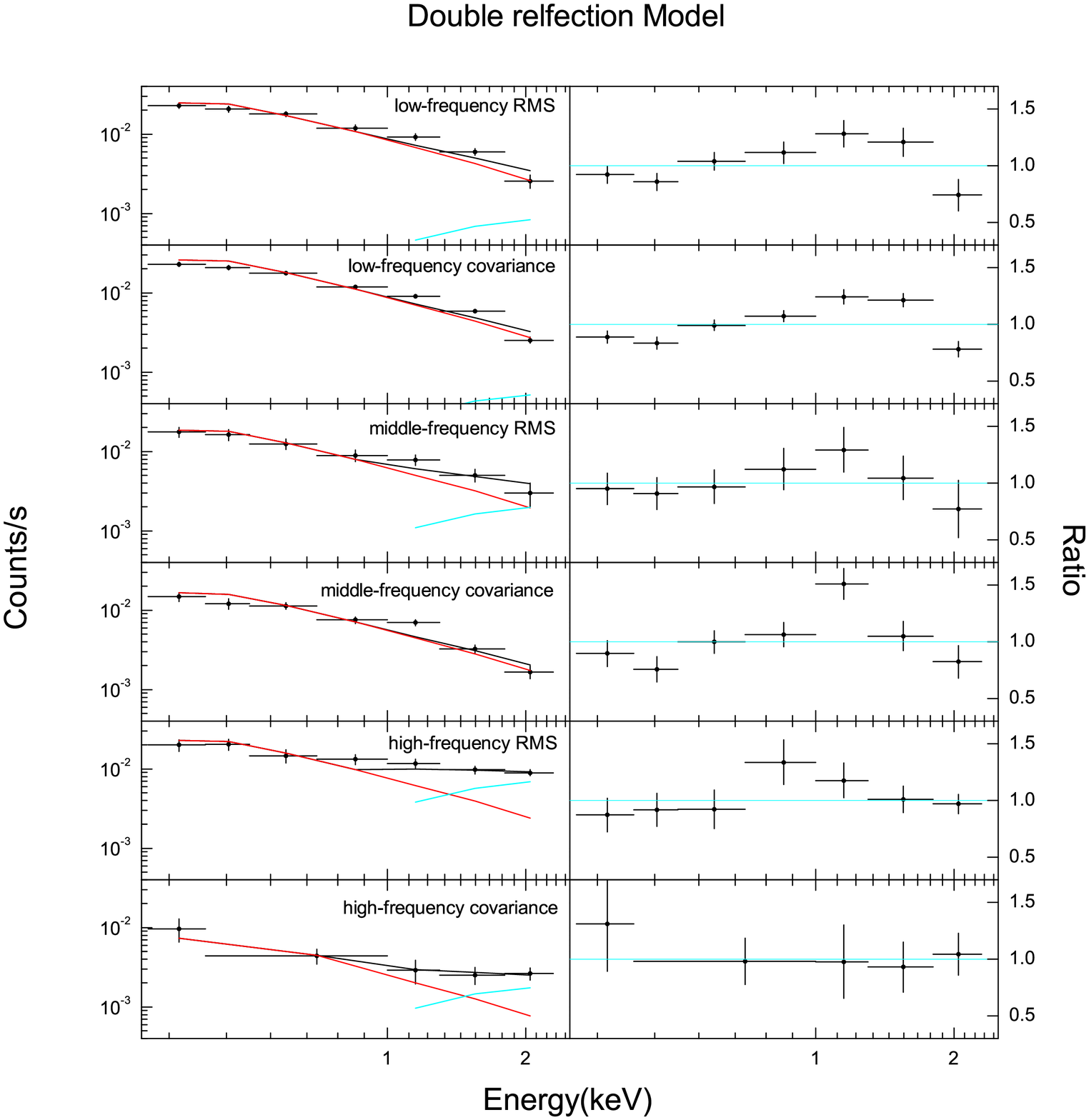}}
  \caption{The fitting results of the the RMS and covariance spectrum shown in the left panel by using the Double Reflection model. The black, red and cyan line represent the fitting results, the comptonisation radiation of hot corona and the reflection of distant material. The data-to-model ratio is shown in the right panel.}
\end{figure}

\subsubsection{the first step}
The Chi-square values/d.o.f of fitting result is 1058/954 (reduced $\chi^2=1.10$), which cannot be distinguished easily compared with the Warm corona and Reflection model in the first step. The hydrogen column density of the host galaxy absorption has upper limit value, $N_{H}(10^{20}cm^{-2})<1.23$. The photon index of hot corona is $\Gamma=1.71\pm0.02$ which is consistent with the typical value. The black hole spin also has a high value about 0.99. The ionization degree has a high value $log\xi=2.54\pm0.98$ for the first reflection and a moderate values, $log\xi=1.38\pm0.14$ for the second reflection. The disk iron abundances is less than one but with a high error.
\subsubsection{the second step}
The Chi-square values/d.o.f of fitting result is 1189/966 (reduced $\chi^2=1.23$), which is worse than that of the first step. In the right panel of Fig.16, we add the covariance spectrum of low, middle and high frequency into plot for comparison with the time-average spectrum. However, the parameters obtained by the fitting result are consistent with that of the first step. The deviation mainly comes from the RMS and covariance spectrum fitting (Fig.17). The two relativistic reflection components do not contribute to the variability spectrum. The radiation of hot corona can also not interpret the variability spectrum although it has a high contribution at the low frequency.
\subsection{Double Warm Corona Model}
Double Warm Corona model is made of two warm corona components, which was used to interpret the origin soft X-ray excess and X-ray Quasi-Periodic Oscillation for the famous Seyfert 1 galaxy, RE J1034+396 in the work of  \citet{2021MNRAS.500.2475J}. We also explore the model for ESO 362-G18. The seed photon temperature of the two warm coronas are fixed at 3 eV. The electron temperature of the hot corona is fixed at 300 keV. The fitting results are shown in Table 7. and Fig.18.

\begin{table*}
\begin{center}
\begin{tabular}{ccccc}
  \hline
  \hline
  Description            &Component     &Parameter                   &Step 1             &Step 2  \\
  \hline
  Cross-normalization    &$constant$    &NuSTAR FPMA                 &$1.29\pm0.01$      &$1.29\pm0.01$ \\
                         &              &NuSTAR FPMB                 &$1.31\pm0.01$      &$1.31\pm0.01$ \\
  \hline
  Host galaxy Absorption &$ztbabs$      &$N_{H}(10^{20}cm^{-2})$     &$<1.16$            &$<0.82$  \\
  \hline
  Warm Absorption        &$zipcf$       &$N_{H}(10^{22}cm^{-2})$     &$0.55\pm0.38$      &$0.55\pm0.13$ \\
                         &$zipcf$       &$log\xi$                    &$-0.22\pm0.11$     &$-0.25\pm0.09$\\
                         &$zipcf$       &$C_{F}$                     &$0.54\pm0.23$      &$0.54\pm0.07$\\
  \hline
  Hot Corona             &$nthcomp1$    &$\Gamma$                    &$1.67\pm0.03$      &$1.67\pm0.03$ \\
                         &$nthcomp1$    &$kT_{e}(keV)$               &$300$              &$300$  \\
  \hline
  Warm Corona            &$nthcomp2$    &$\Gamma$                    &$1.81\pm0.83$      &$1.81\pm0.06$   \\
                         &$nthcomp2$    &$kT_{e}(keV)$               &$0.17\pm0.13$      &$0.17\pm0.01$   \\
                         &$nthcomp3$    &$\Gamma$                    &$1.00^{*}$         &$1.01^{*}$   \\
                         &$nthcomp3$    &$kT_{e}(keV)$               &$0.30^{*}$         &$0.30\pm0.10$   \\
  \hline
  Distant Reflection     &$boru12$     &$log[N_{H,tor}/cm^{-2}]$     &$23.75\pm0.20$     &$23.75\pm0.21$ \\
                         &$boru12$     &$C_{tor}$                    &$0.58\pm0.32$      &$0.59\pm0.36$ \\
                         &$boru12$     &$A_{Fe}$                     &$0.69\pm0.16$      &$0.71\pm0.16$  \\
  \hline
  $\chi^2/d.o.f$         &              &                            &$1072/957$        &$1085/969$     \\
  \hline
\end{tabular}
\caption{Fitting results of Double Warm Corona model for step 1 and 2. The asterisk represents that the parameter can't be constrained in our fitting and the value is pegged.}
\end{center}
\end{table*}

\begin{figure}
\centerline{
  \includegraphics[scale=0.3,angle=0]{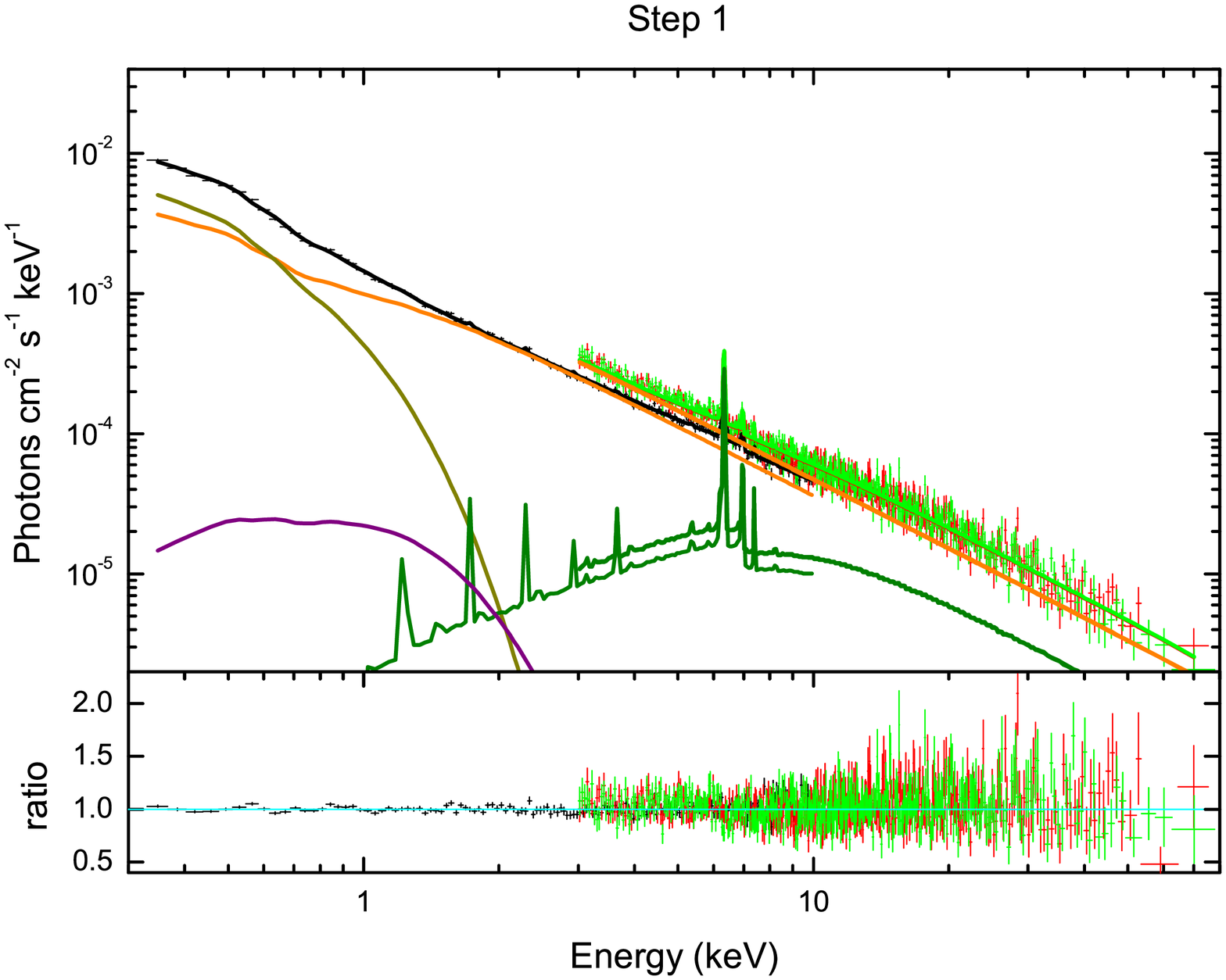}
  \includegraphics[scale=0.3,angle=0]{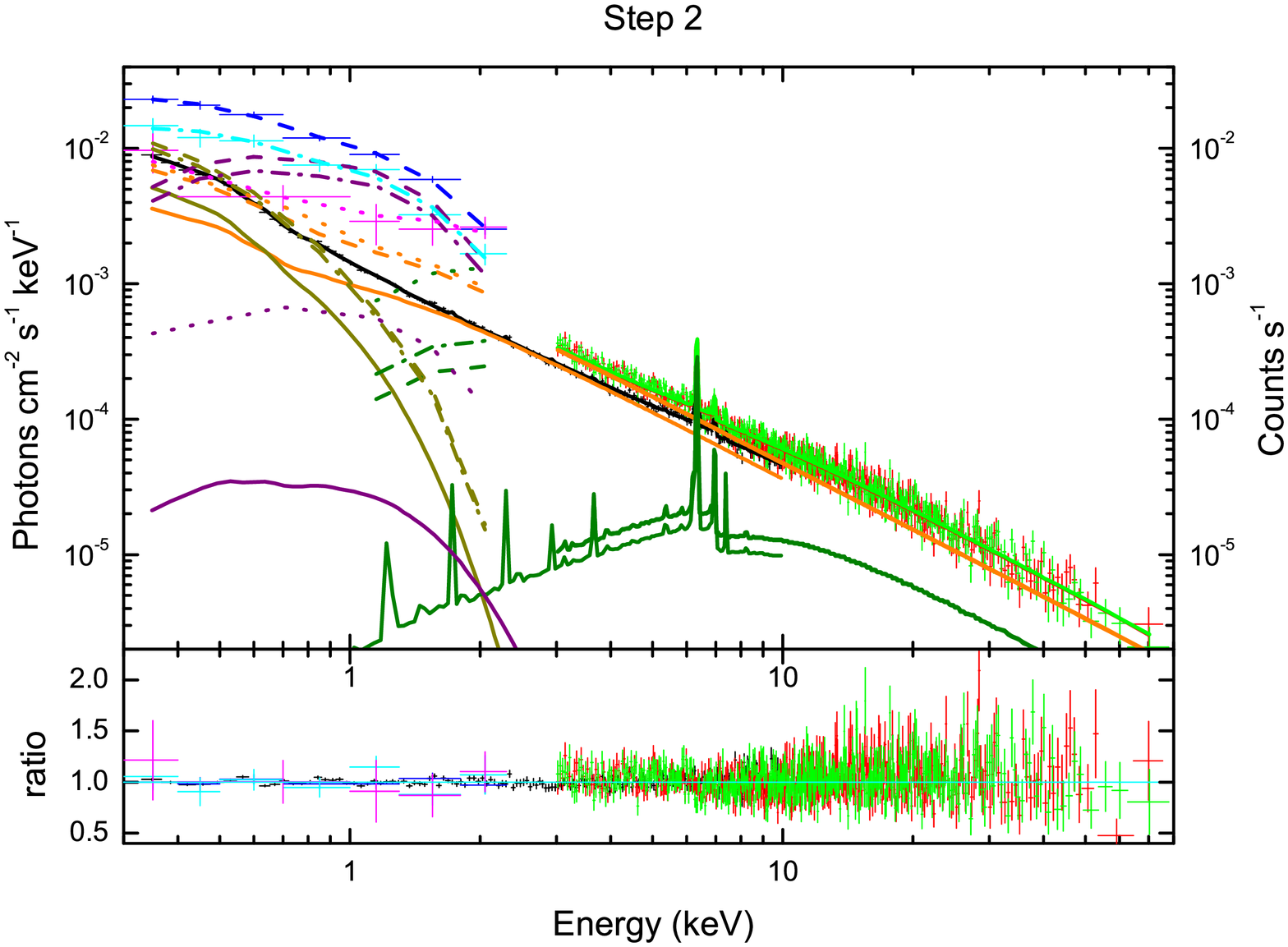}}
  \caption{The fitting results of the Double Warm Corona model for step 1 and 2 shown in left and right panel, respectively. Black, red and green data points are the time-average spectrum of XMM-Newton, NuSTAR FPMA, and FPMB data, respectively. Blue, cyan and magenta data points are the covariance spectrum of low, middle and high frequency, respectively. The corresponding colour lines are the best fitting results. Orange, dark yellow, olive and purple lines represent the radiation of hot corona, the first warm corona, the reflection of distant material and the second warm corona. The dash, dash-dot and dot lines represent the fitting results of covariance spectrum of low, middle and high frequency with each model component which is as same color as the corresponding solid lines. The photon flux unit, photons cm$^{-2}$ s$^{-1}$ keV$^{-1}$, is used for the time-average spectra and the count rate unit, counts s$^{-1}$, is used for the covariance spectrum in the right panel. The bottom panel is the data-to-model ratio. }
\end{figure}

\begin{figure}
\centerline{
  \includegraphics[scale=0.4,angle=0]{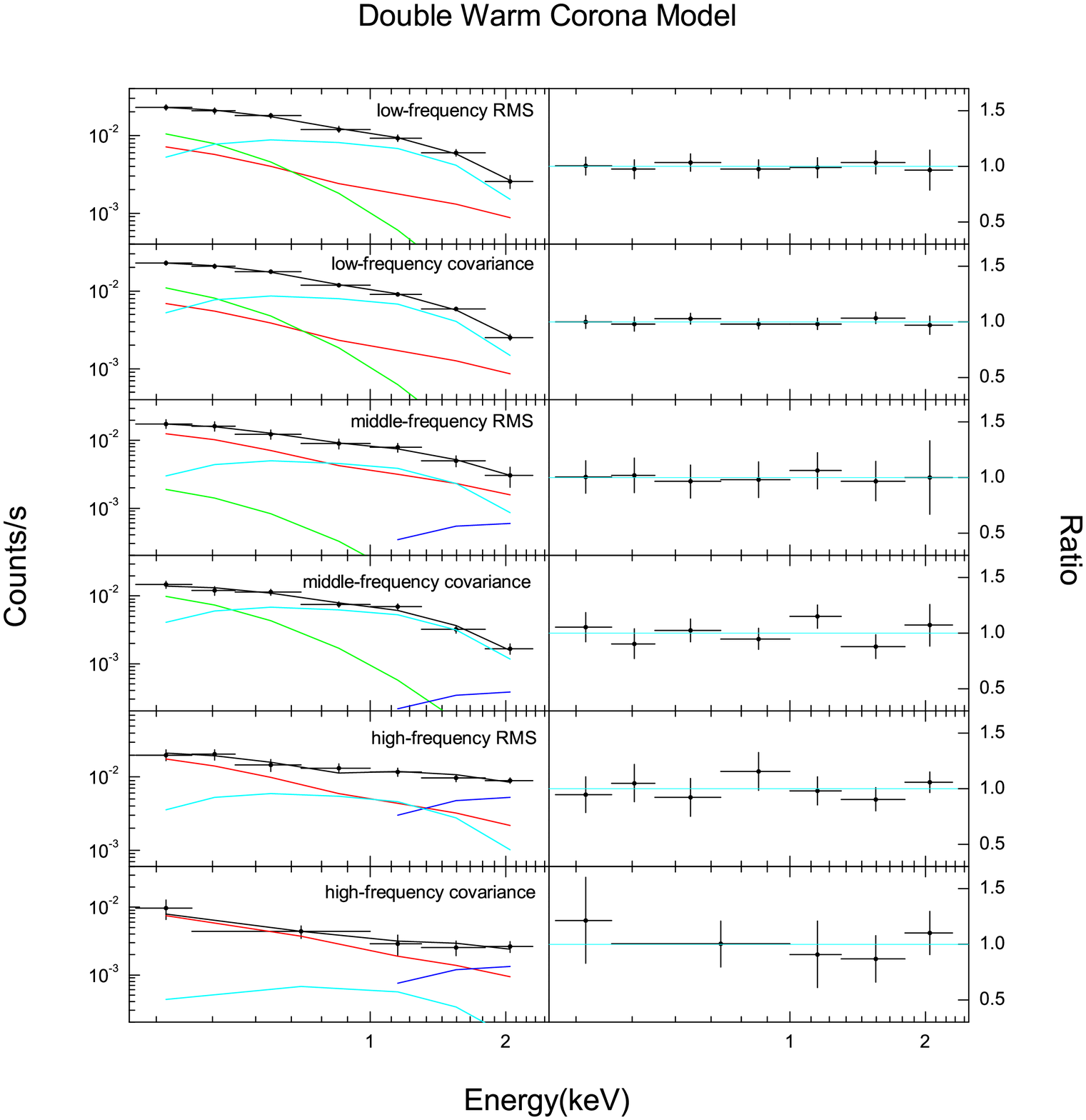}}
  \caption{The fitting results of the the RMS and covariance spectrum shown in the left panel by using the Double Warm Corona model. The black, red, green, cyan and blue line represent the fitting results, the comptonisation radiation of hot corona, the first warm corona, the second warm corona, and the reflection of distant material. The data-to-model ratio is shown in the right panel.}
\end{figure}

\subsubsection{the first step}
The Chi-square values/d.o.f of fitting result is 1072/957 (reduced $\chi^2=1.12$).  The hydrogen column density of the host galaxy absorption has upper limit value, $N_{H}(10^{20}cm^{-2})<1.16$. The photon index of hot corona has a low value compared with the typical value, $\Gamma=1.67\pm0.03$. The first warm corona has a soft spectrum of $\Gamma=1.81$, and the electron temperature is 0.17 keV, it corresponds the optical depth to be about 53. However, the second warm corona cannot constrain the spectrum index and the electron temperature, and it is a reasonable result because the second warm corona is weak compared with other components in the first step (left panel of Fig.18) and the data cannot constrain the component effectively.
\subsubsection{the second step}
The Chi-square values/d.o.f of fitting result is 1085/969 (reduced $\chi^2=1.12$), which is the best fitting result of all of the second step mentioned above and means that the double warm corona model could interpret the variability spectrum well (Fig.19). We find that the second warm is dominant at low frequency, the hot corona and the distant reflection increase their contribution at high frequency. The photon index of hot corona, $\Gamma=1.67\pm0.03$, is consistent with that of the first step. The first warm corona's photon index is $1.81\pm0.06$ and the electron temperature is 0.17 keV, which corresponds the optical depth to be about 53. The second warm corona's photon index is pegged a very low value which corresponds the optical depth to be a very high value.
\subsubsection{constrain the optical depth}
Through the above analysis, the double warm corona model could reproduct both of the time-average spectrum and the variability spectrum. However, the fitting result shows that the optical depth of the second warm corona are too high to be impenetrable based on previous work. The thermally comptonized continuum model, $nthcomp$, has two main free parameters such as the photon index, $\Gamma$, and the electron temperature, $kT_{e}$. The optical depth is inferred from two parameters by the formula as follows \citep{1996MNRAS.283..193Z}:
\begin{equation}
\tau=\sqrt{2.25+\frac{3}{(kT_{e}/m_{e}c^{2})[(\Gamma+0.5)^{2}-2.25]}}-1.5,
\end{equation}
where $\tau$ is the optical depth, $m_{e}$ is the electron mass and $m_{e}c^{2}$ is 511 keV, $\Gamma$ is the photon index. Therefore, the optical depth could be evaluated from above formula easily. But, the optical depth is not constrained easily in fitting process by using the $nthcomp$ model in XSPEC. In order to solve the problem, we adopt the comptonization of soft photons in a hot plasma model, $comptt$, to fit the soft X-ray excess and replace $nthcomp$. $comptt$ also have two main free parameters such as the electron temperature, $kT_{e}$ and optical depth, $\tau$. So, we can constrain the optical depth directly in the fitting process. The fitting results are shown in Table 8. and Fig.20 just for the second step. Chi-square values/d.o.f of fitting result is 1081/969 (reduced $\chi^2=1.11$). The photon index has a low value, $\Gamma=1.66\pm0.03$ compared with the typical value. The optical depths of two warm corona are constrained well due to fitting with the variability spectrum, which is agreed with the prediction of \citet{2020A&A...634A..92U}. The second warm corona component contributes for the fitting statistics as $\Delta \chi^{2}_{2thWC}/d.o.f=220/3$, which implies the second warm corona component is significant. Fig.21 shows the detailed fitting result of the variability spectrum, which displays the second warm corona component to be the main generation mechanism, and the hot corona and distant reflection dominant the radiation gradually at the high frequency.

\begin{table*}
\begin{center}
\begin{tabular}{cccc}
  \hline
  \hline
  Description            &Component     &Parameter                   &Step 2         \\
  \hline
  Cross-normalization    &$constant$    &NuSTAR FPMA                 &$1.29\pm0.01$  \\
                         &              &NuSTAR FPMB                 &$1.31\pm0.01$  \\
  \hline
  Host galaxy Absorption &$ztbabs$      &$N_{H}(10^{20}cm^{-2})$     &$<0.06$  \\
  \hline
  Warm Absorption        &$zipcf$       &$N_{H}(10^{22}cm^{-2})$     &$0.66\pm0.10$   \\
                         &$zipcf$       &$log\xi$                    &$-0.27\pm0.09$  \\
                         &$zipcf$       &$C_{F}$                     &$0.62\pm0.06$   \\
  \hline
  Hot Corona             &$nthcomp1$    &$\Gamma$                    &$1.66\pm0.03$   \\
                         &$nthcomp1$    &$kT_{e}(keV)$               &$300$       \\
  \hline
  Warm Corona            &$comptt1$    &$kT_{e}(keV)$                &$0.18\pm0.01$       \\
                         &$comptt1$    &$\tau$                       &$26.40\pm3.65$        \\
                         &$comptt2$    &$kT_{e}(keV)$                &$0.34\pm0.02$    \\
                         &$comptt2$    &$\tau$                       &$29.08\pm3.43$    \\
  \hline
  Distant Reflection     &$boru12$     &$log[N_{H,tor}/cm^{-2}]$     &$23.58\pm0.15$  \\
                         &$boru12$     &$C_{tor}$                    &$0.80\pm0.14$    \\
                         &$boru12$     &$A_{Fe}$                     &$0.89\pm0.26$   \\
  \hline
  $\chi^2/d.o.f$         &              &                            &$1081/969$     \\
  \hline
\end{tabular}
\caption{Fitting results of Double Warm Corona model just for step 2 by using $comptt$.}
\end{center}
\end{table*}

\begin{figure}
\centerline{
  \includegraphics[scale=0.4,angle=0]{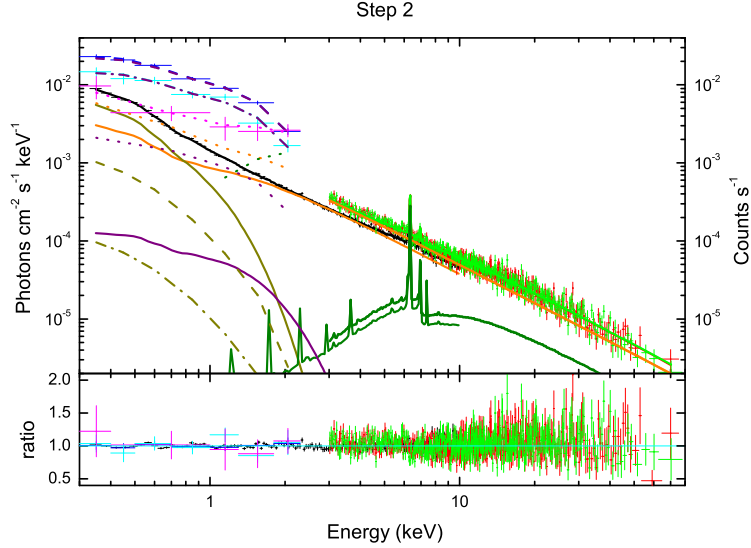}}
  \caption{The fitting results of the double warm corona model by using $comptt$ for step 2 only. Black, red and green data points are the XMM-Newton, NuSTAR FPMA, and FPMB data, respectively. Blue, cyan and magenta data points are the covariance spectrum of low, middle and high frequency, respectively. The corresponding colour lines are the best fitting results. Orange, dark yellow, purple and olive lines represent the radiation of hot corona, the first warm corona, the second warm corona and the reflection of distant material. The dash, dash-dot and dot lines represent the fitting results of covariance spectrum of low, middle and high frequency with each model component which is as same color as the corresponding solid lines. The photon flux unit, photons cm$^{-2}$ s$^{-1}$ keV$^{-1}$, is used for the time-average spectra and the count rate unit, counts s$^{-1}$, is used for the covariance spectrum. The bottom panel is the data-to-model ratio.}
\end{figure}

\begin{figure}
\centerline{
  \includegraphics[scale=0.4,angle=0]{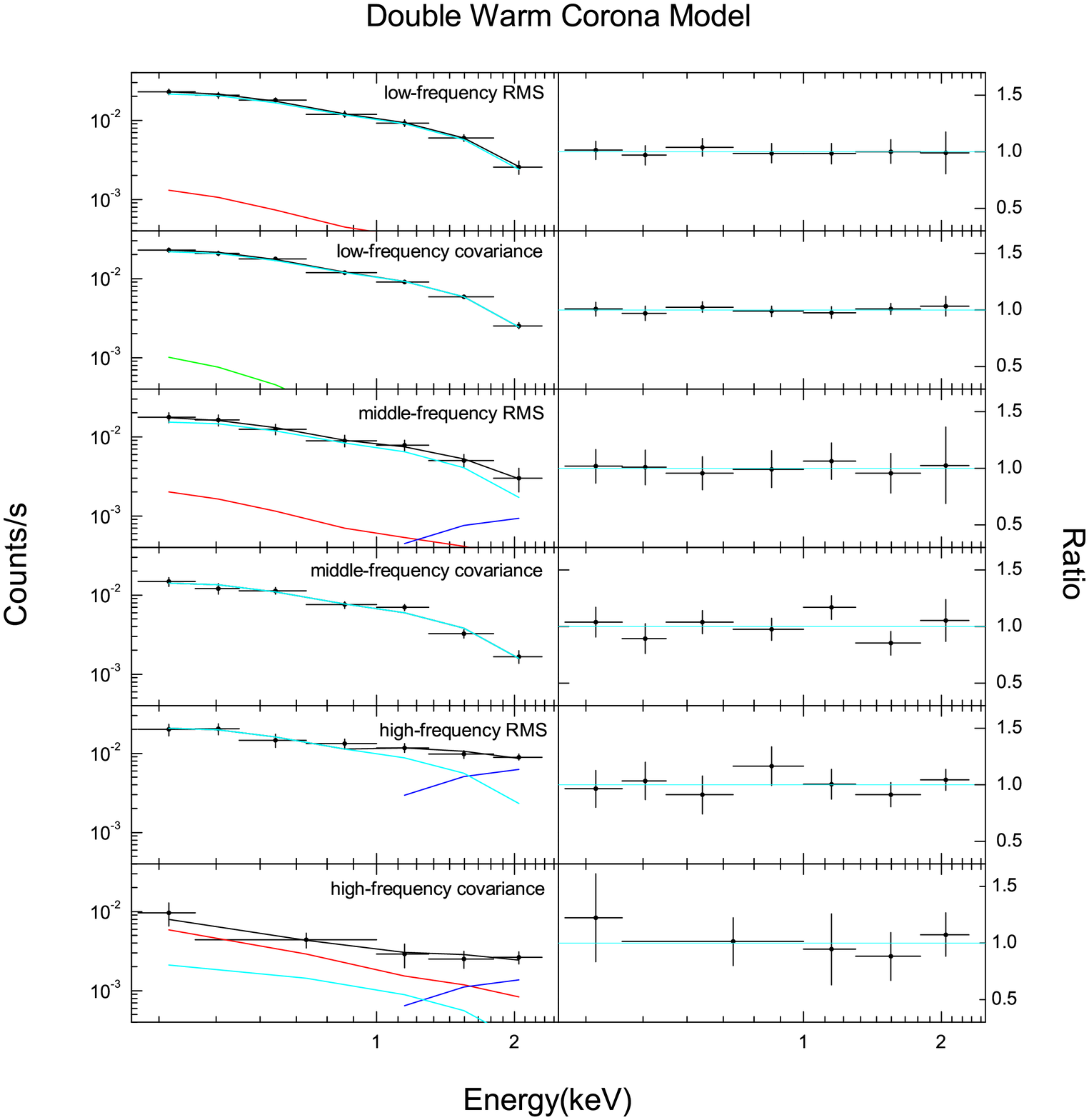}}
  \caption{The fitting results of the the RMS and covariance spectrum shown in the left panel by using the double warm corona model which adopt the $comptt$ model. The black, red, green, cyan and blue lines represent the fitting results, the comptonisation radiation of hot corona, the first warm corona, the second warm corona, and the reflection of distant material. The data-to-model ratio is shown in the right panel.}
\end{figure}

\section{Discussion and Conclusions}
\subsection{Comparing with Xu21}
ESO 362-G18 had been explored in detail by Xu21 using warm corona and relativistic reflection models, and has the same data just like we adopt. We use the same model of Xu21 to review the source by employing the XSPEC software with our data set. We fix all of parameters obtained by Xu21 and the Chi-square values/d.o.f are 1317/1032 for the warm corona model (Model A from the table 2 of Xu21), 1712.86/1030 for the relativistic reflection model (Model B from the table 2 of Xu21) and 1183/1030 for the hybrid model (Model C from the table 2 of Xu21). Even so, we also obtain a compact hot corona ($h\sim3r_g$) and a rapidly spinning black hole consistent with Xu21. However, the relativistic reflection cannot interpret the variability spectrum well in our work. Therefore, the spectral analysis combining with average-time spectrum and variability spectrum could help to decouple the different components which cannot be distinguished easily based on only time-averaged spectra fit statistics.

In the part 3 of Xu21, they test the neutral reflection model, $xillver$ \citep{2010ApJ...718..695G}, and find the Fe K emission is overpredicted. So they instead use the $borus12$ model like as we used in this paper. We also test the neutral reflection model replaced the $borus12$ model in the Double Warm Corona model. Fig.22 is shown the fitting result of the scenario in the second step and the first warm corona component has little contribution for the spectrum. We find the neutral reflection model cannot interpret the covariance spectrum in the high frequency, which implies the neutral reflection doesn't include the cold disk reflection radiation in the soft energy band in ESO 362-G18.

\begin{figure}
\centerline{
  \includegraphics[scale=0.4,angle=0]{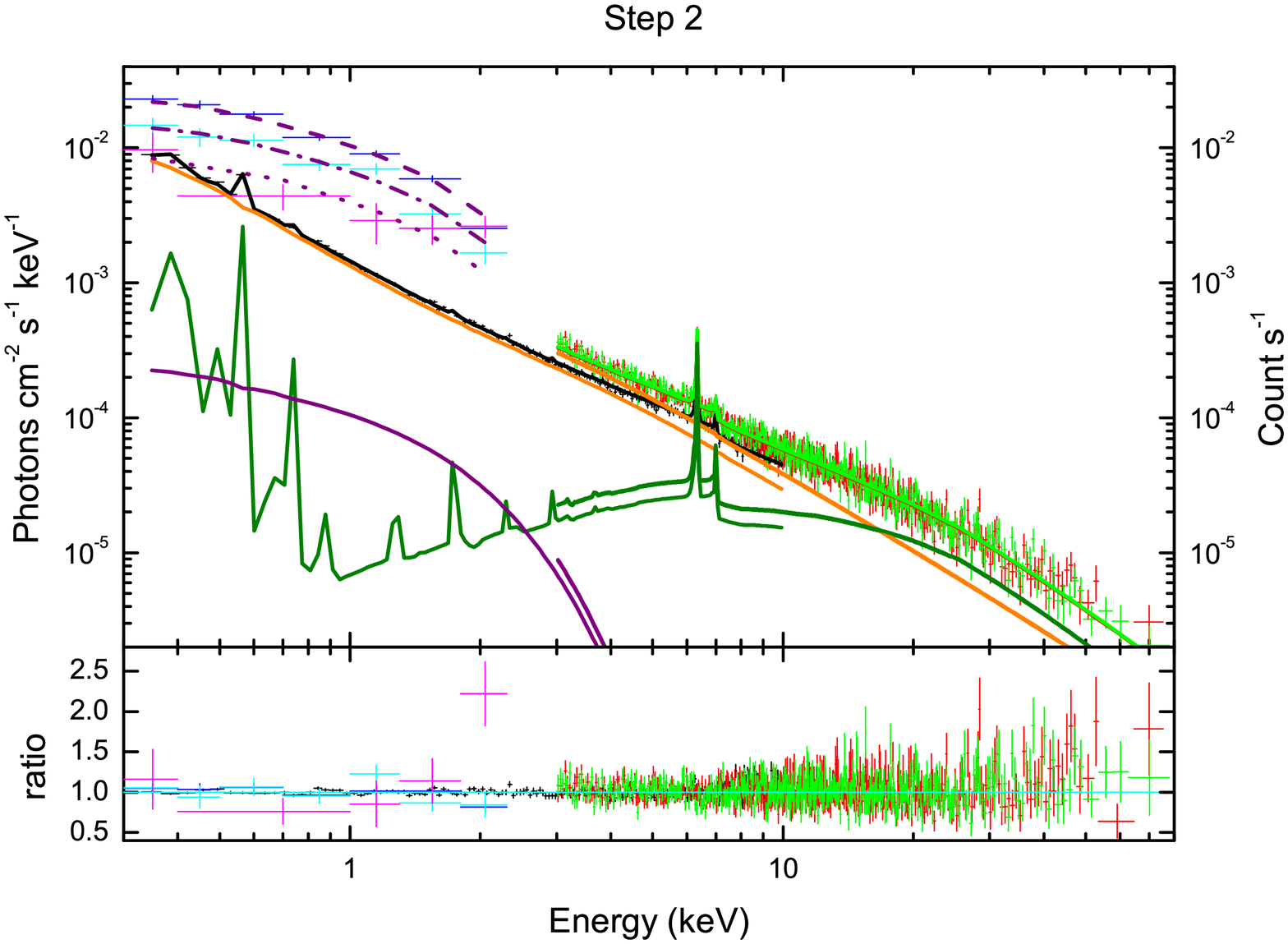}}
  \caption{The fitting results of the double warm corona model and replace $borus12$ with the neutral reflection model, $xillver$ for step 2 only. Black, red and green data points are the XMM-Newton, NuSTAR FPMA, and FPMB data, respectively. Blue, cyan and magenta data points are the covariance spectrum of low, middle and high frequency, respectively. The corresponding colour lines are the best fitting results. Orange, purple and olive lines represent the radiation of hot corona, the second warm corona and the neutral reflection, the first warm corona component has little contribution for the spectrum. The dash, dash-dot and dot lines represent the fitting results of covariance spectrum of low, middle and high frequency with each model component which is as same color as the corresponding solid lines. The photon flux unit, photons cm$^{-2}$ s$^{-1}$ keV$^{-1}$, is used for the time-average spectra and the count rate unit, counts s$^{-1}$, is used for the covariance spectrum. The bottom panel is the data-to-model ratio.}
\end{figure}

\subsection{The photon index of hot corona interpretation}
From spectral analysis mentioned above, we find the that Double Warm Corona model is the best interpretation for both of time-average spectrum and variability spectrum. However, the photon index of the X-ray continuum from the hot corona has a lower value of about $1.65$ than the typical value, $\sim1.8$. The radiation mechanism of the hot corona is unknown so far. The magnetic field could play a crucial role to transport substantial amounts of energy from disk to the corona vertically (e.g. \citet{2006ApJ...640..901H}). The differential rotation of accretion disk and the movement of the plasma will produce a turbulent environment, which could generate the magnetic field reconnection and accelerate the electrons. Electrons are accelerated by the scattering of charged particles from randomly moving plasma clouds which act as magnetic mirrors that reflect the electrons elastically. Those electrons could gain energy by a head-on collision with plasma clouds, this is so called as the first-order Fermi acceleration \citep{1949PhRv...75.1169F}. The electron spectrum obtained from the first-order Fermi acceleration driven by magnetic reconnection can be written as \citep{2012MNRAS.422.2474D}
\begin{equation}
N(\gamma)\sim \gamma^{-n},
\end{equation}
where the index $n$ is given by
\begin{equation}
n=\frac{C+2}{C-1}, C=\frac{\rho_{2}}{\rho_{1}},
\end{equation}
where $C$ is the compression. $\rho_{1}$ is the density of upstream plasma of shock generated by magnetic reconnection and $\rho_{2}$ is the downstream density. Then the soft photons radiated from the disk will take place the inverse Compton scattering with these electrons. The Compton spectrum for incident photons (with energy frequency $\epsilon_{i}$) can be written as \citep{1970RvMP...42..237B}
\begin{equation}
P(\epsilon)=8\pi r^{2}_{0}ch\int f(\epsilon/4\gamma^{2}\epsilon_{i})n_{ph}(\epsilon_{i})d\epsilon_{i},
\end{equation}
where

\begin{eqnarray}
f(x)&=&x+2x^{2}lnx+x^{2}-2x^{3}, 0<x<1; \nonumber \\
    &=&0, x>1.
\end{eqnarray}
$r_{0}=e^{2}/m_{e}c^{2}$ is the classical electron radius, $c$ is the light velocity, $h$ is Planck constant, $\epsilon$ is the energy frequency of scattered photon, $\gamma$ is the Lorentz factor, $n_{ph}$ is the photon number density. Then, we have the Compton spectrum for the power-law electron spectrum (Equation.10):
\begin{equation}
j(\epsilon)=\int P(\epsilon)N(\gamma)d\gamma=8\pi r^{2}_{0}chG,
\end{equation}
where
\begin{equation}
G=\int \int N(\gamma)f(\frac{\epsilon}{4\gamma^{2}\epsilon_{i}})n_{ph}(\epsilon_{i})d\epsilon_{i}d\gamma,
\end{equation}
In generally, the electron spectrum has cutoff energy shape ($\gamma_{1}<\gamma<\gamma_{2}$), if $\epsilon>>\epsilon_{i}$, the lower limit of integral is $(\epsilon/4\epsilon_{i})^{0.5}$ inferred from Equation.13 instead of $\gamma_{1}$. When the upper limit of integral, $\gamma_{2}\rightarrow \infty$, we obtain the Compton spectrum:
\begin{equation}
j(\epsilon)=\pi r^{2}_{0}chN2^{n+3}\frac{n^{2}+4n+11}{(n+3)^{2}(n+1)(n+5)}\epsilon^{-\frac{n-1}{2}}\int \epsilon_{i}^{\frac{n-1}{2}}n_{ph}(\epsilon_{i})d\epsilon_{i}.
\end{equation}
The number density of incident photons from the accretion disk is given by
\begin{equation}
n_{ph}(\epsilon_{i})=\frac{U_{ph}(\epsilon_{i})d\epsilon_{i}}{h\epsilon_{i}}=\frac{8\pi\epsilon_{i}^{2}}{c^{3}}\frac{1}{e^{\frac{h\epsilon_{i}}{kT}}-1}d\epsilon_{i},
\end{equation}
where $U_{ph}$ is the energy density of radiation field. Then, the Compton spectrum can be written as
\begin{equation}
j(\epsilon)=\frac{2r^2_{0}}{\hbar^{2}c^{2}}b(n)N(kT)^{3}(\frac{kT}{h\epsilon})^{\frac{n-1}{2}}.
\end{equation}
where
\begin{equation}
b(n)=\frac{2^{n+3}(n^{2}+4n+11)\Gamma(\frac{n+3}{2})\zeta(\frac{n+3}{2})}{(n+3)^{2}(n+1)(n+5)},
\end{equation}
where $\zeta(x)$ is the Riemann $\zeta$ function and $\Gamma(x)$ is the $\gamma$ function. We find the index of the Compton spectrum is $(n-1)/2$ obtained by Equation.18. Thus, the hardest spectrum could be obtained $\Gamma\sim 0$ when $n\rightarrow 1$, which corresponds the compression, $C$, to be a very large value based on the Equation.11. This, of course, is impossible, but it give a mechanism to obtain a harder spectrum in the hot corona. The photon index of $\Gamma=1.65$ obtained by using the double warm model implies the compression equal to be about 2 , which means the downstream density of shock is two times larger than the upstream density in the magnetic reconnection. It is noted that the hot corona is considered to be made of by thermal electron generally, but we replace the thermal model, $nthcomp$, with a simple power-law model, to obtain a similar photon index, which means that the hot corona radiation could be generated by a non-thermal process.
\subsection{The broad Fe K$\alpha$ line}
If the double warm corona is true, the broad Fe K$\alpha$ line should not be exist because the relativistic reflection component is excluded through the spectral analysis. However, AG14 and Xu21 all reported showing the broad Fe K$\alpha$ line. We also test the Chi-square values of 4-10keV by using the warm corona and reflection model, $\chi^2=425$ for warm corona model and $\chi^{2}=414$ for reflection model, which implies that the relativistic effect should be considered and the broad Fe K$\alpha$ line should be exist. But, the X-ray spectrum of some AGNs with the reported broad Fe K$\alpha$ line had been modeled successfully with narrow lines, and the complexity of reflection continuum could lead to a broad-line interpretation (e.g. \citet{2014ApJ...797...12M, 2016MNRAS.462.4038Y} and the recent work, \citet{2021arXiv210811971T}), which means that the narrow emission is sufficient to model the broad Fe K$\alpha$ line. The origin of the broad Fe K$\alpha$ line is beyond the scope of this paper, so the broad Fe K$\alpha$ line origin of ESO 362-G18 need to be studies deeply in the future work.
\subsection{The X-ray emitting region}
AG14 proposed that the X-ray radiation region locate within50 $r_{g}$  by analyzing the absorption variability with 2005-2010 multi-epoch X-ray observations. The scale of50 $r_{g}$ corresponds to $10^{4}$ light-seconds and the corresponding frequency is $10^{-4}$ Hz which just is within the middle frequency mentioned in the Section.3. From the middle panel of Fig.6, there are no clear lags below 3keV of middle frequency, but it has some leading at 3-5keV with about 2ks and then has some lag at a high energy, about 8-10 keV with about 2ks too, which is the reverberation characteristics because 3-5 keV energy band represent the X-ray continuum radiated from hot corona to illuminate the disk and the lag at 8-10 keV could be a tail of 'Compton hump'. The scale corresponding 2 ks is about9 $r_{g}$ which is less than the X-ray radiation region of50 $r_{g}$. The clear hard-lags shown at the low frequency appear to be $\sim10$ ks which corresponds45 $r_g$ consistent with the radiation region predicted by AG14. In the section 4, we find the double warm model could interpret the variability spectrum well. The second warm corona component is the main generation mechanism, and the hot corona and distant reflection dominant the radiation gradually at the high frequency, which means the warm corona component dominate the lager scale of the X-ray emitting region and the hot corona is mainly concentrated in the inner compact region which is consistent with the hypothesis that the hot corona is a compact source. The contribution of distant reflection at the high frequency could be considered as that the clouds in the distant reflection region have small scale which corresponds the high frequency.
\subsection{Exploring the covariance spectrum property by compared with time-average spectra}
The spectrometer of observation device is expected to try to find out the spectrum of a source. However, the spectrometer obtains is not the actual spectrum, but rather photon counts within specific instrument channels. This observed spectrum is related to the actual spectrum of the source as follows:
\begin{equation}
C(I)=\int^\infty_0f(E)R(I,E)dE,
\end{equation}
where $C(I)$ is the photon counts within specific instrument channels, $I$, $f(E)$ is the actual spectrum of the source,  $R(I,E)$ is the instrumental response with units of cm$^2$. However, $f(E)$ can not be determined by inverting Equation.20 due to the unstableness of $C(I)$ \citep{arn96}. The usual alternative is to try to choose a model spectrum, $f_b(E)$, that can be described in terms of a few parameters, and match, or fit it to the data obtained by the spectrometer. Therefore, the actual spectrum of the source depend on the choice of model. Then, we explore the covariance spectrum property by compared with the actual time-average spectra obtained by the Double Warm Corona model shown in Fig.23. The covariance data adopted is same as Fig.9, but is divided by the scale of energy bin, which differs by one normalized factor from time-average spectrum because the response could be assumed a unit diagonal matrix for the covariance spectrum. Therefore, we adjust the normalized factor of covariance spectrum artificially to display the difference from the time-average spectra clearly in Fig.23. The blue solid line represents the continuum radiation of hot corona and the soft X-ray excess is apparent below 2 keV in the time-average spectra. Significantly, the soft X-ray excess of covariance spectrum is more prominent than that of time-average spectra in 0.5-2 keV, in which the second warm corona component is prominent. Below 0.5 keV, the soft X-ray excess of time-average spectra is stronger than that of covariance spectrum, in which the first warm corona component is prominent. So, the soft X-ray excess consist of the two warm corona components, but the variable degree of the first warm corona component is far less than the second. From Fig.23, we can see the soft X-ray excess of covariance spectrum decrease gradually with frequency until none of the soft X-ray excess is shown in the high-frequency in which the variability mainly cames from the hot corona.
\begin{figure}
\centerline{
  \includegraphics[scale=0.23,angle=0]{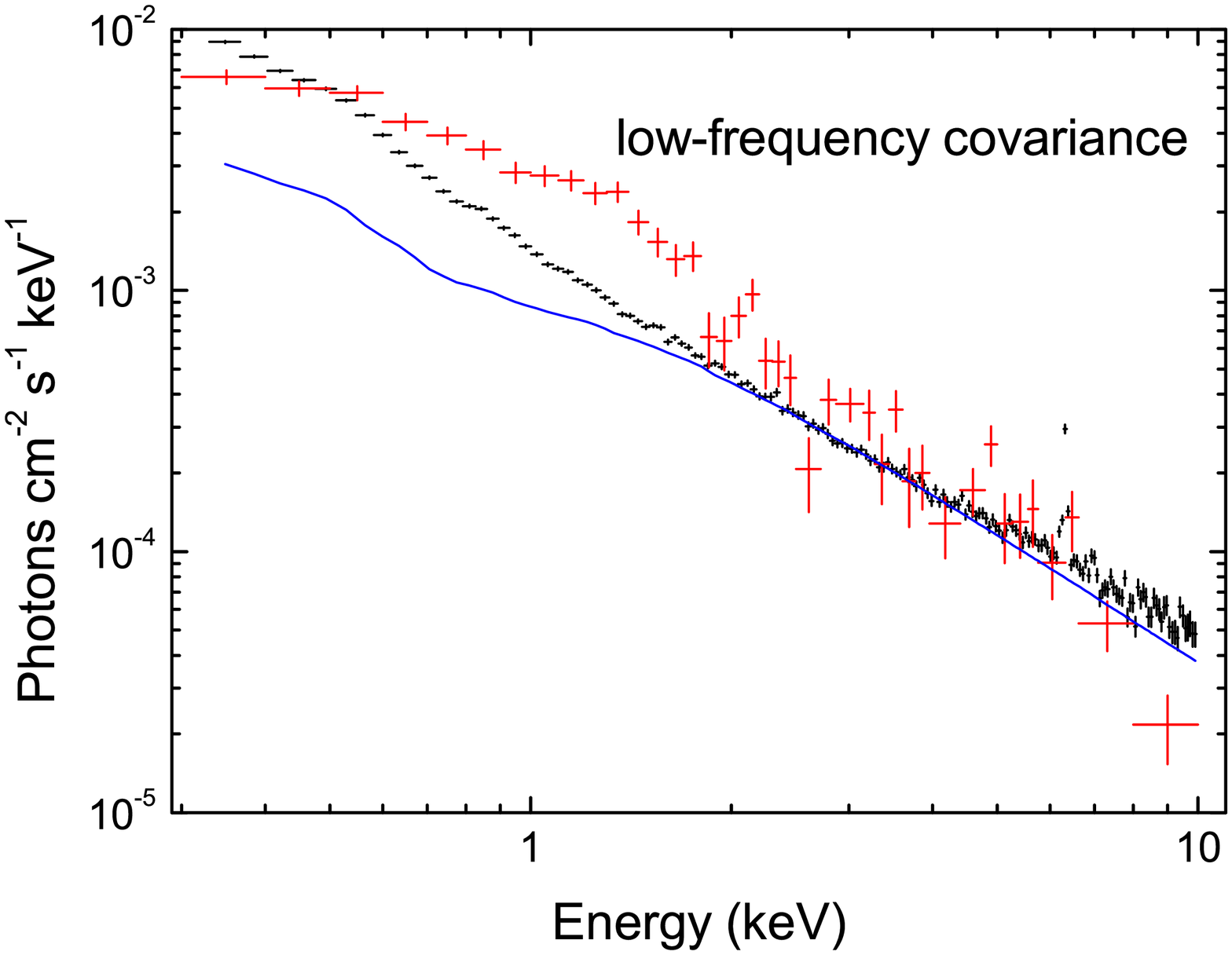}
  \includegraphics[scale=0.23,angle=0]{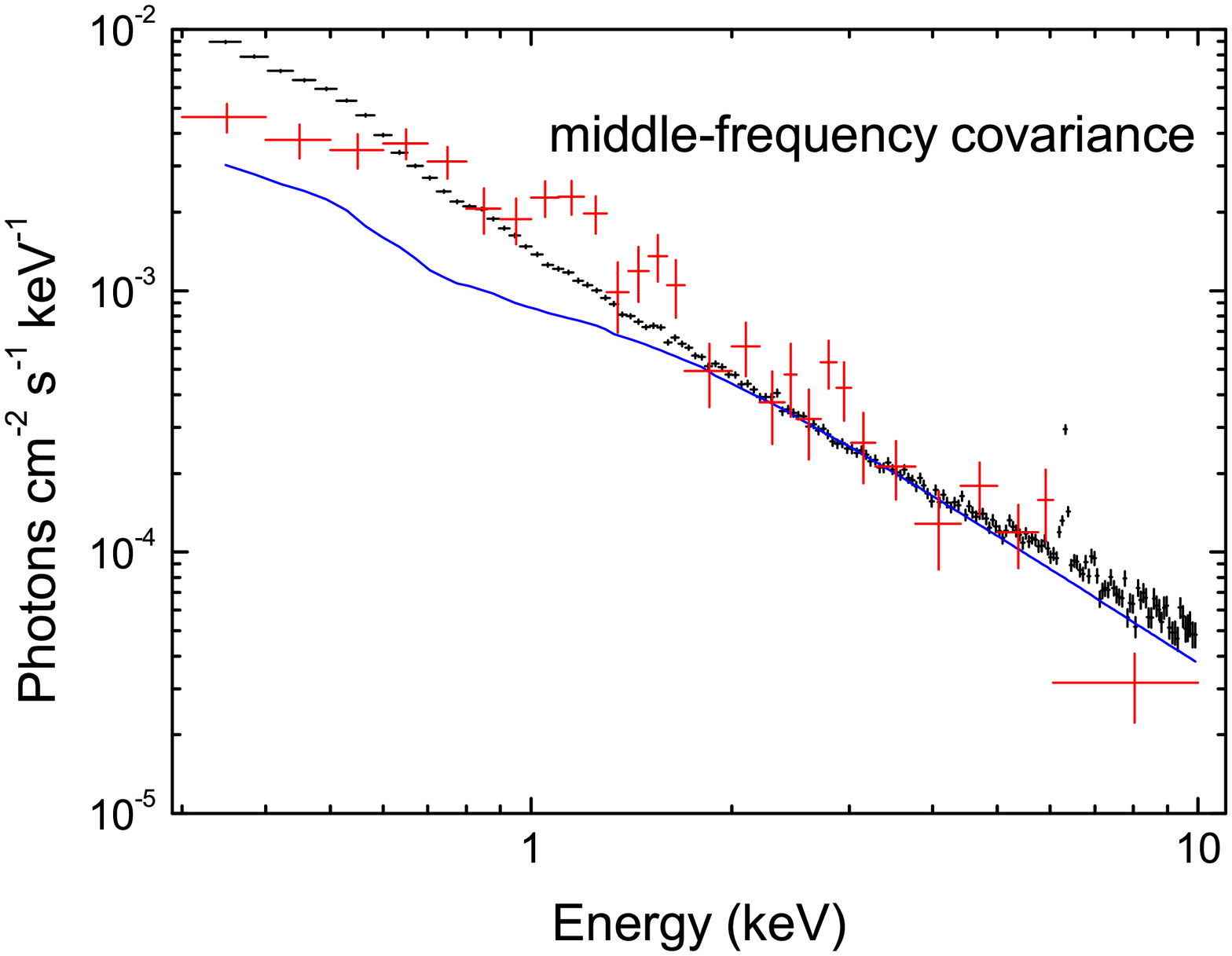}
  \includegraphics[scale=0.23,angle=0]{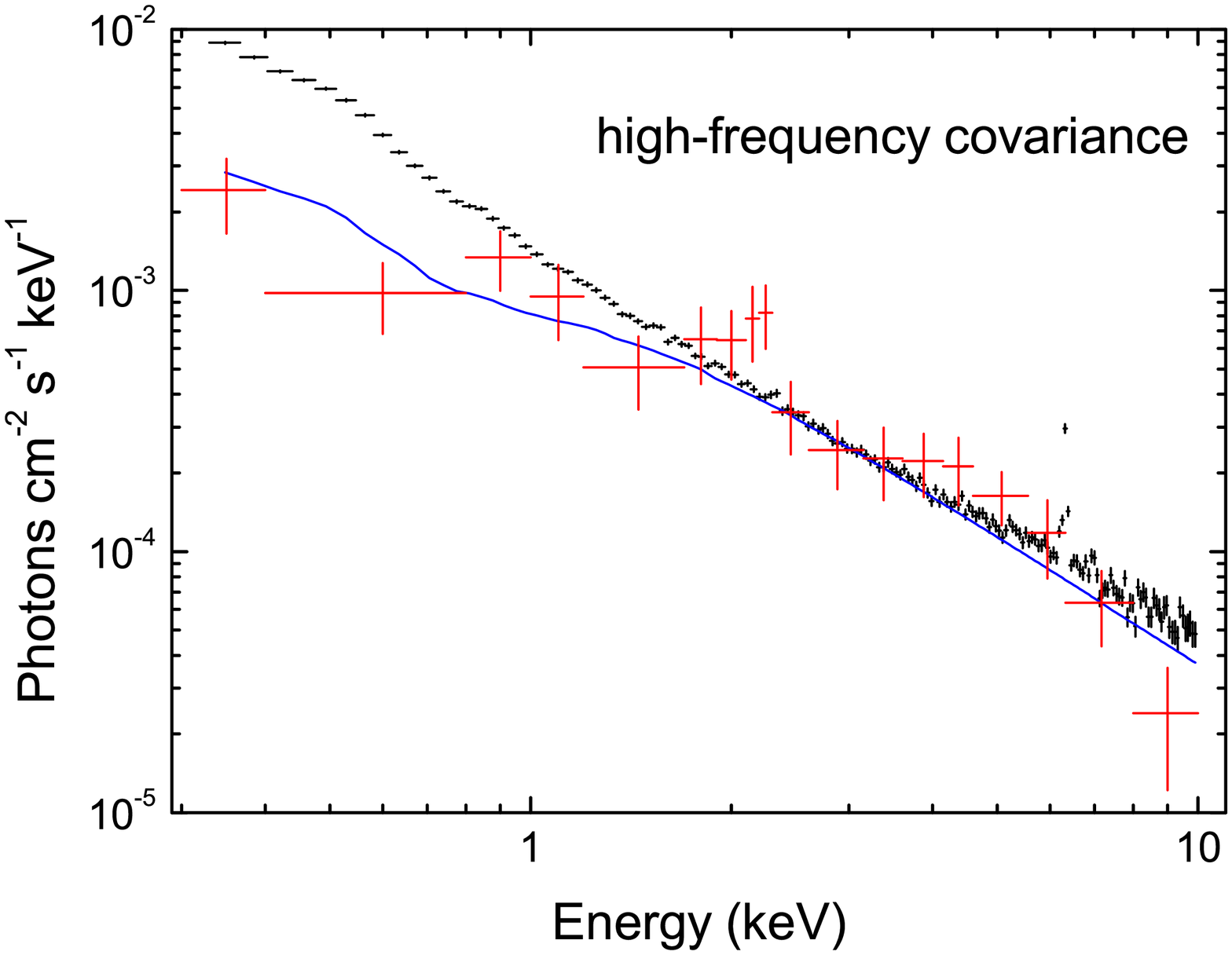}}
  \caption{The energy-dependent covariance of low, middle and high frequencies compared with time-average spectra for each energy bin from left to right panel. The black and red data points are the time-average spectrum and covariance data, respectively. The blue solid line represents the continuum radiation of hot corona.}
\end{figure}
\subsection{Conclusions}
We attempt to distinct between the warm corona and relativistic reflection models for the soft X-ray excess in ESO 362-G18 by using the simultaneous X-ray observation data of NuSTAR and XMM-Newton on Sep. 24th, 2016. We adopt five models including the warm, optically thick corona and relativistically blurred high-density reflection models, or their combination. The conclusions of this paper are as follows:

1. We find that the Double Warm Corona model can interpret both of time-average spectrum and variability spectrum (RMS and covariance spectrum) well, which include the low, middle and high temporal frequencies.

2. The warm corona component is the main dominant mechanism in the low and middle temporal frequencies or a larger scale. The radiation of hot corona and distant reflection is dominant at high temporal frequency or a compacter scale. The distant reflection could be made of some moderate scale neutral clumps which corresponds the high temporal frequency.

3. The radiation mechanism of hot corona maybe comes from the non-thermal comptonized process, such as the first order Fermi acceleration, to interpret the hard spectrum index obtained by the double warm model fitting.

4. If the double warm model is true, the broad Fe K$\alpha$ line is absent because the relativistic reflection component is excluded through the spectral analysis. So the broad Fe K$\alpha$ line origin of ESO 362-G18 need to be studies deeply in the future work.

\section*{Acknowledgments}

Thank the referee for their constructive comments. We acknowledge the financial supports from the National Natural Science Foundation of China 11133006, 11173054, the National Basic Research Program of China (973 Program 2009CB824800), and the Policy Research Program of Chinese Academy of Sciences (KJCX2-YW-T24).


\begin{thebibliography}{}
\bibitem[Ag{\'\i}s-Gonz{\'a}lez et al.(2014)]{2014MNRAS.443.2862A} Ag{\'\i}s-Gonz{\'a}lez, B., Miniutti, G., Kara, E., et al.\ 2014, \mnras, 443, 2862. doi:10.1093/mnras/stu1358

\bibitem[Arnaud(1996)]{arn96} Arnaud K.~A.,\ 1996, ASPC, 101, 17

\bibitem[Ballantyne(2020)]{2020MNRAS.491.3553B} Ballantyne, D.~R.\ 2020, \mnras, 491, 3553. doi:10.1093/mnras/stz3294

\bibitem[Balokovi{\'c} et al.(2019)]{2019RNAAS...3..173B} Balokovi{\'c}, M., Garc{\'\i}a, J.~A., \& Cabral, S.~E.\ 2019, Research Notes of the American Astronomical Society, 3, 173. doi:10.3847/2515-5172/ab578e

\bibitem[Begelman et al.(2015)]{2015ApJ...809..118B} Begelman, M.~C., Armitage, P.~J., \& Reynolds, C.~S.\ 2015, \apj, 809, 118. doi:10.1088/0004-637X/809/2/118

\bibitem[Bennert et al.(2006)]{2006A&A...456..953B} Bennert, N., Jungwiert, B., Komossa, S., et al.\ 2006, \aap, 456, 953. doi:10.1051/0004-6361:20065319

\bibitem[Blumenthal \& Gould(1970)]{1970RvMP...42..237B} Blumenthal, G.~R. \& Gould, R.~J.\ 1970, Reviews of Modern Physics, 42, 237. doi:10.1103/RevModPhys.42.237

\bibitem[\protect\citeauthoryear{Crummy et al.}{2006}]{2006MNRAS.365.1067C} Crummy J., Fabian A.~C., Gallo L., Ross R.~R., 2006, MNRAS, 365, 1067. doi:10.1111/j.1365-2966.2005.09844.x

\bibitem[Czerny \& Elvis(1987)]{1987ApJ...321..305C} Czerny, B. \& Elvis, M.\ 1987, \apj, 321, 305. doi:10.1086/165630

\bibitem[Dauser et al.(2014)]{2014MNRAS.444L.100D} Dauser, T., Garcia, J., Parker, M.~L., et al.\ 2014, \mnras, 444, L100. doi:10.1093/mnrasl/slu125

\bibitem[Dai et al.(2010)]{2010ApJ...709..278D} Dai, X., Kochanek, C.~S., Chartas, G., et al.\ 2010, \apj, 709, 278. doi:10.1088/0004-637X/709/1/278

\bibitem[\protect\citeauthoryear{Done et al.}{2007}]{2007ASPC..373..121D} Done C., Gierli{\'n}ski M., Sobolewska M., Schurch N., 2007, ASPC, 373, 121

\bibitem[Drury(2012)]{2012MNRAS.422.2474D} Drury, L.~O.\ 2012, \mnras, 422, 2474. doi:10.1111/j.1365-2966.2012.20804.x

\bibitem[\protect\citeauthoryear{Fabian et al.}{2002}]{2002MNRAS.331L..35F} Fabian A.~C., Ballantyne D.~R., Merloni A., Vaughan S., Iwasawa K., Boller T., 2002, MNRAS, 331, L35. doi:10.1046/j.1365-8711.2002.05419.x

\bibitem[Fermi(1949)]{1949PhRv...75.1169F} Fermi, E.\ 1949, Physical Review, 75, 1169. doi:10.1103/PhysRev.75.1169

\bibitem[Garc{\'\i}a \& Kallman(2010)]{2010ApJ...718..695G} Garc{\'\i}a, J. \& Kallman, T.~R.\ 2010, \apj, 718, 695. doi:10.1088/0004-637X/718/2/695

\bibitem[Garc{\'\i}a et al.(2014)]{2014ApJ...782...76G} Garc{\'\i}a, J., Dauser, T., Lohfink, A., et al.\ 2014, \apj, 782, 76. doi:10.1088/0004-637X/782/2/76

\bibitem[Garc{\'\i}a et al.(2016)]{2016MNRAS.462..751G} Garc{\'\i}a, J.~A., Fabian, A.~C., Kallman, T.~R., et al.\ 2016, \mnras, 462, 751. doi:10.1093/mnras/stw1696

\bibitem[Garc{\'\i}a et al.(2019)]{2019ApJ...871...88G} Garc{\'\i}a, J.~A., Kara, E., Walton, D., et al.\ 2019, \apj, 871, 88. doi:10.3847/1538-4357/aaf739

\bibitem[\protect\citeauthoryear{Ghosh \& Laha}{2020}]{2020MNRAS.497.4213G} Ghosh R., Laha S., 2020, MNRAS, 497, 4213. doi:10.1093/mnras/staa2259

\bibitem[Gronkiewicz \& R{\'o}{\.z}a{\'n}ska(2020)]{2020A&A...633A..35G} Gronkiewicz, D. \& R{\'o}{\.z}a{\'n}ska, A.\ 2020, \aap, 633, A35. doi:10.1051/0004-6361/201935033

\bibitem[Harrison et al.(2013)]{2013ApJ...770..103H} Harrison, F.~A., Craig, W.~W., Christensen, F.~E., et al.\ 2013, \apj, 770, 103. doi:10.1088/0004-637X/770/2/103

\bibitem[Hirose et al.(2006)]{2006ApJ...640..901H} Hirose, S., Krolik, J.~H., \& Stone, J.~M.\ 2006, \apj, 640, 901. doi:10.1086/499153

\bibitem[Jansen et al.(2001)]{2001A&A...365L...1J} Jansen, F., Lumb, D., Altieri, B., et al.\ 2001, \aap, 365, L1. doi:10.1051/0004-6361:20000036

\bibitem[Jiang et al.(2019a)]{2019MNRAS.489.3436J} Jiang, J., Fabian, A.~C., Dauser, T., et al.\ 2019, \mnras, 489, 3436. doi:10.1093/mnras/stz2326

\bibitem[Jiang et al.(2019b)]{2019MNRAS.484.1972J} Jiang, J., Fabian, A.~C., Wang, J., et al.\ 2019, \mnras, 484, 1972. doi:10.1093/mnras/stz095

\bibitem[\protect\citeauthoryear{Jiang et al.}{2021}]{2021MNRAS.501..916J} Jiang J., Cheng H., Gallo L.~C., Ho L.~C., Buisson D.~J.~K., Fabian A.~C., Harrison F.~A., et al., 2021, MNRAS, 501, 916. doi:10.1093/mnras/staa3737

\bibitem[Jin et al.(2009)]{2009MNRAS.398L..16J} Jin, C., Done, C., Ward, M., et al.\ 2009, \mnras, 398, L16. doi:10.1111/j.1745-3933.2009.00697.x

\bibitem[Jin et al.(2021)]{2021MNRAS.500.2475J} Jin, C., Done, C., \& Ward, M.\ 2021, \mnras, 500, 2475. doi:10.1093/mnras/staa3386

\bibitem[Kalberla et al.(2005)]{2005A&A...440..775K} Kalberla, P.~M.~W., Burton, W.~B., Hartmann, D., et al.\ 2005, \aap, 440, 775. doi:10.1051/0004-6361:20041864

\bibitem[Kallman \& Bautista(2001)]{2001ApJS..133..221K} Kallman, T. \& Bautista, M.\ 2001, \apjs, 133, 221. doi:10.1086/319184

\bibitem[Kara et al.(2013)]{2013MNRAS.434.1129K} Kara, E., Fabian, A.~C., Cackett, E.~M., et al.\ 2013, \mnras, 434, 1129. doi:10.1093/mnras/stt1055

\bibitem[\protect\citeauthoryear{Leighly}{1999}]{1999ApJS..125..317L} Leighly K.~M., 1999, ApJS, 125, 317. doi:10.1086/313287
\bibitem[\protect\citeauthoryear{Liu et al.}{2020}]{2020ApJ...896..160L} Liu H., Wang H., Abdikamalov A.~B., Ayzenberg D., Bambi C., 2020, ApJ, 896, 160. doi:10.3847/1538-4357/ab917a

\bibitem[Merloni et al.(2000)]{2000MNRAS.313..193M} Merloni, A., Fabian, A.~C., \& Ross, R.~R.\ 2000, \mnras, 313, 193. doi:10.1046/j.1365-8711.2000.03226.x

\bibitem[Mosquera et al.(2013)]{2013ApJ...769...53M} Mosquera, A.~M., Kochanek, C.~S., Chen, B., et al.\ 2013, \apj, 769, 53. doi:10.1088/0004-637X/769/1/53

\bibitem[Murphy \& Nowak(2014)]{2014ApJ...797...12M} Murphy, K.~D. \& Nowak, M.~A.\ 2014, \apj, 797, 12. doi:10.1088/0004-637X/797/1/12

\bibitem[Nandra et al.(1991)]{1991MNRAS.248..760N} Nandra, K., Pounds, K.~A., Stewart, G.~C., et al.\ 1991, \mnras, 248, 760. doi:10.1093/mnras/248.4.760

\bibitem[Nowak et al.(1999)]{1999ApJ...510..874N} Nowak, M.~A., Vaughan, B.~A., Wilms, J., et al.\ 1999, \apj, 510, 874. doi:10.1086/306610

\bibitem[Pal et al.(2016)]{2016MNRAS.457..875P} Pal, M., Dewangan, G.~C., Misra, R., et al.\ 2016, \mnras, 457, 875. doi:10.1093/mnras/stw009

\bibitem[Petrucci et al.(2001)]{2001MNRAS.328..501P} Petrucci, P.~O., Merloni, A., Fabian, A., et al.\ 2001, \mnras, 328, 501. doi:10.1046/j.1365-8711.2001.04897.x

\bibitem[\protect\citeauthoryear{Petrucci et al.}{2018}]{2018A&A...611A..59P} Petrucci P.-O., Ursini F., De Rosa A., Bianchi S., Cappi M., Matt G., Dadina M., et al., 2018, Astronomy and Astrophysics, 611, A59. doi:10.1051/0004-6361/201731580

\bibitem[Petrucci et al.(2020)]{2020A&A...634A..85P} Petrucci, P.-O., Gronkiewicz, D., Rozanska, A., et al.\ 2020, \aap, 634, A85. doi:10.1051/0004-6361/201937011

\bibitem[\protect\citeauthoryear{Porquet et al.}{2004}]{2004A&A...422...85P} Porquet D., Reeves J.~N., O'Brien P., Brinkmann W., 2004, Astronomy and Astrophysics, 422, 85. doi:10.1051/0004-6361:20047108

\bibitem[Reeves et al.(2008)]{2008MNRAS.385L.108R} Reeves, J., Done, C., Pounds, K., et al.\ 2008, \mnras, 385, L108. doi:10.1111/j.1745-3933.2008.00443.x

\bibitem[Tomsick et al.(2018)]{2018ApJ...855....3T} Tomsick, J.~A., Parker, M.~L., Garc{\'\i}a, J.~A., et al.\ 2018, \apj, 855, 3. doi:10.3847/1538-4357/aaaab1

\bibitem[Tzanavaris et al.(2021)]{2021arXiv210811971T} Tzanavaris, P., Yaqoob, T., LaMassa, S., et al.\ 2021, arXiv:2108.11971

\bibitem[Ursini et al.(2020)]{2020A&A...634A..92U} Ursini, F., Petrucci, P.-O., Bianchi, S., et al.\ 2020, \aap, 634, A92. doi:10.1051/0004-6361/201936486

\bibitem[Vaughan \& Nowak(1997)]{1997ApJ...474L..43V} Vaughan, B.~A. \& Nowak, M.~A.\ 1997, \apjl, 474, L43. doi:10.1086/310430

\bibitem[Vaughan et al.(2003)]{2003MNRAS.345.1271V} Vaughan, S., Edelson, R., Warwick, R.~S., et al.\ 2003, \mnras, 345, 1271. doi:10.1046/j.1365-2966.2003.07042.x

\bibitem[Verner et al.(1996)]{1996ApJ...465..487V} Verner, D.~A., Ferland, G.~J., Korista, K.~T., et al.\ 1996, \apj, 465, 487. doi:10.1086/177435

\bibitem[Wilkinson \& Uttley(2009)]{2009MNRAS.397..666W} Wilkinson, T. \& Uttley, P.\ 2009, \mnras, 397, 666. doi:10.1111/j.1365-2966.2009.15008.x

\bibitem[Wilms et al.(2000)]{2000ApJ...542..914W} Wilms, J., Allen, A., \& McCray, R.\ 2000, \apj, 542, 914. doi:10.1086/317016

\bibitem[Xu et al.(2021)]{2021ApJ...913...13X} Xu, Y., Garc{\'\i}a, J.~A., Walton, D.~J., et al.\ 2021, \apj, 913, 13. doi:10.3847/1538-4357/abf430

\bibitem[Yaqoob et al.(2016)]{2016MNRAS.462.4038Y} Yaqoob, T., Turner, T.~J., Tatum, M.~M., et al.\ 2016, \mnras, 462, 4038. doi:10.1093/mnras/stw1824

\bibitem[Zdziarski et al.(1996)]{1996MNRAS.283..193Z} Zdziarski, A.~A., Johnson, W.~N., \& Magdziarz, P.\ 1996, \mnras, 283, 193. doi:10.1093/mnras/283.1.193

\bibitem[Zoghbi et al.(2010)]{2010MNRAS.401.2419Z} Zoghbi, A., Fabian, A.~C., Uttley, P., et al.\ 2010, \mnras, 401, 2419. doi:10.1111/j.1365-2966.2009.15816.x

\bibitem[Zhong \& Wang(2013)]{2013ApJ...773...23Z} Zhong, X. \& Wang, J.\ 2013, \apj, 773, 23. doi:10.1088/0004-637X/773/1/23

\bibitem[{\.Z}ycki et al.(1999)]{1999MNRAS.309..561Z} {\.Z}ycki, P.~T., Done, C., \& Smith, D.~A.\ 1999, \mnras, 309, 561. doi:10.1046/j.1365-8711.1999.02885.x

\end{thebibliography}
\end{document}